\documentclass[%
 reprint,  
nofootinbib,
 amsmath,amssymb,
 aps,
 prd,
modulo,superscriptaddress
]{revtex4-2}

\usepackage[table,svgnames,dvipsnames]{xcolor}
\usepackage[mathlines]{lineno}
\usepackage[normalem]{ulem}
\usepackage{aas_macros}
\usepackage{subfigure}
\usepackage{graphicx}
\usepackage{graphics}
\usepackage{dcolumn}
\usepackage{bm}
\usepackage{cases}
\usepackage{booktabs}
\usepackage{comment}
\usepackage{multirow}
\usepackage{makecell}
\usepackage{siunitx}
\usepackage{tabularx}
\usepackage{xspace}
\usepackage{soul} 
\graphicspath{{figures/}}
\usepackage{fontawesome}
\usepackage{hyperref} 
\hypersetup{colorlinks=true,citecolor=blue,filecolor=blue,urlcolor=blue,linkcolor=blue}

\usepackage{orcidlink} 
\usepackage{hanging} 

\usepackage{adjustbox}

\renewcommand{\emph}[1]{\textit{#1}}
\definecolor{RoyalBlue}{rgb}{0.25,.41,.88}
\definecolor{WildStrawberry}{HTML}{EE2967}
\definecolor{RedWine}{rgb}{0.743,0,0}
\definecolor{bittersweet}{rgb}{1.0, 0.44, 0.37}
\definecolor{burntorange}{rgb}{0.8, 0.33, 0.0}
\definecolor{midnightgreen}{rgb}{0.0, 0.29, 0.33}
\definecolor{otherblue}{rgb}{0.20, 0.73, 0.92}
\definecolor{UltraViolet}{HTML}{6433FF}
\definecolor{seagreen}{HTML}{2E8B57}

\newcommand{\alphaiso}{{\alpha_{\rm iso}}}
\newcommand{\alphaap}{{\alpha_{\rm AP}}}
\newcommand{\alphaper}{\alpha_{\perp}}
\newcommand{\alphapar}{\alpha_{\parallel}}
\newcommand{\bgs}{{\tt BGS}\xspace}
\newcommand{\elgo}{{\tt ELG1}\xspace}
\newcommand{\elgt}{{\tt ELG2}\xspace}
\newcommand{\elgs}{{\tt ELG}s\xspace}
\newcommand{\elg}{{\tt ELG}\xspace}
\newcommand{\lrgo}{{\tt LRG1}\xspace}
\newcommand{\lrgt}{{\tt LRG2}\xspace}
\newcommand{\lrgth}{{\tt LRG3}\xspace}
\newcommand{\lrg}{{\tt LRG}\xspace}
\newcommand{\lrgs}{{\tt LRG}s\xspace}

\newcommand{\lrgelg}{{\tt LRG3$+$ELG1}\xspace}
\newcommand{\qso}{{\tt QSO}\xspace}
\newcommand{\desidrone}{{DESI DR1}\xspace}
\newcommand{\desidrtwo}{{DESI DR2}\xspace}
\newcommand{\abacussecond}{{\tt Abacus-2}\xspace}
\newcommand{\rascalc}{{\textsc {RascalC}}\xspace}

\usepackage[nameinlink,noabbrev]{cleveref}
\crefname{equation}{Eq.}{Eqs.}
\crefname{section}{Section}{Sections}
\crefname{figure}{Figure}{Figures}
\crefname{table}{Table}{Tables}
\crefname{appendix}{Appendix}{Appendices}
\Crefname{figure}{Figure}{Figures}
\Crefname{equation}{Equation}{Equations}
\Crefname{section}{Section}{Sections}
\Crefname{table}{Table}{Tables}

\newcommand{\lya}{Ly$\alpha$\xspace}

\newcommand{\DVrd}{D_\mathrm{V}/r_\mathrm{d}}

\newcommand{\DM}{D_\mathrm{M}}
\renewcommand{\DH}{D_\mathrm{H}}

\newcommand{\rd}{r_\mathrm{d}}

\newcommand{\hinvmpc}{\,h^{-1}{\rm Mpc}}
\newcommand{\hmpcinv}{\,h\,{\rm Mpc^{-1}}}

\newcommand{\Mpch}{\,h^{-1} \mathrm{Mpc}}

\begin{document}

\title{Validation of the DESI DR2 Measurements of Baryon Acoustic Oscillations from Galaxies and Quasars}


\author{U.~Andrade\orcidlink{0000-0002-4118-8236}}
\affiliation{Leinweber Center for Theoretical Physics, University of Michigan, 450 Church Street, Ann Arbor, Michigan 48109-1040, USA}
\affiliation{Department of Physics, University of Michigan, 450 Church Street, Ann Arbor, MI 48109, USA}

\author{E.~Paillas\orcidlink{0000-0002-4637-2868}}
\affiliation{Steward Observatory, University of Arizona, 933 N, Cherry Ave, Tucson, AZ 85721, USA}

\author{J.~Mena-Fern\'andez\orcidlink{0000-0001-9497-7266}}
\affiliation{Laboratoire de Physique Subatomique et de Cosmologie, 53 Avenue des Martyrs, 38000 Grenoble, France}

\author{Q.~Li\orcidlink{0000-0003-3616-6486}}
\affiliation{Department of Physics and Astronomy, The University of Utah, 115 South 1400 East, Salt Lake City, UT 84112, USA}

\author{A.~J.~Ross\orcidlink{0000-0002-7522-9083}}
\affiliation{Center for Cosmology and AstroParticle Physics, The Ohio State University, 191 West Woodruff Avenue, Columbus, OH 43210, USA}
\affiliation{Department of Astronomy, The Ohio State University, 4055 McPherson Laboratory, 140 W 18th Avenue, Columbus, OH 43210, USA}
\affiliation{The Ohio State University, Columbus, 43210 OH, USA}

\author{S.~Nadathur\orcidlink{0000-0001-9070-3102}}
\affiliation{Institute of Cosmology and Gravitation, University of Portsmouth, Dennis Sciama Building, Portsmouth, PO1 3FX, UK}

\author{M.~Rashkovetskyi\orcidlink{0000-0001-7144-2349}}
\affiliation{Center for Astrophysics $|$ Harvard \& Smithsonian, 60 Garden Street, Cambridge, MA 02138, USA}

\author{A.~P\'{e}rez-Fern\'{a}ndez\orcidlink{0009-0006-1331-4035}}
\affiliation{Max Planck Institute for Extraterrestrial Physics, Gie\ss enbachstra\ss e 1, 85748 Garching, Germany}

\author{H.~Seo\orcidlink{0000-0002-6588-3508}}
\affiliation{Department of Physics \& Astronomy, Ohio University, 139 University Terrace, Athens, OH 45701, USA}

\author{N.~Sanders\orcidlink{0009-0008-0020-2995}}
\affiliation{Department of Physics \& Astronomy, Ohio University, 139 University Terrace, Athens, OH 45701, USA}

% ...

\collaboration{DESI Collaboration}


\begin{abstract}

The Dark Energy Spectroscopic Instrument (DESI) data release 2 (DR2) galaxy and quasar clustering data represents a significant expansion of data from DR1, providing improved statistical precision in BAO constraints across multiple tracers, including bright galaxies (BGS), luminous red galaxies (LRGs), emission line galaxies (ELGs), and quasars (QSOs). In this paper, we validate the BAO analysis of DR2. We present the results of robustness tests on the blinded DR2 data and, after unblinding, consistency checks on the unblinded DR2 data. All results are compared to those obtained from a suite of mock catalogs that replicate the selection and clustering properties of the DR2 sample. We confirm the consistency of DR2 BAO measurements with DR1 while achieving a reduction in statistical uncertainties due to the increased survey volume and completeness.  We assess the impact of analysis choices, including different data vectors (correlation function vs. power spectrum), modeling approaches and systematics treatments, and an assumption of the Gaussian likelihood, finding that our BAO constraints are stable across these variations and assumptions with a few minor refinements to the baseline setup of the DR1 BAO analysis \cite{DESI2024.III.KP4}. We summarize a series of pre-unblinding tests that confirmed the readiness of our analysis pipeline, the final systematic errors, and the DR2 BAO analysis baseline. The successful completion of these tests led to the unblinding of the DR2 BAO measurements, ultimately leading to the DESI DR2 cosmological analysis, with their implications for the expansion history of the Universe and the nature of dark energy presented in the DESI key paper\cite{DESI.DR2.BAO.cosmo}.

\end{abstract}

\maketitle

\tableofcontents

\section{Introduction}\label{sec:intro}

Baryon acoustic oscillations (BAO) have emerged as one of the most robust and reliable probes for studying the expansion history of the Universe. These oscillations, imprinted in the large-scale distribution of galaxies and quasars, provide a cosmic standard ruler, enabling precise distance measurements across vast cosmic epochs \cite{1998ApJ...496..605E}. Their characteristic physical scale is precisely determined by Cosmic Microwave Background (CMB) measurements, allowing BAO to serve as a powerful tool for mapping the Universe’s expansion history. 
Over the past two decades, BAO measurements have become indispensable in cosmology, providing key constraints on the parameters governing the standard cosmological model, including the nature of dark energy \cite{2005NewAR..49..360E, 2010deot.book..246B,2013PhR...530...87W}. The first detection of the BAO peak by the Sloan Digital Sky Survey (SDSS) \cite{2005ApJ...633..560E} and the Two-degree Field Galaxy Redshift Survey (2dFGRS) \cite{2001MNRAS.327.1297P, 20052dFBAO} initiated the progress toward BAO becoming a cornerstone of observational cosmology. 

Subsequent advancements have further refined the precision of BAO as a cosmological probe, driven by both theoretical and observational progress. On the theoretical side, improvements in BAO reconstruction techniques \cite{Seo:2005ys, 2007ApJ...664..675E, 2012MNRAS.427.2132P} have significantly enhanced our ability to extract the primordial BAO signal. On the observational side, major spectroscopic surveys—including the WiggleZ Dark Energy Survey \cite{2011MNRAS.415.2892B, 2011MNRAS.418.1707B, 2014MNRAS.441.3524K}, the Baryon Oscillation Spectroscopic Survey (BOSS) \cite{2012MNRAS.427.3435A, 2014MNRAS.441...24A, 2017MNRAS.470.2617A}, and the extended BOSS (eBOSS) \cite{2018MNRAS.473.4773A, 2021PhRvD.103h3533A}—have steadily improved BAO constraints, solidifying its role as a foundational probe in cosmology.
 Beyond galaxy and quasar clustering, alternative approaches have been developed to measure BAO at higher redshifts and in different observational regimes. The \lya\ forest absorption in quasars spectra provides a means to probe BAO at $z > 2$, as demonstrated in \cite{White2003,2007PhRvD..76f3009M}. Meanwhile, photometric surveys have enabled transverse BAO measurements \cite{DESBAO2024}, though spectroscopic redshifts provide higher precision for a given number of tracers. These complementary techniques extend BAO constraints across a broader redshift range, enhancing our ability to probe cosmic expansion.

The Dark Energy Spectroscopic Instrument (DESI) represents a significant advancement in large-scale structure surveys, aiming to map the 3D distribution of galaxies and quasars over an unprecedented volume \cite{Snowmass2013.Levi}. Over its five-year survey (2021–2026), DESI plans to obtain spectra of about 40 million galaxies and quasars across 14,000 square degrees, covering a redshift range up to \( z \sim 3.5 \) \cite{DESI2016a.Science, LS.Overview.Dey.2019, DESI2023b.KP1.EDR}. DESI released its first cosmology analysis in April 2024 using Data Release 1 (DR1) \cite{DESI2024.I.DR1}, which included BAO measurements from galaxies and quasars \cite{DESI2024.III.KP4}, \lya\ forest \cite{DESI2024.IV.KP6} and its cosmological interpretation \cite{DESI2024.VI.KP7A}. This was followed by the release of full-shape measurements \cite{DESI2024.V.KP5} and their corresponding cosmological analysis \cite{DESI2024.VII.KP7B} in November. The DR1 BAO analysis introduced several advancements, including a catalog-level blinded analysis, a unified framework for all tracers, extensive systematic tests, and improvements in reconstruction and modeling. These measurements provided new constraints on the cosmic expansion history, reinforcing the robustness of BAO as a cosmological probe.

Building on this foundation, DESI Data Release 2 (DR2) \cite{DESI.DR2.DR2} provides an even more powerful dataset, expanding the survey area and achieving higher completeness. These improvements significantly enhance the statistical precision of BAO measurements. The combined precision of BAO constraints across six redshift bins improves from $\sim 0.52\%$ in DR1 to $\sim 0.24\%$ in DR2, more than doubling the measurement precision. However, this increased precision also heightens sensitivity to systematic effects, requiring even more rigorous validation procedures to ensure robust cosmological constraints.

This paper presents the validation process for the DR2 BAO measurements, detailing the unblinding tests, systematic quantifications, and finalization of the baseline analysis setup. While the general framework follows the DR1 BAO analysis, several refinements were introduced based on the tests conducted in this study. A critical aspect of this validation is the blinding scheme, which prevents confirmation bias. We employed the DR1 blinding pipeline \cite{KP3s9-Andrade, Brieden:2020, KP3s10-Chaussidon}, ensuring that key results remained blinded until all predefined validation steps were completed. The unblinded DR2 BAO constraints and their cosmological implications are presented in the DESI key paper \cite{DESI.DR2.BAO.cosmo}.

The structure of this paper is as follows: In \cref{sec:data}, we describe the DESI DR2 dataset, detailing the sample selection, clustering estimators, and blinding procedure. \cref{sec:mocks} introduces the mock catalogs used to validate the analysis. \cref{sec:modeling} outlines the modeling framework, including BAO reconstruction, parameter inference, and the construction of covariance matrices. \cref{sec:changes_y3} presents key methodological updates in DR2 and the treatment of systematic uncertainties. \cref{sec:results} details our main findings, comparing DR1 and DR2 BAO constraints, assessing robustness across multiple tests, and evaluating systematic effects. \cref{sec:post_unblinding} explores post-unblinding tests, including correlated systematics and the impact of fiducial cosmology assumptions. Finally, in \cref{sec:conclusions}, we summarize our findings and confirm the robustness of the DESI DR2 BAO analysis, establishing its readiness for cosmological interpretation in \cite{DESI.DR2.BAO.cosmo}.

\section{Data}\label{sec:data}

\begin{table*}
\begin{ruledtabular}
\centering
\begin{tabular}{lccrccc}
Tracer & No. of redshifts & Redshift range & $z_{\rm eff}$ & Area [deg$^2$] & $P_0(k=0.14)$ & $V_{\rm eff}$ (Gpc$^3$) \\\hline 
\texttt{BGS}  & 1,188,526  & $0.1 < z < 0.4$ & 0.295 & 12,355 & 7000  & 3.8 \\
\texttt{LRG1}  & 1,052,151 & $0.4< z < 0.6$ & 0.510 & 10,031 & 10000 & 4.9 \\
\texttt{LRG2}  & 1,613,562 & $0.6< z < 0.8$ & 0.706 & 10,031 & 10000 & 7.6 \\
\texttt{LRG3}  & 1,802,770 & $0.8< z < 1.1$ & 0.922 & 10,031 & 10000 & 9.8 \\
\texttt{ELG1}  & 2,737,573  & $0.8< z < 1.1$ & 0.955 & 10,352 & 4000  & 5.8 \\
\texttt{ELG2}  & 3,797,271  & $1.1< z < 1.6$ & 1.321 & 10,352 & 4000  & 2.7 \\
\texttt{QSO}   & 1,461,588  & $0.8< z < 2.1$ & 1.484 & 11,181 & 6000  & 2.7 \\
\end{tabular}
\end{ruledtabular}
\caption{Statistics for each DESI tracer type and redshift bin used for BAO measurements. Columns show the number of good redshifts, the redshift range, the effective redshift $z_{\rm eff}$, the survey area, the adopted value of $P_0(k=0.14)$ for each tracer, and the corresponding effective volume $V_{\rm eff}$. 
The effective redshift $z_{\rm eff}$ represents the weighted mean redshift of a sample, indicating where the measurement is most sensitive. The parameter $P_0(k=0.14)$ is a fiducial power spectrum value used to optimally weight the clustering signal for each tracer. The effective volume $V_{\rm eff}$ quantifies the statistical power of a given tracer sample in measuring BAO, accounting for the survey volume and the signal-to-noise contribution from the tracer number density and clustering amplitude. 
The survey area differs for each tracer due to priority vetoes (e.g., a \qso target can remove sky area from lower-priority samples) and small variations in imaging vetoes. The effective volume is computed following Eq. 2.2 in \cite{DESI2024.III.KP4}.}

\label{tab:sample}
\end{table*}

\subsection{DESI DR2}
The DESI Data Release 2 (DR2) \cite{DESI.DR2.DR2} dataset represents the culmination of nearly three years of observations using the DESI instrument \cite{DESI2016b.Instr,DESI2022.KP1.Instr}, from 14 May 2021 to 9 April 2024. Conducted at the Nicholas U. Mayall Telescope on Kitt Peak National Observatory, Arizona, DESI observes the spectra of 5,000 targets \cite{TS.Pipeline.Myers.2023} simultaneously within a 7 deg$^2$ field of view \cite{Corrector.Miller.2023}, utilizing robotic positioners \cite{FocalPlane.Silber.2023} to align optical fibers \cite{FiberSystem.Poppett.2024} with celestial coordinates. These fibers channel light to ten climate-controlled spectrographs, enabling precise redshift measurements critical for cosmological studies.

DESI observations use a dynamic time allocation strategy, which divides observing time into `bright time' and `dark time' programs based on observing conditions \cite{SurveyOps.Schlafly.2023}, with distinct target classes defined for each program. This ensures optimal data quality for different target classes, including galaxies, quasars, and stars. The DR2 dataset contains 6,671 dark time and 5,172 bright time tiles, each corresponding to specific sky positions and associated target sets. The sky coverage of these tiles can be seen in the top panel of Figure 2 in our companion paper \cite{DESI.DR2.BAO.cosmo}. 

In this work, we use large-scale structure (LSS) catalogs that were constructed based on the results of the DESI spectroscopic reduction \cite{Spectro.Pipeline.Guy.2023} and redshift estimation (\texttt{Redrock}; \cite{Redrock.Bailey.2024,Anand24redrock}) pipelines applied to the DR2 dataset in a homogeneous processing run denoted as `Loa'. The DESI LSS catalog pipeline is detailed in \cite{KP3s15-Ross}, with the specific choices (e.g., mask definitions, completeness weights, treatment for imaging systematics) applied to the version used for DESI DR2 BAO measurements mostly matching those described in \cite{DESI2024.II.KP3}. All new choices are described in section II.A of the companion paper \cite{DESI.DR2.BAO.cosmo}. 

The LSS catalogs are split into four distinct tracer types that apply different target selection criteria: the Bright Galaxy Sample (BGS;\cite{BGS.TS.Hahn.2023}), Luminous Red Galaxies (LRG; \cite{LRG.TS.Zhou.2023}), Emission Line Galaxies (ELG; \cite{ELG.TS.Raichoor.2023}), and Quasars (QSO; \cite{QSO.TS.Chaussidon.2023}). For BAO measurements, the samples are split into the same redshift bins as applied to DR1 \cite{DESI2024.II.KP3}. Basic details on the sample size in each redshift bin are provided in \cref{tab:sample}. Additionally, the data from the third redshift bin of the LRGs and the first one of the ELGs are combined into a single $0.8<z<1.1$ LRG+ELG sample, applying weights that optimally balance the contribution from each target type \cite{KP4s5-Valcin}. The DR2 LSS catalogs used in this work will be released publicly with DR2 version \texttt{v1.1/BAO}. Particular details on the characteristics of each sample and how this informs the tests presented throughout this work are introduced in the next subsection.

\subsection{Sample Characteristics and Splits}
\label{subsec:samples_and_splits}
 
\begin{figure*}
    \centering
    \includegraphics[width=0.9\textwidth]{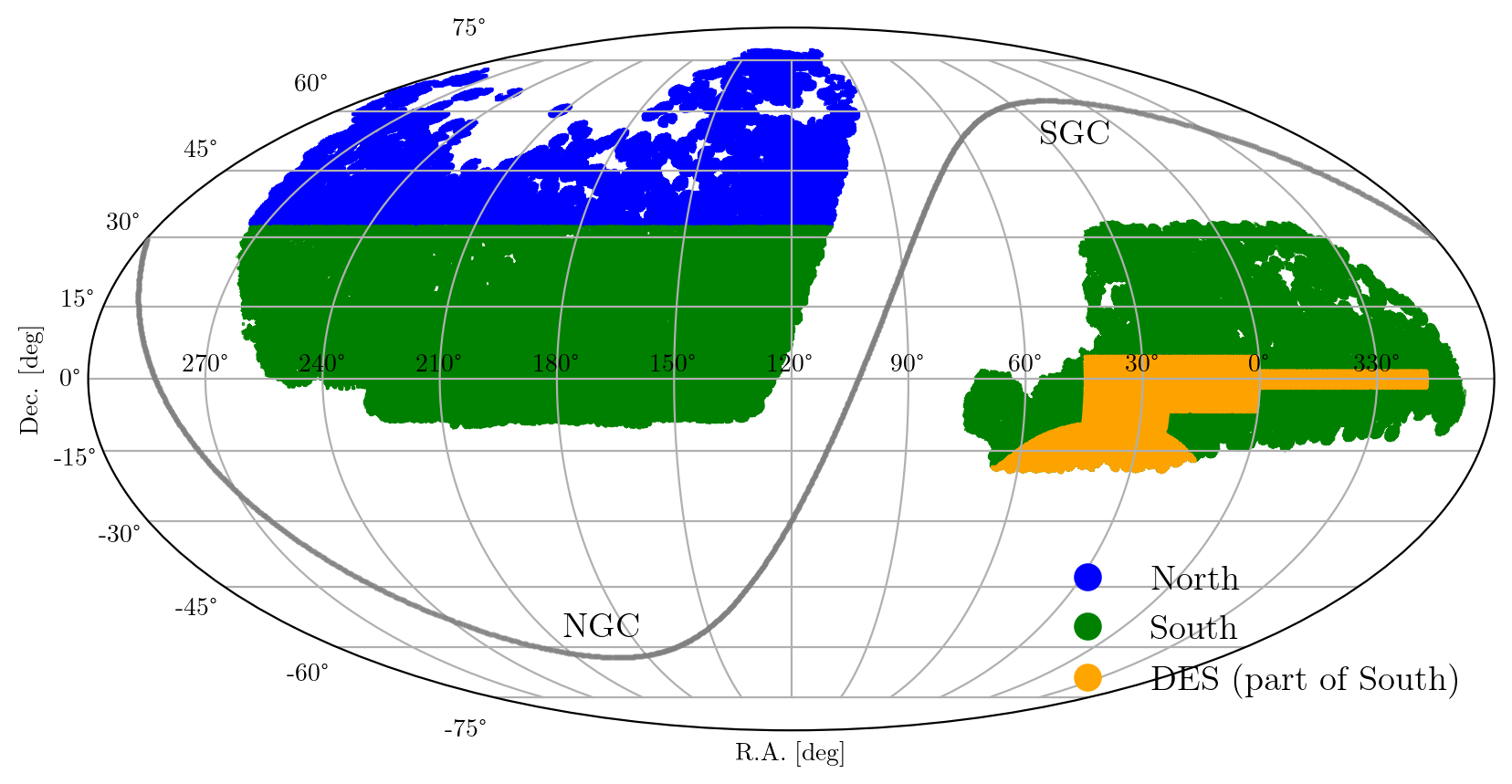} 
    \caption{
    The DESI DR2 footprint used in the large-scale structure (LSS) analysis, highlighting different survey regions. The survey is divided into the {North Galactic Cap (NGC)} and {South Galactic Cap (SGC)}, separated by the Galactic Equator (gray curve). Within these caps, the imaging footprint is further divided into {North} (blue) and {South} (green) regions based on the sources of imaging data. The {Dark Energy Survey (DES)} region (yellow) is a subset of the South region, characterized by deeper imaging and smaller point-spread functions compared to DECaLS (see text). These divisions play a key role in our robustness tests, ensuring that BAO measurements remain consistent across different survey regions.}
    \label{fig:footprint_regions}
\end{figure*}

The DESI survey includes four primary tracer samples: BGS, LRGs, ELGs and QSOs. Each of these tracers has distinct characteristics in terms of number density, redshift distribution, and clustering amplitude, which influence their roles in cosmological analyses.

\textbf{BGS:} This sample is sub-selected from the DESI \texttt{BGS\_BRIGHT} selection, which is flux-limited at $r < 19.5$. Due to this flux limit, the nominal \texttt{BGS\_BRIGHT} sample exhibits a number density that varies significantly with redshift. To mitigate this redshift evolution and obtain a more uniform sample, we apply an absolute magnitude cut of $M_r < -21.35$. This adjustment ensures a more consistent number density across the redshift range. For additional details, see the companion paper \cite{DESI.DR2.BAO.cosmo}.

\textbf{LRG:} This sample maintains an approximately constant number density of $5 \times 10^{-4}\, h^{3}\, {\rm Mpc}^{-3}$ in the redshift range $0.4 < z < 0.8$. Beyond $z = 0.8$, the number density decreases due to a $z$-band flux limit. Due to its large survey volume and strong clustering signal relative to shot noise, the LRG sample provides the highest effective volume ($V_{\rm eff} = 9.8$ Gpc$^3$), making it a key contributor to the constraining power of cosmological analyses.

\textbf{ELG:} This sample has a comparable number density to the LRGs for $z = 0.8$\footnote{The relative number density between ELGs and LRGs depends on the completeness in a given area, leading to variations in observed values across different regions of the survey footprint.}, where the LRG number density begins to sharply drop and extends up to $z=1.6$. However, the clustering amplitude of ELGs is approximately one-third that of LRGs, which reduces their constraining power despite their greater abundance. The ELG sample is also more susceptible to systematic variations in target density caused by imaging systematics. Nevertheless, it has been demonstrated that these systematics do not significantly affect BAO measurements \cite{KP3s2-Rosado}. Imaging systematics are mitigated by applying weights to the galaxy samples to nullify trends with imaging properties.

\textbf{QSO:} The quasar sample is the sparsest among the DESI tracers, with a number density of approximately $2.5 \times 10^{-5}\, h^{3}\, {\rm Mpc}^{-3}$. This low number density results in the sample being shot-noise-dominated, which limits the precision of clustering measurements. Despite this, QSOs provide valuable information for probing the large-scale structure due to the large cosmic volume spanned by their high redshift range.

\subsubsection{Spatial and Imaging-Based Splits}

Beyond these intrinsic characteristics, the spatial distribution of the samples across the survey footprint introduces additional factors to account for. For all tracer types, there are distinct survey regions of interest, as shown in \cref{fig:footprint_regions}. These regions are defined by two key divisions:  
\begin{itemize}
    \item {The Galactic Cap (GC) division}: Separates the {North Galactic Cap (NGC)} and {South Galactic Cap (SGC)} based on spatial location relative to the Galactic Equator.  
    \item {The Imaging-Based Division}: Defines the {North} and {South} regions based on the sources of imaging data used for DESI targeting.
\end{itemize}

We describe these divisions in more detail below.

\textbf{The Galactic Cap Division:} The DESI large-scale structure (LSS) catalogs are divided into {NGC} and {SGC} for convenience, as these regions are {spatially disjoint and sufficiently separated} across the Galactic Equator. Due to this separation, clustering measurements can be computed {independently} for each region and then combined into a fiducial measurement {without loss of information}.  

The NGC footprint contains {70\% of the DR2 bright-time (BGS) sample} and {66\% of the dark-time (LRGs, ELGs, QSOs) sample}. One of our robustness tests evaluates the impact of restricting the clustering analysis to the NGC footprint.

\textbf{The Imaging-Based Division:} The optical imaging data used for DESI targeting comes from two distinct sources, defining the {North} and {South} imaging regions:
\begin{itemize}
    \item The {North} region contains photometric data from the {BASS+MzLS} surveys \citep{BASS.Zou.2017}. All NGC data with declination greater than {32.375 degrees} is part of the North imaging region (\cref{fig:footprint_regions}, blue region).
    \item The {South} region is entirely observed with {DECam} \cite{decam}. Approximately {$1130$ deg$^2$} in the SGC is targeted by the Dark Energy Survey (DES \cite{DES_overview} (\cref{fig:footprint_regions}, yellow region). The remaining area ({$\approx 3580$ deg$^2$} in the SGC and {$\approx 5770$ deg$^2$} in the NGC) is covered by the {DECaLS} survey \cite{LS.Overview.Dey.2019} (\cref{fig:footprint_regions}, green region)
    \item Within the South region, the DES forms a distinct subset. DES typically provides {deeper imaging and smaller point-spread functions} compared to DECaLS.
\end{itemize}

The North imaging region covers {31\% of the DR2 bright-time footprint} and {20\% of the dark-time footprint}, while the DES region covers {7\% of the bright-time footprint} and {9\% of the dark-time footprint}.

Since these imaging differences may introduce {systematic effects}, our analysis includes robustness tests to evaluate the consistency of BAO measurements across these imaging regions. Thus, we will examine the effects of excluding the DES and North regions on the clustering measurements across all samples. See \cite{KP3s15-Ross} for more details on how the LSS catalogs are processed accounting for the details of these imaging regions.

\subsubsection{Intrinsic Galaxy Property Splits}

In addition to regions-based splits, we perform robustness tests based on two types of intrinsic properties: \textit{stellar mass splits} and \textit{magnitude splits}. These tests help assess potential biases in BAO measurements arising from the dependence of galaxy clustering on stellar mass and luminosity, as both properties are correlated with galaxy bias, which affects how galaxies trace the underlying large-scale structure.

Stellar mass splits are based on estimates from the stellar mass catalog described in Appendix C of \cite{LRG.TS.Zhou.2023}. Stellar masses are derived using $g-r$, $r-z$, $z-W_1$, and $W_1-W_2$ colors from the Legacy Survey DR9, processed with a Random Forest model \cite{2001MachL..45....5B} trained on the S82-MGC galaxy sample \cite{2015ApJS..221...15B}. A key update from \cite{LRG.TS.Zhou.2023} is the use of {\texttt{kibo-v1} spectroscopic redshifts} instead of photometric redshifts, significantly improving the accuracy of the mass estimates. For LRGs with DESI redshifts, the stellar mass uncertainty calibrated against the training sample is $\sigma_{\rm NMAD}=0.086$, making them well-suited for mass-split analyses. BGS are likely to be similar, so we perform stellar mass splits for BGS and LRGs. In contrast, the scatter for ELGs is significantly larger, reducing the reliability of their stellar mass estimates; thus, we do not apply stellar mass splits to the ELG sample. We divide the BGS sample into two stellar mass bins while dividing the LRG sample into three bins.

Magnitude splits are applied to tracers with sufficiently high number densities, including the BGS, LRG and ELG samples. 

For BGS, LRGs and ELGs, magnitude splits are performed using extinction-corrected $r$-band, $W_1$-band and $g$-band magnitudes, respectively. 
We divide the BGS and ELG samples into two magnitude bins while dividing the LRGs into three bins. These splits help to probe the clustering dependence on galaxy brightness, particularly in cases like ELGs where stellar mass splits are unreliable.

\begin{figure}
    \centering
    \includegraphics[width=1\linewidth]{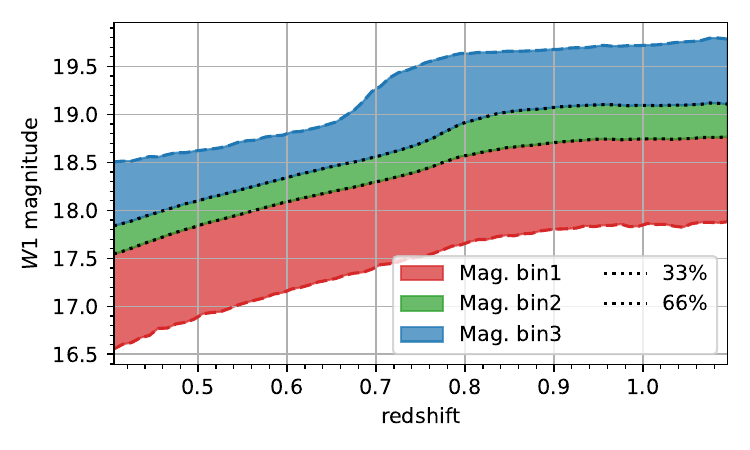}
    \caption{Illustration of the magnitude-based sample split applied to the DESI Large-Scale Structure galaxy samples. The plot shows the W1-band magnitudes as a function of redshift, with dashed lines indicating the 33\% and 66\% percentile thresholds used to define three magnitude bins. This binning strategy helps assess the impact of galaxy brightness on clustering measurements while ensuring a consistent redshift distribution within each bin.}
    \label{fig:split}
\end{figure}

To ensure a consistent redshift distribution across subsamples in all stellar mass and magnitude splits, we apply percentile-based thresholds within small redshift intervals. Specifically, for the BGS and ELG samples, we use the median (50\%) value of stellar mass and extinction-corrected magnitude in each $\Delta z = 0.01$ interval to define the splits. For LRGs, we refine this approach by using the 33\% and 66\% percentiles, dividing the sample into three bins, as shown in \cref{fig:split}. Additionally, the same binning strategy is applied to the corresponding random catalogs to maintain a matched spatial and redshift distribution, ensuring that systematic effects do not bias the clustering measurements.

The primary motivation behind these splits is to assess and mitigate any potential biases in the DESI DR2 BAO measurements. Spatial and imaging-based splits help evaluate the impact of survey selection effects, while intrinsic property splits provide insight into galaxy bias variations within the same sample. By performing these robustness tests, we ensure that our DESI DR2 BAO constraints remain unbiased and reliable for cosmological interpretation.

\subsection{Clustering Measurements}

To quantify the large-scale structure of the Universe, we measure the two-point correlation function and the power spectrum, which provide complementary descriptions of galaxy clustering. The correlation function characterizes clustering in configuration space, while the power spectrum offers a Fourier-space perspective, with both methods capturing the same underlying information but differing in sensitivity to systematics and scale-dependent features. 
To maintain consistency with previous analyses, we follow the same code settings as in DR1 \citep{DESI2024.II.KP3}, applying well-tested estimators to both correlation function and power spectrum measurements. Additionally, various observational weights are applied to correct for systematics, including imaging systematics weights, redshift failure weights, and the Feldman-Kaiser-Peacock (FKP) weight, which optimally down-weights galaxies in high-density regions to reduce sample variance \citep{FKP}. These weights are consistently incorporated in both configuration-space and Fourier-space clustering measurements.

\subsubsection{Correlation Function Estimator}
We use the Landy-Szalay estimator \cite{Landy-Szalay:1993} to compute the two-point correlation function, which measures the excess probability of finding two galaxies separated by a distance $s$ and cosine of the angle relative to the line of sight, $\mu$. The estimator is defined as:
\begin{equation} \label{eq:landy-szalay}
\hat{\xi}(s,\mu) = \frac{DD(s,\mu) - 2DR(s,\mu) + RR(s,\mu)}{RR(s,\mu)},
\end{equation}
where:
\begin{itemize}
\item $DD(s,\mu)$ represents the weighted number of galaxy-galaxy pairs,
\item $DR(s,\mu)$ is the number of galaxy-random pairs,
\item $RR(s,\mu)$ denotes the number of random-random pairs.
\end{itemize}
The random catalog is constructed to match the survey footprint and the selection function to mitigate systematic effects. From the correlation function, we compute multipole moments (monopole $\ell=0$, quadrupole $\ell=2$, hexadecapole $\ell=4$) using Legendre polynomials:
\begin{equation}
\hat{\xi}_\ell(s) = \frac{2\ell + 1}{2} \int_{-1}^{1} d\mu \, \hat{\xi}(s,\mu) \mathcal{L}_\ell(\mu).
\end{equation}
For post-reconstruction measurements (\cref{subsec:reconstruction}), we use a modified version of the Landy-Szalay estimator, as in \cite{Padmanabhan2012}. We perform the correlation function measurements using a modified version of the \texttt{CorrFunc} pair counting code \citep{Sinha:2019reo}, implemented in \texttt{pycorr}\footnote{\url{https://github.com/cosmodesi/pycorr/}}. We use a bin width of $4 \hinvmpc$ in $s$ and 240 $\mu$ bins from -1 to 1.

\subsubsection{Power Spectrum Estimator}
In Fourier space, we estimate the power spectrum using the FKP estimator \citep{FKP}, which accounts for the effects of survey geometry. The weighted galaxy fluctuation field is given by:
\begin{equation}
F(\mathbf{r}) = n_d(\mathbf{r}) - \alpha n_r(\mathbf{r}),
\end{equation}
where $n_d(\mathbf{r})$ and $n_r(\mathbf{r})$ are the weighted galaxy and random number densities, with the latter having a total weighted number $1/\alpha$ times that of the data catalog. This expression is appropriately modified for post-reconstruction measurements as in \cite{White:2015eaa}.

Power spectrum multipoles are computed using the Yamamoto estimator \citep{Yamamoto2006}, which efficiently handles line-of-sight variations across the survey volume:
\begin{align}
\hat{P}_{\ell}(k) &= \frac{2 \ell + 1}{A N_{k}} \sum_{\Vec{k} \in k} \sum_{\Vec{r}_{1}} \sum_{\Vec{r}_{2}} F(\Vec{r}_{1}) F(\Vec{r}_{2}) \mathcal{L}_{\ell}(\hat{k} \cdot \hat{\eta}) \nonumber \\ 
&\quad \times e^{i \Vec{k} \cdot (\Vec{r}_{2} - \Vec{r}_{1})} - \mathcal{N}_{\ell}.
\end{align}
The summations extend over all galaxy pairs with positions \(\Vec{r}_{1}\) and \(\Vec{r}_{2}\), as well as over wavevectors \(\Vec{k}\) within the given \(k\)-bin. \(\hat{\eta}\) is the line-of-sight direction. \(\mathcal{N}_{\ell}\) accounts for the shot-noise correction applied to the monopole component, while \(A\) serves as the normalization factor. \(N_k\) corresponds to the total number of modes contributing to the estimation in the given \(k\)-bin.

To efficiently implement the Yamamoto estimator using Fast Fourier Transforms (FFTs), we follow the approach described in \cite{Bianchi:2015oia}, which allows for rapid computation of the power spectrum multipoles in large-volume surveys. This methodology is incorporated into the \texttt{nbodykit} framework \citep{Hand:2017pqn} and is further optimized in \texttt{pypower}\footnote{\url{https://github.com/cosmodesi/pypower}}, which we use for our analysis.

We follow the same code settings as in DR1 \citep{DESI2024.II.KP3}. While our fiducial results are based on the correlation function, following the choice of DR1 based on the performance of the analytical covariance matrix, the power spectrum analysis serves as a cross-check, and its results are presented in \cref{app:fourier}.

\subsection{Blinding}\label{sec:blinding}

Blinding is an integral part of the validation process, designed to prevent confirmation bias. Following the DESI DR1 blinding scheme \cite{KP3s9-Andrade}, DESI DR2 BAO measurements were kept blinded during the validation process, including the determination of the systematic error budget (see \cref{sec:sys_error}). Unblinding was only performed after completing all pre-defined validation tests listed in \cref{tab:unblinding_tests}. For completeness, we briefly summarized the blinding scheme, for further details see \cite{KP3s9-Andrade, Brieden:2020, KP3s10-Chaussidon}.

We adopt a catalog-level blinding that modifies galaxy redshifts and weights to mimic the effects of a different underlying cosmology. This approach allows collectively blinding three key observables: BAO, redshift-space distortions (RSD) and primordial non-Gaussianity (PNG).

\begin{itemize}

    \item \textbf{Blinding the expansion rate:} Galaxy redshifts are modified by first converting the observed redshifts to comoving distances using a blind cosmology, then reconverting them to blinded redshifts based on the fiducial cosmology. This process introduces an unknown dilation of all scales, in both the radial and transverse directions. As a result, both the isotropic and anisotropic BAO signals are blinded, ensuring that the inferred expansion rate remains concealed during validation.

    \item \textbf{Blinding RSD (RSD shift):} Galaxy redshifts are perturbed based on the local density and peculiar velocity field, mimicking changes to the growth rate of structure. 
    This will distort anisotropies in the power spectrum, therefore blinding RSD.

    \item \textbf{Blinding PNG (Scale-Dependent Bias):} The scale-dependent bias signature of PNG \citep{Dalal:2007cu} is mimicked by applying an additional weight to each galaxy. This weight is computed from the real-space density field ($\delta^r$), reconstructed from the observed galaxy distribution, and incorporates a blinded value of \( f_{\mathrm{NL}} \). This modification induces a scale-dependent bias effect, ensuring that PNG constraints remain blinded in the large-scale power spectrum.

\end{itemize}

We adopt the Planck-2018 results \citep{Planck-2018-cosmology} as our fiducial cosmology. The blind cosmology is randomly selected following the methodology outlined in \cite{KP3s9-Andrade}, introducing small shifts in the BAO scaling parameters, the growth of structure, and the PNG parameter \( f_{\mathrm{NL}} \). These controlled modifications ensure that cosmological constraints remain blinded throughout validation, preventing any confirmation bias in the analysis. The fiducial parameters are:

\begin{equation}\nonumber
\begin{split}
\omega_b &= 0.02237, \quad \omega_{\rm cdm} = 0.12, \quad h = 0.6736, \\
A_s &= 2.083 \times 10^{-9}, \quad n_s = 0.9649, \quad \quad N_{\rm ur} = 2.0328, \\
N_{\rm ncdm} &= 1.0, \quad \omega_{\rm ncdm} = 0.0006442, \quad w_0 = -1, \quad w_a = 0.
\end{split}
\label{eq:fiducial-cosmo}
\end{equation}

Here, \(\omega_b = \Omega_b h^2\) and \(\omega_{\rm cdm} = \Omega_{\rm cdm} h^2\) are the physical baryon and cold dark matter densities, where \(\Omega_b\) and \(\Omega_{\rm cdm}\) represent the corresponding density parameters. $A_s$ and $n_s$ are the amplitude and tilt of the primordial power spectrum, $N_{\rm ur}$ is the effective number of relativistic species, $N_{\rm ncdm}$ and $\omega_{\rm ncdm}$ are the number and physical density of massive neutrinos, and $w_0$ and $w_a$ give the present-day value and time evolution of the dark energy equation of state under the CPL parameterization \citep{Chevallier:2001,Linder2003}. These values define the baseline cosmological model used in our analysis.

The final covariance matrices for the BAO analysis were computed after unblinding by applying the same predefined procedure to the unblinded catalogs and clustering measurements. This step was necessary because accurately calibrating the covariance matrices requires knowledge of the actual clustering signal in the data. While the blinding procedure ensured that all systematic and methodological choices were made independently of the final BAO results, refining the covariance after unblinding allowed us to incorporate the actual statistical properties of the DESI DR2 dataset. This ensured that the final measurements remained robust and appropriately reflected the uncertainties in the analysis.

\section{Mocks}\label{sec:mocks}

Mock catalogs are synthetic datasets that simulate the large-scale distribution of galaxies and quasars in the universe based on theoretical models. These mocks are designed to closely replicate real observations, reproducing the expected clustering statistics, survey geometry, and observational effects. By incorporating known cosmological parameters and realistic noise, mocks provide a controlled environment to test analysis pipelines, evaluate systematic uncertainties, and ensure the robustness of measurements. They play a critical role in large-scale structure surveys like DESI, where the precision of the data demands rigorous validation before drawing scientific conclusions.

For the DESI DR2 BAO analysis, we relied on the \texttt{AbacusSummit} 2nd-generation mocks (\abacussecond DR2 mocks). The `cutsky' mocks we used, which include angular sky coordinates and redshifts over the entire (planned) DESI footprint, are the same as used for DR1 analyses and are described in \cite{DESI2024.II.KP3}. For this analysis, we have applied the `altmtl' method described in \cite{DESI2024.II.KP3, KP3s7-Lasker} to match them to the DESI DR2 footprint and observational completeness. We then passed the outputs through the DESI LSS pipeline to match the selection properties of the different target samples, following the \texttt{kibo-v1} specification. Percent-level adjustments to the assumed redshift failure fractions were applied to better match the final $n(z)$ to the DR2 data. A more in-depth description of the methodology will be provided in \cite{KP3s8-Zhao}. 

The input boxes for these mocks are based on halo catalogs from the \texttt{AbacusSummit} suite of base simulations \citep{AbacusSummit}, populated with galaxies using a halo occupation distribution (HOD) framework, following the methodology of \citep{EDR_HOD_LRGQSO2023} for dark-time tracers and \cite{Smith:2023jqs} for BGS. The HOD models used in these mocks were calibrated to the DESI SV3 data, as detailed in \cite{rocher2023desi,yuan2024desi,smith2024generating}. 

For the dark-time tracers, the DR2 mocks are identical to those used in DR1, except for updates in the footprint mask to reflect the DR2 survey geometry. 

A detailed comparison between the clustering of these mocks and that of DESI tracers is presented in \cite{DESI2024.II.KP3}, where it was found that the agreement is at or better than 2\% in the inferred real-space over-density field for all tracers and redshift bins.

The BGS mocks, in contrast, required changes for DR2 related to the fact that we have changed the absolute magnitude selection for the DR2 BGS sample. The BGS mocks include absolute magnitudes and simulate the entire \texttt{BGS\_BRIGHT} and \texttt{BGS\_FAINT} samples. However, it was found that no absolute magnitude cut applied to the mock \texttt{BGS\_BRIGHT} sample could reproduce the observed number density above $z=0.35$ (whereas the absolute magnitude cut applied to DR1 does produce a good match between data and mock so this was not an issue in DR1). To match the number density (and/or the total number of included redshifts), it was thus required to include data from the mock \texttt{BGS\_ANY} sample with a redshift-dependent cut on the absolute magnitude. The resulting sample matches the DR2 sample well in terms of the redshift distribution (and total number), but has a $\sim10\%$ higher clustering amplitude than observed in the data. While this discrepancy does not affect the BAO scale measurement—since the amplitude difference is absorbed into the free bias parameters of the BAO model—it is an important consideration for studies beyond BAO. A detailed examination of the clustering comparison between BGS mocks and DR2 data is ongoing, and upcoming analyses will incorporate improved mocks that better match the observed data, addressing previous discrepancies. A full description of these updated mocks will be presented in a forthcoming study.

The mock catalogs enabled comparisons between the observed clustering and theoretical expectations, allowing us to validate the reconstruction methodology and assess systematic uncertainties in the BAO fitting process. A total of 25 realizations were analyzed for all tracers. These realizations were created from 25 realizations of simulation boxes with side lengths of 2 $h^{-1}$ Gpc. For the BGS mocks, the entire DR2 sample fits within the simulation volume. However, for the dark time tracers, replication of the box is required. The replicated boxes are the same as those used in DR1, which simply tiled the boxes 3x3x3 to produce boxes with side lengths 6 $h^{-1}$ Gpc. Thus, some added correlation is expected both for the clustering measurements within any redshift bin and between redshift bins. This may decrease the expected precision of the results from any individual realization (as the unique volume should determine this) and does decrease our ability to consider the results from all redshift bins together in an ensemble sense. The results from the mocks remain highly valuable, however, for testing for any biases in the recovery of BAO parameters and examining where DESI results fall within the distribution of the 25 mocks. Even with the replication issues, recovering a result that is within the distribution implies it is at least a few per cent likely.

The mocks provided an essential benchmark for exploring potential systematic effects, ensuring that the final results were unbiased.

\section{Modeling}\label{sec:modeling}

Accurately extracting the BAO signal from the clustering of galaxies and quasars requires a comprehensive modeling framework that accounts for various physical and observational effects. This process involves combining theoretical predictions, observational data, and statistical methods to interpret the observed signal while minimizing systematic biases. Key components of the modeling include the treatment of geometric distortions, the impact of the comoving sound horizon as a standard ruler, and the role of reconstruction in mitigating non-linear effects.

To capture the dependence of the observed clustering on these effects, the signal is modeled in terms of dilation parameters that encapsulate the effects of geometric distortions and sound horizon rescaling. These parameters, defined in the subsequent sections, allow the BAO feature to be extracted and interpreted in a manner that is robust to the choice of fiducial cosmology.

\subsection{BAO dilation parameters}

In spectroscopic surveys like DESI, angular positions and redshifts are used to infer the comoving coordinates of galaxies and quasars, assuming a fiducial cosmology. If the true cosmology differs from the fiducial one, the reconstructed coordinates will exhibit both isotropic and anisotropic distortions, the latter quantified by the Alcock-Paczynski (AP) effect \cite{1979Natur.281..358A}. These distortions cause the observed modes along and perpendicular to the line of sight to be remapped as 
\begin{equation}
k_{\parallel}^{\rm true} = \frac{D^{\rm fid}_{\rm H}(z)}{D_{\rm H}(z)} k_{\parallel}, \quad {\bf k}_{\perp}^{\rm true}
= \frac{D^{\rm fid}_{\rm M}(z)}{D_{\rm M}(z)} {\bf k}_{\perp},
\end{equation}
where $D_{\rm H}(z) = c/H(z)$ is the Hubble distance, and $D_{M}(z)$ is the comoving angular diameter distance, both evaluated at a redshift $z$.

The template power spectrum is the other key aspect to consider.
In the BAO fitting framework, rather than constructing a power spectrum model for each different cosmology we test, we assume that the BAO pattern can be mapped by a simple rescaling of a given fiducial template power spectrum to match the BAO feature in the observed power spectrum: $k^\prime = \frac{\rd}{\rd^{\rm temp}}k^{\rm true}$.\footnote{
Here, we use the notation $\rd^{\text{temp}}$ rather than $\rd^{\text{fid}}$ to distinguish the fiducial sound horizon assumed in the template cosmology from that used in the grid cosmology, which maps sky coordinates to physical distances. For simplicity, we will now refer to both collectively as the fiducial cosmology, denoted as``\texttt{fid}," following the nomenclature used in the DESI key paper \cite{DESI.DR2.BAO.cosmo}.}

Connecting both points motivates the introduction of two free dilation parameters, which rescale the BAO portion of the power spectrum along and across the line of sight, as ${\bf k} \rightarrow ({\bf k}_{\perp}/\alpha_{\perp}, ~k_{\parallel}/\alpha_{\parallel})$, and leads to their physical interpretation
\begin{equation}
\alpha_{\parallel} = \frac{D_{\rm H}(z) / \rd}{D^{\rm fid}_{\rm H}(z)/\rd^{\rm temp}}, \quad \alpha_{\perp} = \frac{D_{\rm M}(z) / \rd}{D^{\rm fid}_{\rm M}(z)/\rd^{\rm temp}}.
\end{equation}

Alternatively, as is done in this work, these parameters can be translated into an isotropic/anisotropic basis
\begin{equation}
    \alpha_{\rm iso} = (\alpha_{\parallel} \alpha_{\perp}^2)^{1/3}, \quad \alpha_{\rm AP} = \frac{\alpha_{\parallel}}{{\alpha_{\perp}}},
\end{equation}
leading to a coordinate transformation of the form
\begin{align}
k^\prime &= \frac{\alpha_{\text{AP}}^{1/3}}{\alpha_{\text{iso}}} 
\left[ 1 + \mu^2 \left( \frac{1}{\alpha_{\text{AP}}^2} - 1 \right) \right]^{1/2}k, \\
\mu^\prime &= \frac{\mu}{\alpha_{\text{AP}} \left[ 1 + \mu^2 \left( \frac{1}{\alpha_{\text{AP}}^2} - 1 \right) \right]^{1/2}}.
\end{align}
Here the unprimed coordinates represent the \textit{observed} coordinates and the primed coordinates ($k^\prime$, $\mu^\prime$) are the coordinates at which the model is evaluated. With this, the BAO portion of the observed galaxy power spectrum $P_{\text{wg, obs}}(k, \mu)$ can then formally be related to the BAO feature in the model power spectrum $P_{\rm wg}(k^\prime, \mu^\prime)$ using these parameters:
\begin{equation}
P_{\rm{wg, obs}}(k, \mu) = \frac{1}{\alpha_{\rm{iso}}^3} P_{\rm wg}\left(k^\prime, \mu^\prime; \alpha_{\rm{iso}}, \alpha_{\rm{AP}}\right).
\end{equation}

Note that $\alpha_{\rm iso}$ and $\alpha_{\rm AP}$  therefore relates to the distances as: 
\begin{equation}
\alpha_{\rm iso} = \frac{D_{\rm V}(z) / \rd}{D^{\rm fid}_{\rm V}(z)/\rd^{\rm temp}}, \quad \alpha_{\rm AP} = \frac{D_{\rm H}(z) / D^{\rm fid}_{\rm H}(z) }{D_{\rm M}(z)/D^{\rm fid}_{\rm M}(z)
},
\label{eq:alphaiso_alphaap}
\end{equation}
where $D_{\rm V}(z)$ is the spherically averaged distance defined as:
\begin{equation}
D_{\rm V}(z) = \left[z D_{\rm H}(z)D_{\rm M}^2(z)\right]^{1/3}.
\end{equation}

\subsection{Reconstruction}
\label{subsec:reconstruction}

Reconstruction techniques aim to recover the linear BAO signal from nonlinear degradations caused by gravitational growth, peculiar velocities (RSD), and galaxy bias \cite{Eisenstein2007b}. This process involves displacing galaxies along the inferred large-scale displacement field, effectively `undoing' the effects of bulk flows. By mitigating these nonlinear degradations, the reconstructed density field not only enhances the amplitude of the BAO peak, bringing it closer to the linear case and improving measurement precision but also helps remove nonlinear shifts in the location of the BAO feature that could potentially bias the cosmological constraints \cite{Eisenstein2007b,KP4s2-Chen}.

Under the plane-parallel approximation and assuming the linear continuity equation, the displacement field can be derived from the galaxy overdensity field $\delta_{g}(k)$ as \citep{KP4s3-Chen}:
\begin{equation} \label{eq:displacement}
\mathbf{D}(\mathbf{k}) = -\frac{i\mathbf{k}}{k^2} \frac{S(k) \delta_{g}(\mathbf{k})}{b(1 + \beta \mu^2)},
\end{equation}
where $S(k)$ is a smoothing kernel applied to filter out small scales dominated by highly non-linear dynamics and shot noise, $b$ is the linear galaxy bias, $f$ is the growth rate of structure, and $\beta=f/b$. The term $b(1 + \beta \mu^2)$ accounts for the contributions of linear bias and large-scale redshift-space distortions (RSD) in the observed galaxy density field and ensures that the derived displacement field reflects the underlying matter displacement field in real space.

In practice, the plane-parallel approximation is not valid when analyzing survey data, where we cannot assume the same line of sight for all pairs of objects. However, \cref{eq:displacement} serves to illustrate the point that the displacement calculation requires knowledge of the galaxy bias and the linear growth rate of structure, as well as prescriptions for defining the smoothing kernel and for estimating the galaxy overdensity field from the discrete galaxy positions.

Beyond the plane-parallel approximation, different reconstruction algorithms offer numerical solutions that allow for the estimation of the displacement field for varying lines of sight. As in DR1, our default choice of algorithm for the DR2 clustering analysis is the \texttt{iterative FFT reconstruction} \cite{Burden:2015pfa}, which solves redshift-space linearized continuity equation in Fourier space by iteratively removing RSD. This algorithm has been shown to be robust and efficient when compared against other popular algorithms in the literature \citep{KP4s3-Chen}.

\begin{table}[h]
    \centering
    \caption{Reconstruction parameters for each tracer, including the smoothing scale $\Sigma_{\rm sm}$, linear galaxy bias $b$, and growth rate of structure $f$ at the effective redshift $z_{\rm eff}$.}
    \label{tab:recon_params}
    \begin{ruledtabular}
    \begin{tabular}{lccc}
        Tracer & $\Sigma_{\rm sm}$ [$h^{-1}$Mpc] & $b$ & $f(z_{\rm eff})$  \\
        \hline
        BGS  & 15  & 1.5 & 0.69 \\
        LRG  & 15  & 2.0 & 0.83 \\
        LRG+ELG  & 15  & 1.6 & 0.86 \\
        ELG  & 15  & 1.2 & 0.90 \\
        QSO  & 30  & 2.1 & 0.93 \\
    \end{tabular}
    \end{ruledtabular}
\end{table}

We reconstruct the DR2 catalogs using \texttt{pyrecon}\footnote{\url{https://github.com/cosmodesi/pyrecon}}, adopting the same baseline settings as in DR1, which were stress-tested on DESI mock catalogs in \cite{KP4s4-Paillas}. The overdensity field is painted on a grid with a cell size of $4 \hinvmpc$, and is then smoothed by a Gaussian kernel with characteristic width $\Sigma_{\rm sm}$, which is taken to be $15 \hinvmpc$ (BGS, LRGs, ELGs) or $30 \hinvmpc$ (QSO), as summarized in \cref{tab:recon_params}. The linear galaxy bias was determined by \cite{KP4s11-Garcia-Quintero, KP4s10-Mena-Fernandez}, and corresponds to 1.5 for BGS, 2.0 for LRGs, 1.2 for ELGs, and 2.1 for QSO. The growth rate of structure is predicted from our fiducial cosmology at the effective redshift of each tracer. We also studied the impact of varying both the bias and the growth rate of structure on the reconstruction process. \cite{KP4s9-Perez-Fernandez} studied the impact of the choice of fiducial cosmology on the reconstruction process, and the propagation of this effect on the BAO constraints is included in the systematic error budget.

We performed a series of robustness tests for reconstruction, where we added a redshift padding to the galaxy catalogs before reconstructing them (allowing for a more appropriate estimation of the velocity field at the survey edges), and using a redshift-dependent value for the growth rate of structure and the linear galaxy bias. In all cases, we found minimal effects on the resulting clustering measurements and BAO fits.

\subsection{BAO Model}

\begin{table}
    \renewcommand{\arraystretch}{1.3}
    \centering
    \begin{tabular}{ l  c  c c  c c  c }
        \hline
        \hline
        Parameter & Reconstruction & \bgs & \lrgs & \elgs & \qso \\
        \hline
        $\Sigma_{\perp}^{\rm fid}\, [\Mpch]$ & Pre & $6.5$ & $4.5$ & $4.5$ & $3.5$\\
        $\Sigma_{\parallel}^{\rm fid}\, [\Mpch]$ & Pre & $10.0$ & $9.0$ & $8.5$ & $9.0$\\
        $\Sigma_s^{\rm fid}\, [\Mpch]$ & Pre & $2.0$ & $2.0$ & $2.0$ & $2.0$\\
        \hline
        $\Sigma_{\perp}^{\rm fid}\, [\Mpch]$ & Post & $3.0$ & $3.0$ & $3.0$ & $3.0$\\
        $\Sigma_{\parallel}^{\rm fid}\, [\Mpch]$ & Post & $8.0$ & $6.0$ & $6.0$ & $6.0$\\
        $\Sigma_s^{\rm fid}\, [\Mpch]$ & Post & $2.0$ & $2.0$ & $2.0$ & $2.0$\\
        \hline
        \hline
    \end{tabular}
    \caption{Gaussian priors on the non-linear BAO damping parameters used for pre- and post-reconstruction BAO fits across tracers. The values in the table represent the mean ($\mu$) of the Gaussian priors, with uncertainties set to $\pm 1\Mpch$ for $\Sigma_{\perp}$ and $\pm 2\Mpch$ for $\Sigma_{\parallel}$ and $\Sigma_s$, consistent with the priors used in the DR1 BAO analysis \cite{DESI2024.III.KP4}.}
    \label{tab:damping_params}
\end{table}

With all the necessary components in place, we can now construct the BAO model used to fit the observations. The galaxy power spectrum in redshift space can be modeled as a combination of smooth and oscillatory components to capture the BAO signal. Following \cite{KP4s2-Chen}, the model for the galaxy power spectrum is expressed as:
\begin{equation} \label{eq:generic_pk_model}
P_{g}(k, \mu) = B(k, \mu) P_{\text{nw}}(k) + C(k, \mu) P_{\text{w}}(k) + D(k),
\end{equation}
where:
\begin{itemize}
    \item $P_{\text{nw}}(k)$ and $P_{\text{w}}(k)$ are the smooth (no-wiggle) and oscillatory (wiggle) components of the linear power spectrum $P_{\text{lin}}(k)$, respectively, obtained using the \textit{peak average} method \cite{Brieden:2022lsd}.
    \item $B(k, \mu)$ encapsulates the broadband clustering shape, incorporating the effects of RSD and galaxy bias.
    \item $C(k, \mu)$ isolates the BAO feature while accounting for anisotropic damping caused by non-linear growth and peculiar velocities.Therefore $C(k, \mu)P_w(k)$ is the only term that is dilated by the \{$\alphaiso$, $\alphaap$\}.
    \item $D(k)$ models any deviations from the linear theory in the broadband shape of the power spectrum multipoles.
\end{itemize}

The broadband term is expressed as:
\begin{equation} \label{eq:damping_fog}
B(k, \mu) = \left(b_1 + f \mu^2 [1 - s(k)]\right)^2 F_{\text{FoG}},
\end{equation}
where:
\begin{itemize}
    \item $b_1$ is the linear galaxy bias.
    \item $f$ is the linear growth rate of structure.
    \item $s(k)$ is the smoothing kernel, which is set to zero for unreconstructed catalogs and the \texttt{RecSym} reconstuction convention \citep[our default; see][]{KP4s4-Paillas}, and as $s(k) = \exp[-(k \Sigma_{\text{sm}})^2 / 2]$ for the \texttt{RecIso} convention, where $\Sigma_{\text{sm}}$ represents the smoothing scale used in reconstruction.
    \item $F_{\text{FoG}} = \left[1 + \frac{1}{2} k^2 \mu^2 \Sigma_s^2\right]^{-2}$ accounts for the damping from the Fingers-of-God (FoG) effect, with $\Sigma_s$ as the free parameter controlling virial motions.
\end{itemize}

The oscillatory component is damped anisotropically:
\begin{equation}
C(k, \mu) = \left(b_1 + f \mu^2 \right)^2 \exp\left[-\frac{1}{2} k^2 \left(\mu^2 \Sigma_{\parallel}^2 + (1-\mu^2) \Sigma_{\perp}^2\right)\right],
\end{equation}
where:
\begin{itemize}
    \item $\Sigma_{\parallel}$ and $\Sigma_{\perp}$ are the damping scales parallel and perpendicular to the line of sight, respectively. 
\end{itemize}
The exponential factor models the non-linear damping of the BAO feature, which reconstruction reduces by mitigating bulk motions. (see \cref{tab:damping_params} for the values used pre- and post-reconstruction BAO fits.)

To account for additional broadband contributions beyond the linear theory prediction, we use a spline-based approach to introduce a flexible modeling term:
\begin{equation} \label{eq:spline_pk}
D_{\ell}(k) = \sum_{n=-1}^{n_{\text{max}}} a_{\ell, n} W_3\left(\frac{k}{\Delta} - n\right),
\end{equation}
where:
\begin{itemize}
    \item $W_3$ is a piecewise cubic spline kernel that ensures smooth modeling of broadband deviations.
    \item $\Delta$ is the spacing between spline nodes, chosen to avoid overlap with the BAO signal, typically set to twice the BAO wavelength (\(\sim 0.06 \, h \, \text{Mpc}^{-1}\)).
    \item $n_{\text{max}}$ determines the number of spline terms, sufficient to span the analyzed \(k\)-range (\(0.02 \, h \, \text{Mpc}^{-1} < k < 0.30 \, h \, \text{Mpc}^{-1}\)) without reproducing the oscillatory features.  Based on these criteria, \( n_{\text{max}} \) is set to 7 in this work, the same as in DR1.
\end{itemize}

The model multipoles are convolved with the data window function to ensure accurate comparison with the observed power spectrum. This step enables precise modeling over the $ k$ range used for BAO fitting. Specifically, since we fit the power spectrum monopole and quadrupole ($P_0(k)$ and $P_2(k)$), rather than the full anisotropic power spectrum $P_{\rm obs}(k, \mu)$, we integrate over the Legendre polynomials, $\mathcal{L}_\ell(\mu)$, to obtain the model multipoles:

\begin{equation}
    P_{\ell}(k) = \frac{2\ell + 1}{2} \int_{-1}^{1} d\mu \, P_{\rm obs}(k, \mu) \mathcal{L}_\ell(\mu).
\end{equation}

To predict the multipoles of the galaxy correlation function in configuration space, we start from the multipoles of the power spectrum of \cref{eq:generic_pk_model}, \textit{excluding} the $D_{\ell}(k)$ factor, and Hankel transform them as
\begin{equation}
    \xi_\ell(s) = \frac{i^\ell}{2\pi^2} \int_0^\infty \mathrm{d}k\, k^2 j_\ell (ks) P_\ell (k) \,,
\end{equation}
where $j_\ell$ are the spherical Bessel functions. To parameterize the remaining broadband part, one would Hankel transform the same power spectrum spline functions from \cref{eq:spline_pk}. However,  \cite{KP4s2-Chen} showed that, for our choice of fitting scales in configuration space ($60 \hinvmpc < s < 150 \hinvmpc$), these functions quickly approach zero on large scales. We, therefore, exclude these terms, except for the $n= 0, 1$ terms of the quadrupole, as done in \cite{DESI2024.III.KP4}. In addition to these two terms, in the configuration space BAO fitting. We also introduce two additional nuisance parameters for each multipole, aimed to control potential large-scale systematics in the data\footnote{In Fourier space, these large-scale systematics can be confined below a certain $k_{\rm min}$ and are therefore more easily mitigated by simply truncating the data vector.}:
\begin{equation}
    \tilde{D}_\ell(s) = b_{\ell, 0} + b_{\ell,2}\left( \frac{s k_{\rm min}}{2\pi} \right)^2 ,
\end{equation}
with $k_{\rm min} = 0.02 \hmpcinv$.

\begin{table*}
\centering
\begin{tabular}{c|c|c|c}
\hline
\hline
\textbf{Parameter} & \textbf{\( P(k) \) Prior} & \textbf{\( \xi(r) \) Prior} & \textbf{Description} \\
\midrule
$\alpha_{\text{iso}}$ & $[0.8, 1.2]$ & $[0.8, 1.2]$ & Isotropic BAO dilation \\
$\alpha_{\text{AP}}^*$ & $[0.8, 1.2]$ & $[0.8, 1.2]$ & Anisotropic (AP) BAO dilation \\
$\Sigma_\perp$ & $\mathcal{N}(\Sigma_\perp^{\text{fid}}, 1.0)$ & $\mathcal{N}(\Sigma_\perp^{\text{fid}}, 1.0)$ & Transverse BAO damping $[h^{-1} \text{Mpc}]$ \\
$\Sigma_{\parallel}$ & $\mathcal{N}(\Sigma_{\parallel}^{\text{fid}}, 2.0)$ & $\mathcal{N}(\Sigma_{\parallel}^{\text{fid}}, 2.0)$ & Line-of-sight BAO damping $[h^{-1} \text{Mpc}]$ \\
$\Sigma_s$ & $\mathcal{N}(2.0, 2.0)$ & $\mathcal{N}(2.0, 2.0)$ & Finger of God damping $[h^{-1} \text{Mpc}]$ \\
$b_1$ & $[0.2, 4]$ & $[0.2, 4]$ & Linear galaxy bias \\
$d\beta^*$ & $[0.7, 1.3]$ & $[0.7, 1.3]$ & Linear RSD parameter \\
$a_{0, n}$ & $\mathcal{N}(0, 10^4)$ & N/A & Spline parameters for the monopole \\
$a_{2, n}^*$ & $\mathcal{N}(0, 10^4)$ & $\mathcal{N}(0, 10^4)$ & Spline parameters for the quadrupole \\
$b_{0, n}$ & N/A & $[-\infty, \infty]$ & Unknown large scale systematics \\
$b_{2, n}^*$ & N/A & $[-\infty, \infty]$ & Unknown large scale systematics \\
Fitting range & $[0.02, 0.3]$ $h \text{Mpc}^{-1}$ & $[58, 152]$ $h^{-1} \text{Mpc}$ & Measurement bin edges \\
Data binning & $0.005$ $h \text{Mpc}^{-1}$ & $4$ $h^{-1} \text{Mpc}$ & Measurement bin width \\
\hline
\hline
\end{tabular}
\caption{The free parameters and their priors for Fourier-space (FS) and configuration-space (CS) analyses. $\mathcal{N}(\mu, \sigma)$ refers to a normal distribution of mean $\mu$ and standard deviation $\sigma$, $[x_1, x_2]$ to a flat distribution between $x_1$ and $x_2$. Parameters marked with * are fixed to the following values when only a 1D fit is performed: $\alpha_{\text{AP}} = 1$, $d\beta = 1$, $a_{2,n} = 0$, $b_{2,n} = 0$.}
\label{tab:prior}
\end{table*}

\subsection{Covariance Matrices}

For the DESI DR2 analysis, we rely exclusively on analytical and semi-analytical methods to estimate covariance matrices. In doing so, we have saved considerable time and effort that would be needed to create a high-precision mock-based covariance matrix via the calibration, generation and processing of a necessarily large suite of approximate simulations. The faster analytical methods also gave us more flexibility to update the covariance matrices several times as the data evolved, in particular, to keep them blinded initially and to unblind the covariances shortly after the correlation function measurements. The covariance matrix pipelines had been fixed before unblinding, only the input data was changed from blinded to unblinded.

Configuration-space covariances are generated using the \textsc{RascalC} semi-analytical code\footnote{\url{https://github.com/oliverphilcox/RascalC}}~\cite{rascal,rascal-jackknife,RascalC,RascalC-legendre-3,2023MNRAS.524.3894R}, which has already been the fiducial method for the DESI DR1 BAO analysis \cite{DESI2024.III.KP4}.
This approach computes the covariance matrices with the empirical two-point correlation function, accurate survey geometry and selection effects, but without the contributions of three-point and connected four-point functions\footnote{Only the 2-point function and the disconnected 4-point function are included.}.
These higher-point contributions to the large-scale covariance matrix appear similar to the effects of shot noise.
As a result, rescaling the shot noise allows to mimic the omitted terms.
The amount of rescaling is calibrated on the jackknife covariance matrix estimate from data.
The up-to-date methodology is comprehensively described in \cite{KP4s7-Rashkovetskyi} along with its validation on DESI DR1 mocks.
The only new challenge for DR2 covariance estimation we identified is the denser BGS sample (which is also harder to model in mocks), which took more time and displayed slightly worse intrinsic precision.
In many other aspects (e.g., footprint boundaries, holes and completeness uniformity) the data has become simpler and more regular than DR1.
The code to generate the final DESI DR2 covariances for the correlation functions is accessible on \texttt{GitHub}\footnote{\url{https://github.com/cosmodesi/RascalC-scripts/tree/DESI-DR2-BAO/DESI/Y3}, {\tt post} and {\tt pre} directories for after and before reconstruction respectively.}.

For the analyses in Fourier space, we use analytical covariance matrices computed with \textsc{TheCov}\footnote{\url{https://github.com/cosmodesi/thecov}}\cite{Wadekar:2019rdu, KP4s8-Alves}. The description of the methodology and a comparison with DESI DR1 mock-based sample covariances are presented in \cite{KP4s8-Alves}. For the purposes of this work, we only compute the disconnected part of the 4-point function (sometimes referred to as the Gaussian covariance), including survey geometry effects, and we use the power spectrum measured directly from the corresponding data set as input. Although the connected term for BAO-reconstructed power spectrum multipoles has been recently studied using perturbation theory \cite{Zhao:2024xit}, it has little impact on BAO fits \cite{KP4s8-Alves} and is therefore neglected in the context of the validation tests presented here.

\subsection{Parameter inference}

By combining all components from the previous subsections, the BAO model systematically accounts for the choice of fiducial cosmology, broadband contamination, anisotropies, and non-linear effects, ensuring accurate recovery of the BAO scale. This approach allows the oscillatory features to serve as a robust standard ruler for cosmological distance measurements.

Unless otherwise noted, our baseline data vector consists of the post-reconstruction monopole and quadrupole moments of the galaxy correlation function (LRGs, ELGs and QSOs), or simply the post-reconstruction monopole in the case of the BGS isotropic BAO fits. A summary of our baseline scale cuts and free parameters of the BAO model is shown in \cref{tab:prior}, along with the prior distributions used during parameter inference.

We sample the posterior distribution with the \texttt{desilike}\footnote{\url{https://github.com/cosmodesi/desilike/}} framework, using a wrapper around the Markov chain Monte Carlo code \texttt{emcee} \cite{emcee}. We assume a Gaussian likelihood and adopt the Gelman-Rubin statistic \cite{GelmanRubin} as the convergence criteria for our chains, demanding that $\textrm{R} - 1 < 0.01$. We also perform maximization using the \texttt{minuit} profiler \citep{iminuit}.

\section{Key Changes in the DESI DR2 Analysis}
\label{sec:changes_y3}

This section summarizes the key updates made in the \desidrtwo\ analysis compared to \desidrone, highlighting improvements in data handling, methodology, and systematic error treatment. While the systematic error budget is largely inherited from the \desidrone BAO analysis, the \desidrtwo analysis introduces tracer-dependent refinements to improve accuracy and robustness. These updates leverage the larger dataset and refined analysis techniques to achieve more precise BAO measurements.

\subsection{Data and Methodology Updates}

\begin{itemize}
    \item \textbf{Magnitude Cut for BGS:} In the \desidrone analysis, the baseline Bright Galaxy Sample (BGS) was defined using a magnitude cut of $M_r < -21.5$ (denoted as BGS-BRIGHT-21.5). For \desidrtwo, a slightly fainter magnitude cut of $M_r < -21.35$ (BGS-BRIGHT-21.35) was adopted as the baseline. This lower threshold maintains an approximately constant number density to $z<0.4$, increasing it from $\sim$0.0005$h^3$Mpc$^{-3}$ to $\sim$0.001$h^3$Mpc$^{-3}$ after applying completeness corrections (see Figure 3 in \cite{DESI.DR2.BAO.cosmo}). The effective volume calculation increases by 17\% compared to the fiducial -21.5 cut (holding $P_0$ fixed), without imparting any additional complexity in modeling the sample's clustering or the covariance of its clustering measurements.

    \item \textbf{Combining LRG and ELG Data:} Unblinding tests now include the combined sample in addition to individual tracers, ensuring robustness across all redshift bins~\cite{KP4s5-Valcin,Y3BAO-Sanders}. The combined sample is used as the default for cosmological inference.

    \item  \textbf{Minimum Scale Cut ($s_{\text{min}}$):} For the \desidrone analysis, the minimum scale used in the BAO fits was $s_{\text{min}} = 50~\Mpch$. The effect of changing $s_{\rm min}$ on the quality of fits to mocks was studied extensively by \cite{KP4s2-Chen}, who showed that the recovered alpha values and their errors are very stable against changes to the minimum scale in the range $50\lesssim s_{\rm min}\lesssim 80\;\Mpch$. Fits to the blinded \desidrtwo data revealed somewhat large $\chi^2$ values—exceeding the upper boundary set by the mock distributions—when using $s_{\rm min}=50\;\Mpch$ for some redshift bins, which improved significantly when changing the scale cuts to $s_{\rm min}=60\;\Mpch$. In both cases of $s_{\rm min}$, the impact on the BAO measurements was negligible. This may suggest that the flexibility of our broadband nuisance parameters is no longer sufficient for the range $50 < s < 60~\Mpch$ for some tracers, given the increased signal-to-noise of \desidrtwo. While this warrants further investigation, the stability of the BAO measurements justifies adopting a more conservative choice of $s_{\text{min}} = 60~\Mpch$ for DR2, ensuring that we robustly isolate the $s$-range of the BAO feature.
    
    \item \textbf{2D BAO Fits:} For \desidrtwo, 2D BAO fits are employed for \elgo\ and \qso, where the increased signal-to-noise ratio in the clustering allows for stable anisotropic BAO measurements, providing additional cosmological information. In these cases, both the monopole and quadrupole moments of the correlation function are fitted simultaneously to constrain $\alpha_{\rm iso}$ and $\alpha_{\rm AP}$.  However, for tracers where the quadrupole signal-to-noise is lower, robust determination of $\alpha_{\rm AP}$ becomes more challenging. In these cases, we perform a 1D fit using only the monopole, which primarily constrains $\alpha_{\rm iso}$, to avoid introducing a weak, non-Gaussian constraint on $\alpha_{\rm AP}$. This is the case for BGS, where weaker Alcock-Paczynski distortions at low redshift and higher correlations between $\alpha_{\rm iso}$ and $\alpha_{\rm AP}$ lead us to conservatively adopt a 1D fit as the default. A detailed discussion on the criteria used to assess 1D vs. 2D fits, along with supporting tests on mocks and data, is provided in \cref{app:1D_vs_2D}.

    \item \textbf{Split Tests:} Additional data-splitting tests have been introduced to further validate the robustness of BAO measurements. These include tests based on imaging survey regions to assess the impact of residual imaging systematics, ensuring that variations in survey depth and observational conditions do not bias the results. Additionally, mass- and magnitude-split tests are performed to verify the stability of BAO measurements across different galaxy populations, probing potential dependencies on intrinsic tracer properties. 

\end{itemize}

\subsection{Systematic Error Treatment}\label{sec:sys_error}

While the core methodology of the analysis remains consistent with \desidrone, the updates implemented for \desidrtwo reflect targeted improvements to better account for the increased data volume and enhanced precision. By incorporating tracer-specific and redshift-dependent systematic error treatments, adjusting scale cuts, and adopting conservative fiducial cosmology assumptions, the \desidrtwo analysis ensures robust and unbiased BAO measurements. This section provides a detailed summary of the systematic error contributions, highlighting the careful refinements applied to each tracer in this updated analysis.

\begin{table*}[t]
    \renewcommand{\arraystretch}{1.3}
    \centering
    \begin{tabular}{ c | c | c | c | c | c }
        \hline
        \hline
        Tracer & Parameter &  Theory (\%) & HOD (\%) &  Fiducial (\%) & Total (\%) \\
        \hline
        \bgs & $\alphaiso$ & 0.1 & No detection & 0.1 & 0.141 \\
        \hline
        \lrgo & $\alphaiso$ & 0.1 & No detection & 0.1 & 0.141 \\
             & $\alphaap$ & 0.2 & 0.19 & 0.18 & 0.329 \\
        \hline
        \lrgt & $\alphaiso$ & 0.1 & No detection & 0.1 & 0.141 \\
            & $\alphaap$ & 0.2 & 0.19 & 0.18 & 0.329 \\
        \hline
        \lrgth & $\alphaiso$ & 0.1 & 0.17 & 0.1 & 0.221 \\
             & $\alphaap$ & 0.2 & 0.19 & 0.18 & 0.329 \\
        \hline
        \lrgelg & $\alphaiso$ & 0.1 & 0.17 & 0.1 & 0.221 \\
         & $\alphaap$ & 0.2 & 0.19 & 0.18 & 0.329 \\
        \hline
        \elgo & $\alphaiso$ & 0.1 & 0.17 & 0.1 & 0.221 \\
         & $\alphaap$ & 0.2 & No detection & 0.1 & 0.224 \\
        \hline
        \elgt & $\alphaiso$ & 0.1 & 0.17 & 0.1 & 0.221 \\
         & $\alphaap$ & 0.2 & No detection & 0.1 & 0.224 \\
        \hline
        \qso & $\alphaiso$ & 0.1 & 0.17 & 0.1 & 0.221 \\
         & $\alphaap$ & 0.2 & 0.19 & 0.18 & 0.329 \\         
        \hline
        \hline
    \end{tabular}
    \caption{Systematic error contributions to $\alpha_{\text{iso}}$ and $\alpha_{\text{AP}}$ for the different DESI tracers. The sources of systematic errors include theoretical systematics, HOD-related systematics, and uncertainties from the fiducial cosmology. The total systematic error for each parameter is obtained by taking the quadratic sum of the individual contributions.}
    
    \label{tab:systematic_errors}
\end{table*}

\begin{itemize}

    \item \textbf{Fiducial Cosmology Systematics:} Fiducial cosmology-related systematic errors were adjusted, increasing the systematic uncertainty on $\alpha_{\rm AP}$ from 0.1\% in \desidrone to 0.18\% in \desidrtwo for the \lrg sample. Consequently, this increase also affects the \lrgth+\elgo combined sample, as the combined sample inherits the largest errors among its components. This adjustment resulted from including the DESI-motivated evolving dark energy model in the fiducial cosmology test \cite{KP4s9-Perez-Fernandez}. As a result, the total systematic uncertainty on $\alpha_{\rm AP}$ increased from 0.3\% to 0.335\%. This change was determined following a reanalysis of the \abacussecond DR1 mocks using the best-fit $w_0w_a$CDM cosmology from DESI DR1 BAO + CMB + SN data \cite{DESI2024.VI.KP7A} as the fiducial cosmology throughout the pipeline. The decision to reassess the systematic error budget was made before unblinding the BAO constraints (item \#9 in our unblinding checklist, reviewed in later sections).
    
    \item \textbf{HOD-Related Systematics:} For \desidrone, although tracer-dependent HOD systematics were determined \cite{KP4s10-Mena-Fernandez,KP4s11-Garcia-Quintero, DESI2024.VI.KP7A}, a conservative approach was taken by adopting the largest detected systematic shift across all tracers as a universal HOD systematic error. For \desidrtwo, we refine this treatment by incorporating tracer dependency, applying different HOD systematic contributions for each tracer type. We refer to \cref{tab:systematic_errors} for a detailed quantification of these values. The updated treatment is as follows:      

    \begin{itemize}
        \item \textbf{BGS:} No HOD-related systematic error is added, as none were detected within the statistical precision of our test in \desidrone.
        \item \textbf{LRG1; LRG2:} The \desidrone values are used, with no systematic error in $\alpha_{\text{iso}}$ and a 0.19\% error in $\alpha_{\text{AP}}$.
        \item \textbf{ELG1; ELG2:} ELG-specific HOD errors are applied, with a 0.17\% error in $\alpha_{\text{iso}}$ and none detected within the statistical precision of our test in $\alpha_{\text{AP}}$.
        \item \textbf{Combined LRG3+ELG1:} The largest of the \lrg and \elg errors is adopted (0.17\% for $\alpha_{\text{iso}}$ and 0.19\% for $\alpha_{\text{AP}}$).
        \item \textbf{LRG3:} \lrgth is treated similarly to the \lrgelg sample, as the redshift distribution $n(z)$ varies significantly across this range, likely accompanied by an evolving bias. Consequently, applying the same HOD systematics as the other \lrg bins would likely underestimate the systematic uncertainties for this redshift bin.

        \item \textbf{QSO:} In \desidrone, no HOD systematics were detected on $\alpha_{\text{iso}}$ in the 1D BAO analysis for QSOs. However, for \desidrtwo, QSOs were upgraded to a 2D BAO analysis. Since we do not have a pre-determined 2D BAO HOD systematic error for this tracer, we adopt a conservative approach by assigning the same systematic error as the \lrgelg combined sample—not just for HOD, but for all systematic contributions. Given the large statistical uncertainty in QSO measurements, the impact of these systematics is expected to be negligible.
        
    \end{itemize}
    
    These refinements ensure a tracer-dependent treatment of HOD-related systematics, moving beyond the conservative universal approach of \desidrone while maintaining robustness in \desidrtwo.
\end{itemize}

Several systematic effects identified and tested in \desidrone were not reassessed in \desidrtwo because they were previously found to be negligible or robust at a precision sufficient for DR2, with no new evidence warranting re-evaluation. For example, the theoretical systematics included in \cref{tab:systematic_errors} remain unchanged from \desidrone, as their impact was thoroughly evaluated in \cite{KP4s2-Chen} with a precision of 0.01–0.1\%, and there is no reason to expect any change in theoretical systematics since we use the same BAO fitting model and method described in \cite{KP4s2-Chen}. Similarly, systematic uncertainties related to reconstruction algorithms were studied in \cite{KP4s3-Chen} using the Y5 footprint, which found no significant contribution to BAO measurements. These findings are consistent with Table 12 of \cite{DESI2024.III.KP4}, which summarizes the systematic effects considered in \desidrone. 

Although there is no evidence requiring re-evaluation, as a sanity check, we allow for theoretical systematics to be correlated across different tracers and redshift bins (see \cref{sec:post_unblinding} for details).

Additional systematic effects, such as fiber assignment and spectroscopic efficiency corrections, were also not explicitly revisited in detail, as they are indirectly tested through our standard validation procedures. The most stringent test of fiber assignment systematics comes from the requirement that BAO fits yield unbiased results when applied to realistic mock catalogs, as presented in Table VI. These tests confirmed that fiber assignment does not introduce significant bias in BAO measurements, validating the robustness of our correction methods. Furthermore, spectroscopic completeness effects were analyzed in DESI DR1 \cite{KP3s3-Krolewski, KP3s4-Yu} and found to have no measurable impact on BAO constraints. Specifically, \cite{KP3s3-Krolewski} examined potential systematics, while \cite{KP3s4-Yu} studied the impact of observational variations on clustering measurements. These DR1 findings provided sufficient evidence that spectroscopic completeness does not bias BAO measurements, so no additional reanalysis was conducted for \desidrtwo.

The calibration accuracy of DESI redshifts has been evaluated to be better than 1 km/s in \cite{Koposov:2024ixy}. We note, however, that the clustering analysis has been performed using redshifts in the frame of the solar barycenter and not the CMB frame (relative speed of 369.82 km/s, see \cite{Planck-2018-overview} Table 3). This difference in reference frame has a negligible impact on the LSS analysis after averaging over the angular distribution of our survey. For the nearest BGS redshift bin, the angular averaged redshift offset is of 0.0002 which results in a negligible correction to the distances.

To summarize, the \desidrtwo analysis builds upon the solid foundation established in \desidrone. This ensures that our BAO constraints remain robust while leveraging the expanded dataset and increased statistical precision of DR2.

\section{Results}
\label{sec:results}

As part of the validation process, we conducted our analysis exclusively on blinded catalogs to prevent potential confirmation biases. \textit{All the plots and statistical checks in this section were initially produced using the blinded data} to verify that they met the blinding criteria outlined in \cref{sec:unblindin_tests}. Once the blinding tests were passed, we regenerated the plots using the final unblinded catalogs. Therefore, the figures presented in this section now reflect the fully validated, unblinded measurements.\footnote{In the following \cref{sec:post_unblinding}, we present what we refer to as post-unblinding tests. These tests were planned before unblinding but were only carried out afterward. While they do not influence our primary analysis choices, they provide additional insights into the stability of our results under different assumptions.}

\subsection{Evolution of BAO Precision: BOSS \& eBOSS to DESI} 

\begin{figure}
    \centering
    \includegraphics[width=1\linewidth]{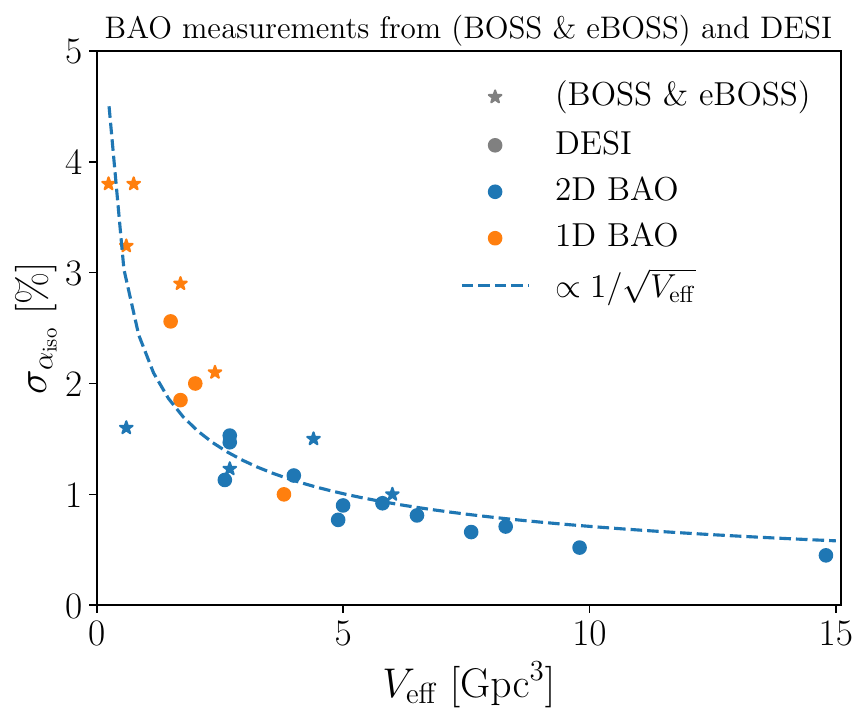}
    \caption{Comparison of BAO measurement precision as a function of effective survey volume, $V_{\rm eff}$, for BOSS $\&$ eBOSS, including points from \cite{2012MNRAS.427.3435A, 2021MNRAS.500..736B} and DESI \cite{DESI2024.III.KP4, DESI.DR2.BAO.cosmo}. The plot illustrates the fractional uncertainty on the BAO scale, $\sigma_{\rm iso}$, highlighting the expected scaling relation $\sigma_{\rm iso} \propto 1/\sqrt{V_{\rm eff}}$ (dashed line). The DESI data points demonstrate a significant reduction in uncertainty compared to BOSS $\&$ eBOSS, benefiting from increased survey volume and improved analysis techniques. The distinction between 1D and 2D BAO fits is also shown, providing historical context for the evolution of BAO precision.}
    \label{fig:bao_precision}
\end{figure}

The precision of BAO measurements has steadily improved with increasing survey volume, advances in data analysis techniques, and improvements in observational hardware. \cref{fig:bao_precision} illustrates the relationship between the effective survey volume, $V_{\rm eff}$, and the fractional uncertainty on the BAO scale, $\sigma_{\rm iso}$, for both BOSS $\&$ eBOSS and DESI. 

The trend shown in the figure follows the expected scaling of $\sigma_{\rm iso} \propto 1/\sqrt{V_{\rm eff}}$, highlighting how larger datasets enable more precise BAO constraints. While early BAO measurements, such as those from BOSS $\&$ eBOSS, were limited by statistical uncertainties due to smaller galaxy samples, DESI has significantly expanded the survey volume, leading to a substantial reduction in $\sigma_{\rm iso}$. While DESI remains statistically limited, the significant increase in dataset size has driven a major improvement in measurement precision, complemented by refined analysis techniques.

Additionally, the figure differentiates between one-dimensional (1D) and two-dimensional (2D) BAO analyses. While 2D BAO fits extract additional information from anisotropic clustering, 1D fits provide robust constraints in lower signal-to-noise regimes, such as for BGS in DESI (see \cref{app:1D_vs_2D} for details). This historical comparison contextualizes the improvements made in BAO analyses and demonstrates DESI's capability to push the limits of precision cosmology through larger survey volumes and improved methodologies.

\subsection{The Unblinding Tests}\label{sec:unblindin_tests}

\begin{table*}
    \centering
    \renewcommand{\arraystretch}{1.4}
    \setlength{\tabcolsep}{6pt} 
    \footnotesize 
    \begin{tabular}{c p{7.5cm} p{7.5cm}} 
        \hline\hline
        \textbf{\#} & \textbf{Test} & \textbf{Result} \\
        \hline
        1 & Are $\chi^2$ reasonable and consistent with mocks? & \textbf{Yes}. Reasonable $\chi^2$ for all tracers, and consistency between data and mocks in all cases. \\
        2 & Are the reconstruction settings appropriate for \desidrtwo? & \textbf{Yes}. Good performance of reconstruction for all tracers, and data errors consistent with the mocks. \\
        3 & Are results robust to imaging systematics? & \textbf{Yes}. BAO constraints remain stable even in the extreme case removing imaging systematics weights from clustering measurements. \\
        4 & Are results robust to data splits? & \textbf{Yes}. BAO constraints show robustness when testing different sky regions and magnitude/mass splits. \\
        5 & Are uncertainties consistent with mocks? & \textbf{Yes}. Errors on the $\alpha$ parameters measured from data fall within the distribution spanned by the mocks for all tracers. \\
        6 & Are the $\alpha$ parameters consistent between pre- and post-reconstruction? & \textbf{Yes}. Pre-reconstruction $\alpha_{\rm iso}$ values show slight deviations relative to mocks but remain within expected variations. Reconstruction improves precision, and the shifts in $\alpha$ values are consistent with the mock distribution. \\
        7 & Are mock fits unbiased? & \textbf{Yes}. Most tracers show biases below $0.5\sigma$. The largest offset is $2.46\sigma$ for post-reconstruction \qso, which remains below our $3\sigma$ threshold for systematic detection. \\

        8 & Are $\alpha$ values consistent between configuration space and Fourier space? & \textbf{Yes}. Results from the correlation function are consistent to within $0.5\sigma$ with those from the power spectrum. \\
        9 & Are we reassessing the systematic error budget? & The systematic error budget is primarily based on \desidrone, with updates as described in \cref{sec:sys_error}. \\
        10 & Are \lrgs and \elgs consistent in the overlapping bin $0.8 < z < 1.1$? & \textbf{Yes}. The $\alphaiso$ offset between \lrgth and \elgo is $0.41\sigma$, consistent with mock expectations. The agreement on $\alphaap$ is within $-1.66\sigma$. \\
        \hline\hline
    \end{tabular}
    \caption{Unblinding checklist. Each test evaluates whether differences observed in the blinded data are consistent with the range spanned by 25 \abacussecond DR2 mocks. A test is considered successfully passed if the observed differences fall within this range.}
    \label{tab:unblinding_tests}
\end{table*}

To ensure the robustness of our \desidrtwo BAO measurements, we conducted a series of pre-unblinding validation checks, which are summarized in \cref{tab:unblinding_tests}. These tests verify the consistency of the clustering measurements across different tracers, data splits, and methodologies while assessing systematic uncertainties. Each test is evaluated by comparing the observed differences in the blinded data to the full range spanned by the 25 \abacussecond DR2 mocks. A test is considered successfully passed if the observed differences fall within this range. We determined the systematic error budget detailed in \cref{tab:systematic_errors} prior to unblinding.

Following DR1, we define a systematic effect as one that exceeds a significance of $3\sigma$, where $\sigma$ refers to the statistical precision associated with the given test.

When introducing our various results in the following subsections, we will discuss how each of them helps to assess one or more items from the unblinding checklist (\cref{tab:unblinding_tests}).

\subsection{Consistency and Precision in \desidrone and \desidrtwo Two-Point Clustering Measurements}

\begin{figure*}
\centering
\begin{tabular}{ccc}
  \includegraphics[width=0.3\linewidth]{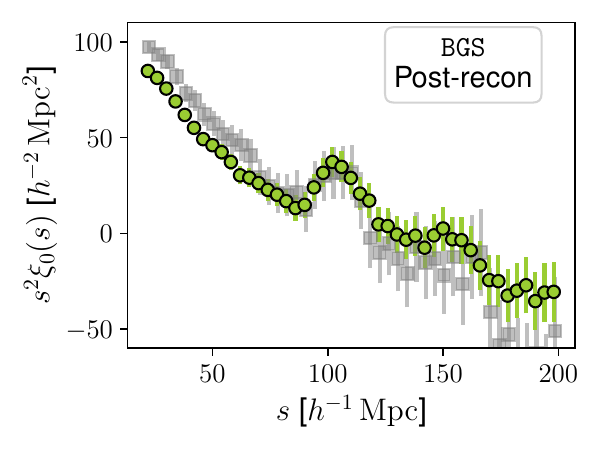} &
  \includegraphics[width=0.3\linewidth]{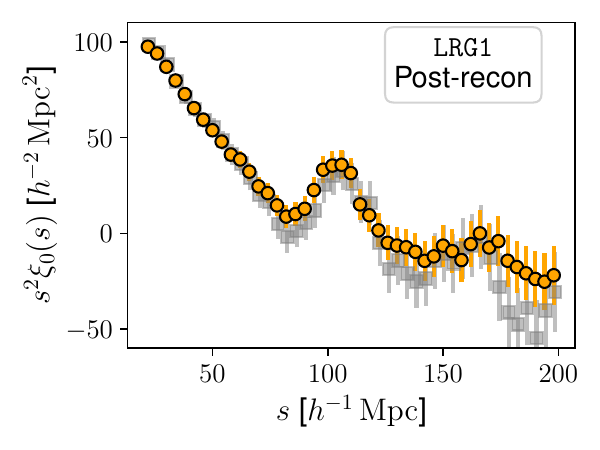} &
  \includegraphics[width=0.3\linewidth]{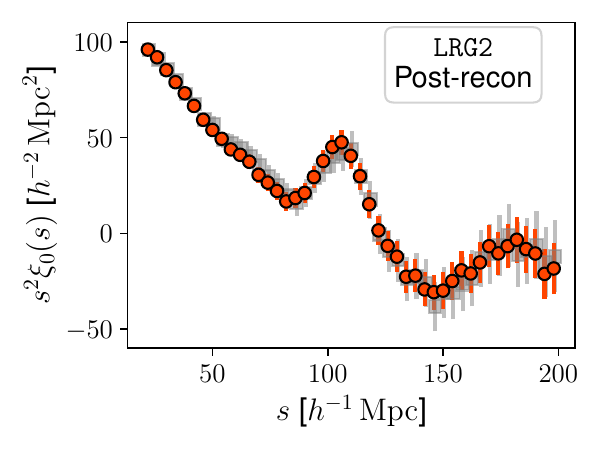} \\

  \includegraphics[width=0.3\linewidth]{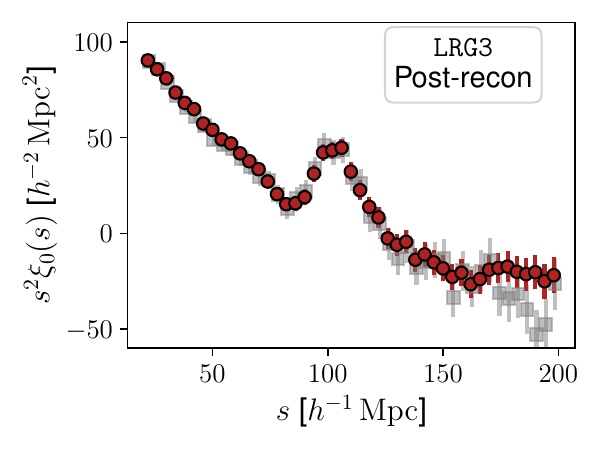} &
  \includegraphics[width=0.3\linewidth]{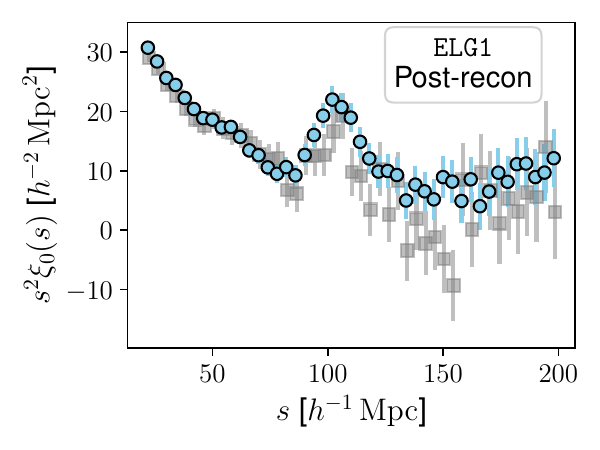} &
  \includegraphics[width=0.3\linewidth]{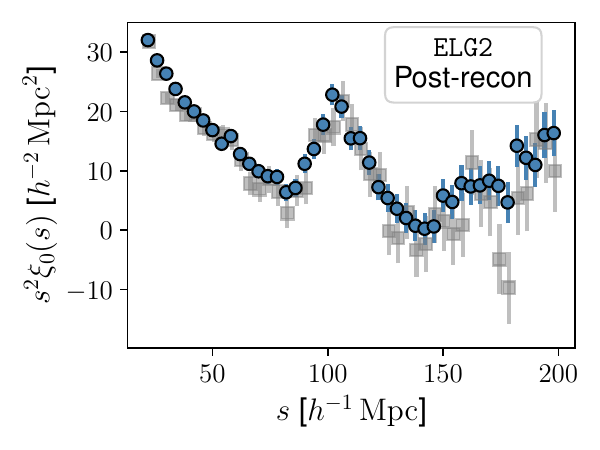} \\

  \includegraphics[width=0.3\linewidth]{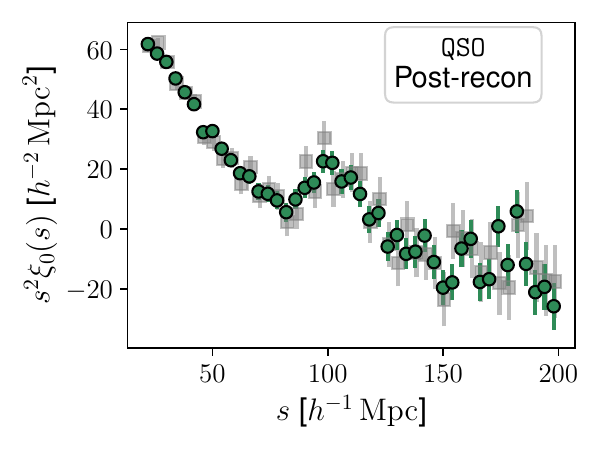} &
  \includegraphics[width=0.3\linewidth]{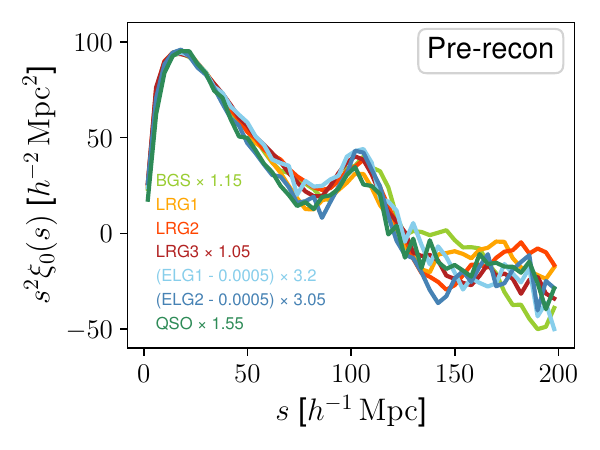} &
  \includegraphics[width=0.3\linewidth]{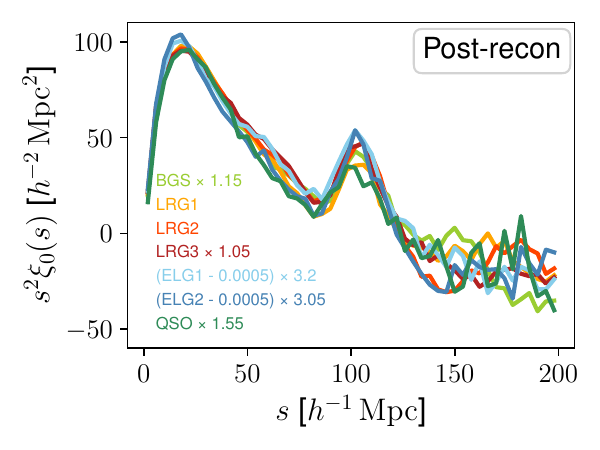} \\
\end{tabular}
\caption{
Comparison of the monopole of the unblinded two-point correlation function, $\xi_0(s)$, between \desidrone (gray) and \desidrtwo (colored) for various tracer samples and redshift ranges. The panels show results for \bgs ($0.1 < z < 0.4$), \lrg ($0.4 < z < 0.6$, $0.6 < z < 0.8$, $0.8 < z < 1.1$), \elg ($0.8 < z < 1.1$, $1.1 < z < 1.6$), and \qso ($0.8 < z < 2.1$). The bottom-middle and bottom-right panels show combined views of all tracers before (Pre-recon) and after (Post-recon) BAO reconstruction, respectively, highlighting the sharpening of the BAO feature after reconstruction. The DR2 measurements exhibit reduced scatter, particularly for the \bgs and \elg samples, reflecting improved precision. Notable differences in amplitude, especially for \bgs at small scales and \elg at large scales, are discussed in the text and are attributed to changes in sample selection and potential residual imaging systematics, respectively.}
\label{fig:two-function}
\end{figure*}

We compare the two-point clustering measurements from \desidrone and \desidrtwo, as shown in \cref{fig:two-function}. This comparison serves two purposes: assessing the consistency between data releases with different footprints and completeness and highlighting improvements in measurement precision. An analogous plot (\desidrone vs. \desidrtwo) but for the mean of the \abacussecond mocks is shown in \cref{app:abacus_DR1_vs_DR2}.

Across all samples, the DR2 data are noticeably smoother, indicating enhanced precision in the clustering measurements. This is particularly evident in the \elgo and \bgs samples, where the BAO feature appears sharper in DR2. The clustering amplitudes between DR1 and DR2 are generally consistent, suggesting that the sample properties have remained stable across the survey footprint. However, the \bgs sample presents a notable exception, with a shift in the small-scale amplitude. This difference arises from the application of a fainter absolute magnitude cut in the DR2 sample (see \cref{sec:changes_y3}), which affects the clustering signal at smaller scales. For the \elg samples—especially \elgo—we observe higher large-scale amplitudes in DR2, which could indicate residual imaging systematics. These two effects are mitigated as follows.

The bottom-middle and bottom-right panels of \cref{fig:two-function} illustrate the impact of potential residual systematics in DR2, showing the clustering measurements across all tracers before (pre-reconstruction) and after (post-reconstruction) BAO reconstruction, respectively. The observed differences between DR1 and DR2 clustering amplitudes can largely be accounted for by an approximate multiplicative factor of 1.15 for \bgs and a constant offset of 0.0005 for \lrg. Importantly, our BAO fitting model includes both a bias multiplicative term and a free constant offset, ensuring that these residual effects do not significantly impact the BAO results. Furthermore, \cite{KP3s2-Rosado} previously demonstrated that DESI BAO measurements from \elg samples in DR1 remained robust even in the absence of explicit imaging systematics corrections. Given that the level of residual contamination in DR2 remains within a comparable range, it does not pose a concern for BAO measurements\footnote{Future work will investigate these residuals further, particularly for models that rely on the broadband clustering amplitude, where such contamination may have a larger effect.}.

\subsection{BAO measurements from DESI DR2}

\begin{figure*}
\centering
\begin{tabular}{cc}
    \hspace{-4cm}
    \includegraphics[width=0.24\linewidth]{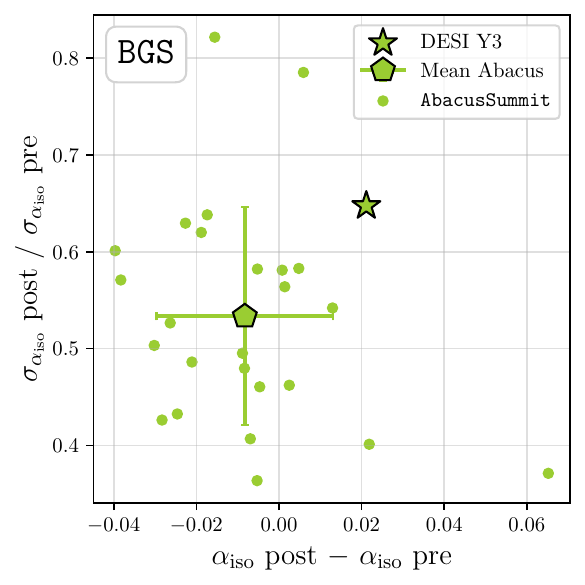}
    \\ 
    \includegraphics[width=0.48\linewidth]{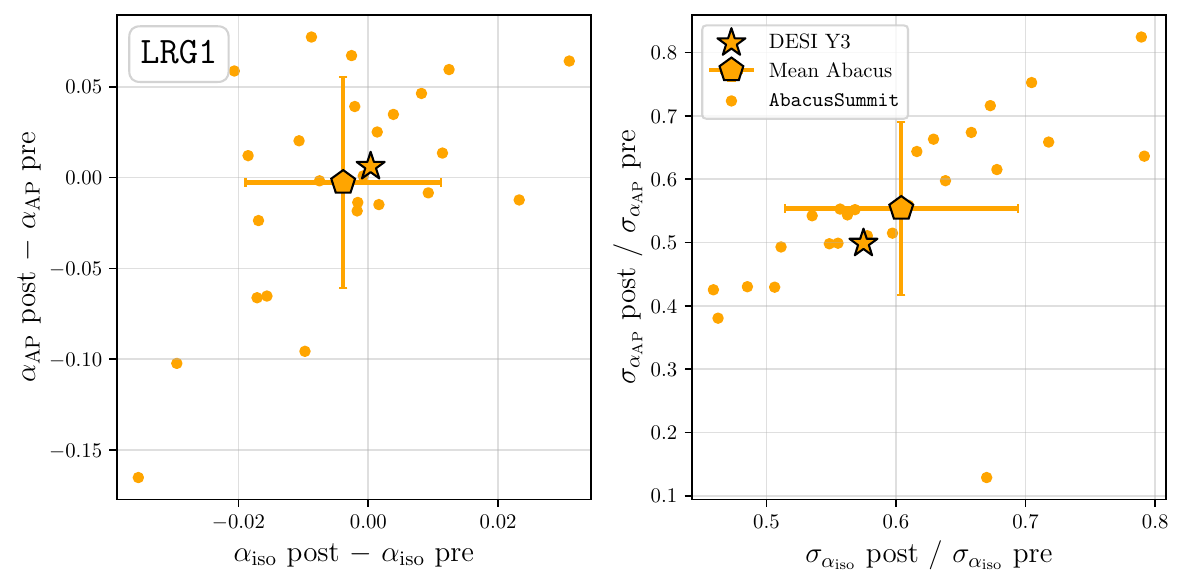} 
    &
    \includegraphics[width=0.48\linewidth]{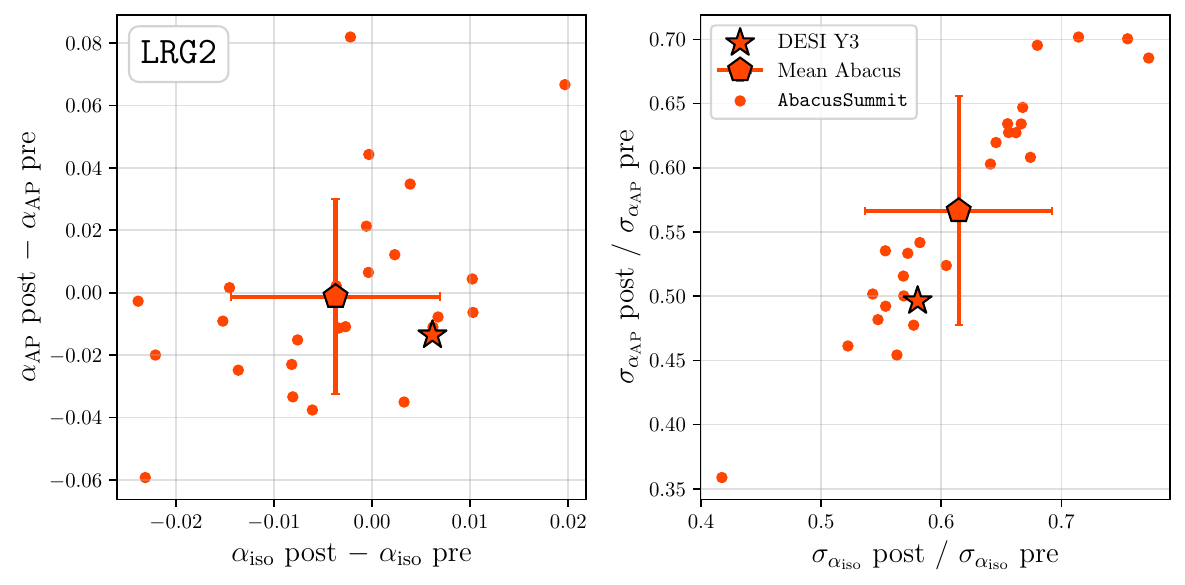}
    \\
    \includegraphics[width=0.48\linewidth]{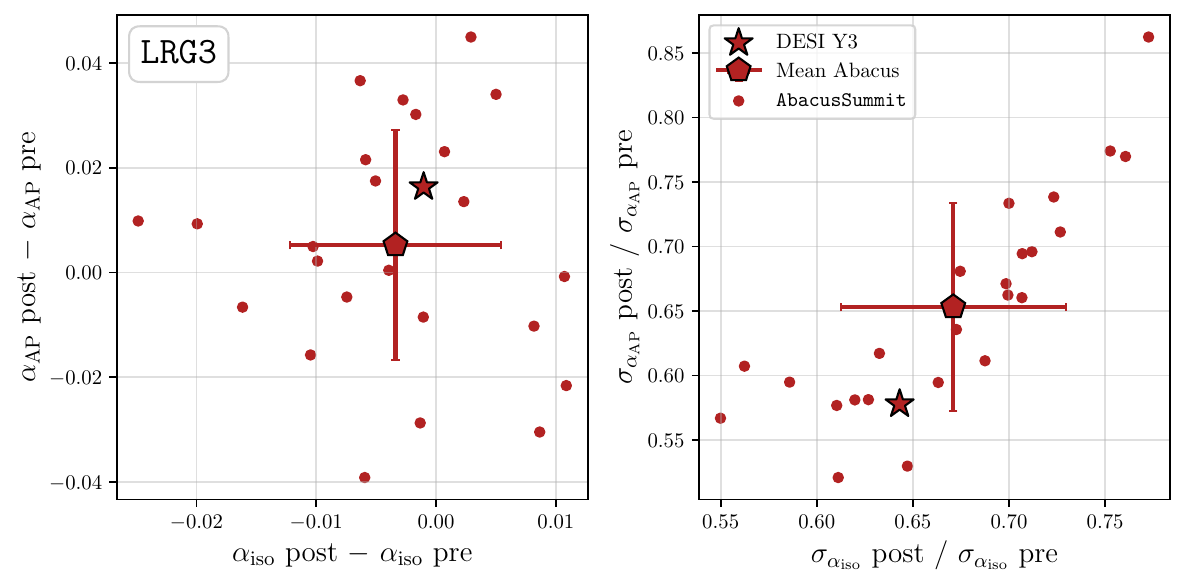}
    & 
    \hspace{-0.25cm}
    \includegraphics[width=0.48\linewidth]{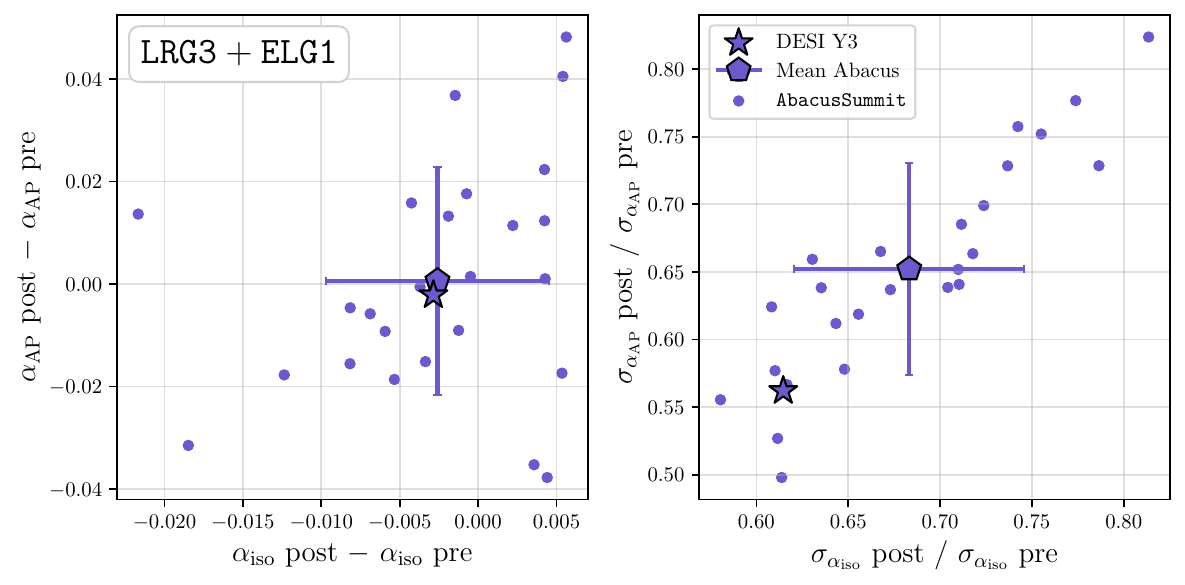}
    \\
    \includegraphics[width=0.48\linewidth]{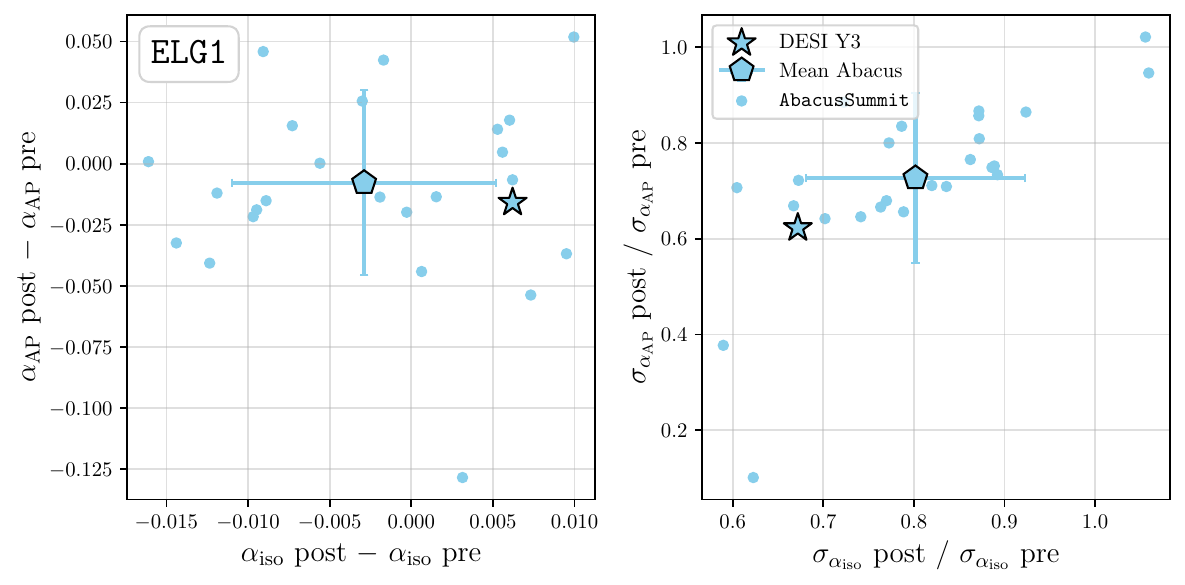}
    &
    \includegraphics[width=0.48\linewidth]
    {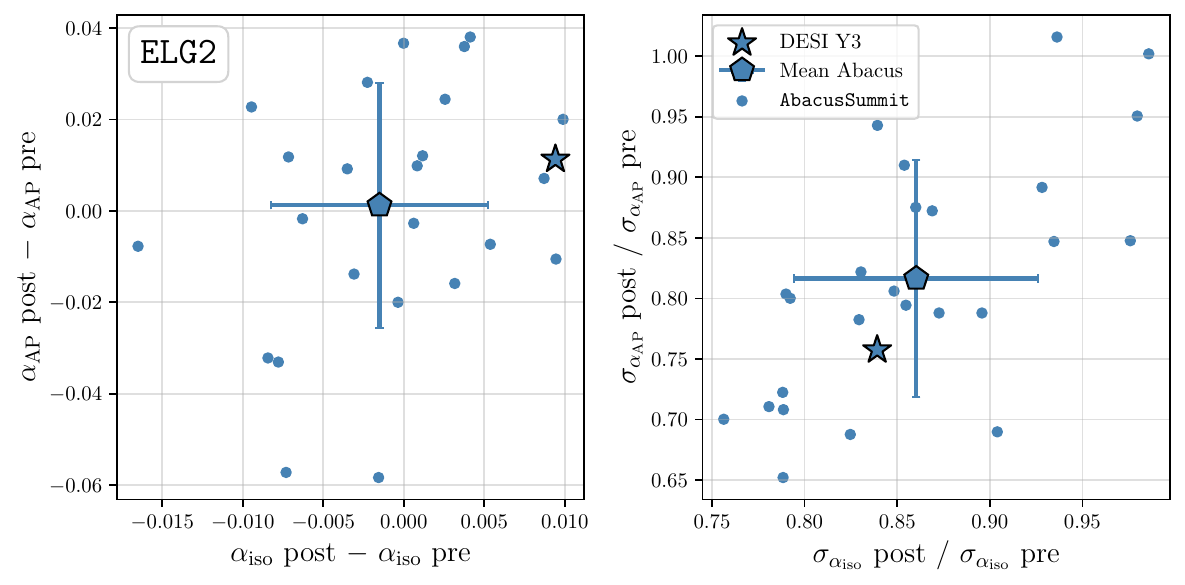}
    \\
    \includegraphics[width=0.48\linewidth]
    {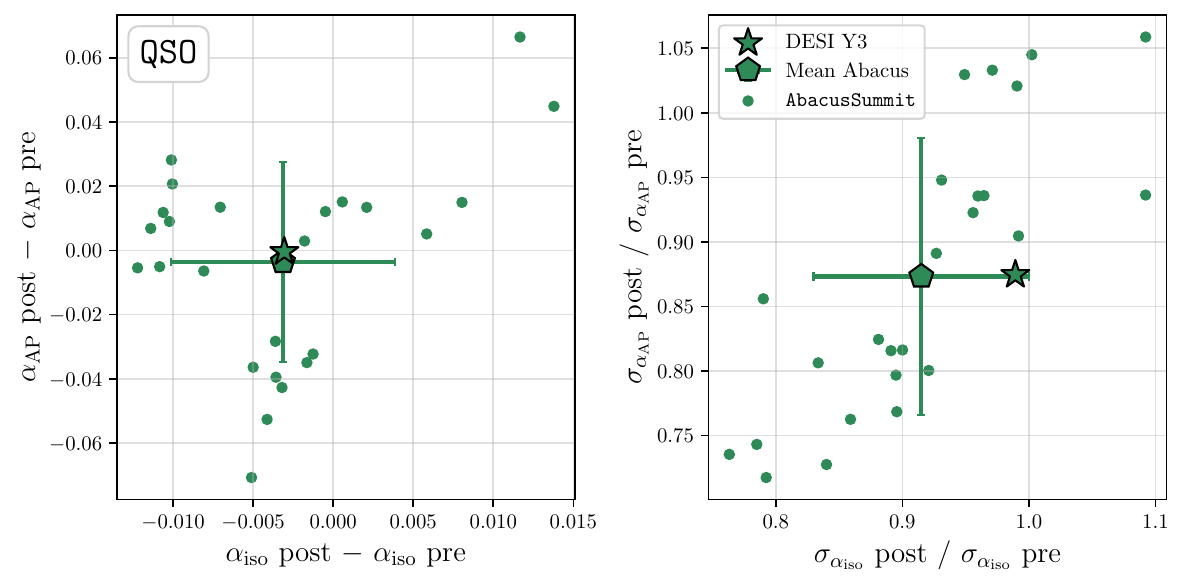}
\end{tabular}
\caption{Pre- and post-reconstruction BAO measurements, comparing \desidrtwo measurements (stars) with the 25 \abacussecond DR2 (the open points with the error bars are the means and the standard deviations around the means of the 25 mocks). All \desidrtwo fits are consistent with what is expected from the mocks.}\label{fig:scatteralpha}
\end{figure*}

\begin{figure*}
\centering
\begin{tabular}{ccc}
    \includegraphics[width=0.3\linewidth]{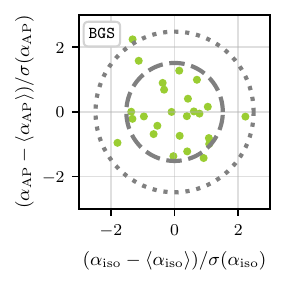}
    & 
    \includegraphics[width=0.3\linewidth]{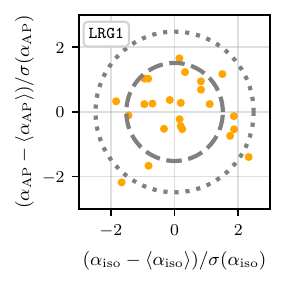} 
    &
    \includegraphics[width=0.3\linewidth]{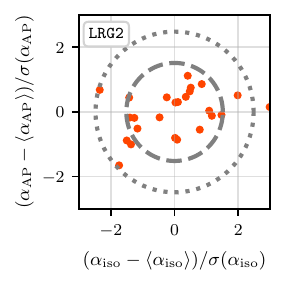}
    \\
    \includegraphics[width=0.3\linewidth]{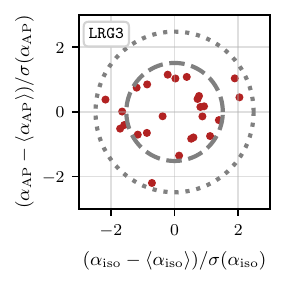}
    & 
    \hspace{-0.25cm}
    \includegraphics[width=0.3\linewidth]{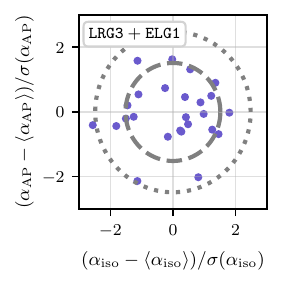}
    &
    \includegraphics[width=0.3\linewidth]{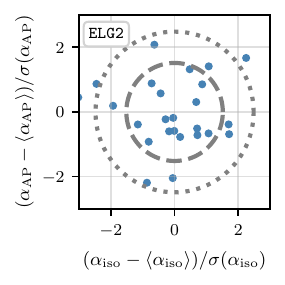}
    \\
    \includegraphics[width=0.3\linewidth]
    {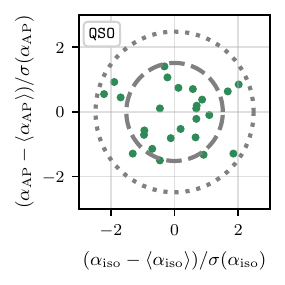}
\end{tabular}

\caption{Comparison of the estimated uncertainties $\sigma_{\alpha_{\rm iso}}$ and $\sigma_{\alpha_{\rm AP}}$ across the 25 \abacussecond DR2 mocks. Scatter points represent individual best-fit realizations, while the dashed circles indicate the expected 1$\sigma$ (68\%) and 2$\sigma$ (95\%) confidence regions of a multivariate Gaussian distribution centered at zero. Each panel corresponds to a different tracer and redshift bin, illustrating the relationship between individual mock uncertainties and the expected statistical spread. The general agreement between the scatter points and the contours indicates that our uncertainty estimates are robust and do not show significant biases.}
\label{fig:scatter_abacus}
\end{figure*}

\begin{table*}
\centering
\renewcommand{\arraystretch}{1.2}
\setlength{\tabcolsep}{6pt}
\begin{tabular}{llccrrrl}
\hline
\hline
 Tracer          & Recon.   & $\alpha_{\rm iso}$   & $\alpha_{\rm AP}$   & $\Delta \alpha_{\rm iso}$   & $\Delta \alpha_{\rm AP}$   & $r$      & $\chi^2 / {\rm dof}$   \\
\hline
 {\tt BGS}       & Post    & $0.9980 \pm 0.0020$  & ---                 & $-0.97$                     & ---                        & ---      & 18.0/16                \\
 {\tt LRG1}      & Post    & $0.9971 \pm 0.0017$  & $0.9941 \pm 0.0060$ & $-1.70\sigma$               & $-0.99\sigma$              & $0.102$  & 30.3/33                \\
 {\tt LRG2}      & Post    & $0.9993 \pm 0.0013$  & $1.0012 \pm 0.0049$ & $-0.49\sigma$               & $0.25\sigma$               & $0.148$  & 24.0/33                \\
 {\tt LRG3}      & Post    & $0.9991 \pm 0.0012$  & $0.9956 \pm 0.0042$ & $-0.76\sigma$               & $-1.05\sigma$              & $0.111$  & 23.2/33                \\
 {\tt ELG1}      & Post    & $1.0009 \pm 0.0021$  & $1.0096 \pm 0.0070$ & $0.43\sigma$                & $1.37\sigma$               & $-0.231$ & 38.1/33                \\
 {\tt ELG2}      & Post    & $1.0004 \pm 0.0015$  & $0.9982 \pm 0.0054$ & $0.26\sigma$                & $-0.33\sigma$              & $-0.135$ & 24.6/33                \\
 {\tt LRG3+ELG1} & Post    & $1.0004 \pm 0.0010$  & $1.0001 \pm 0.0037$ & $0.36\sigma$                & $0.02\sigma$               & $0.082$  & 38.8/33                \\
 {\tt QSO}       & Post    & $0.9943 \pm 0.0023$  & $1.0049 \pm 0.0091$ & $-2.46\sigma$               & $0.54\sigma$               & $-0.022$ & 29.1/33                \\
 \hline
 {\tt BGS}       & Pre     & $1.0094 \pm 0.0041$  & ---                 & $2.29\sigma$                      & ---                        & ---      & 27.0/16                \\
 {\tt LRG1}      & Pre     & $1.0004 \pm 0.0031$  & $0.9898 \pm 0.0113$ & $0.14\sigma$                & $-0.91\sigma$              & $0.327$  & 38.9/33                \\
 {\tt LRG2}      & Pre     & $1.0041 \pm 0.0023$  & $1.0099 \pm 0.0094$ & $1.76\sigma$                & $1.05\sigma$               & $0.471$  & 18.2/33                \\
 {\tt LRG3}      & Pre     & $1.0030 \pm 0.0018$  & $0.9930 \pm 0.0067$ & $1.68\sigma$                & $-1.03\sigma$              & $0.157$  & 28.3/33                \\
 {\tt ELG1}      & Pre     & $1.0043 \pm 0.0026$  & $1.0197 \pm 0.0099$ & $1.63\sigma$                & $1.99\sigma$               & $-0.107$ & 49.3/33                \\
 {\tt ELG2}      & Pre     & $1.0015 \pm 0.0018$  & $0.9951 \pm 0.0067$ & $0.81\sigma$                & $-0.73\sigma$              & $-0.034$ & 19.9/33                \\
 {\tt LRG3+ELG1} & Pre     & $1.0035 \pm 0.0015$  & $1.0021 \pm 0.0059$ & $2.25\sigma$                & $0.36\sigma$               & $0.182$  & 34.9/33                \\
 {\tt QSO}       & Pre     & $0.9979 \pm 0.0025$  & $1.0079 \pm 0.0106$ & $-0.83\sigma$               & $0.74\sigma$               & $0.097$  & 16.2/33                \\
\hline
\hline
\end{tabular}
\caption{Constraints on the BAO scaling parameters from fits to the mean of 25 \abacussecond DR2 mocks, where the average errors have been rescaled by $\sqrt{25}$ so that the errors reflect the combined volume of all mocks. The lower and upper sub-panels show results before and after reconstructing the galaxy catalogs. The fifth and sixth columns show the offset, in terms of the number of standard deviations, from the expectation of $\alphaiso = \alphaap = 1$. $r$ is the cross-correlation coefficient between $\alphaiso$ and $\alphaap$. We also show the $\chi^2$ per degree of freedom at the best fit.}

\label{tab:fit_mocks}
\end{table*}

We now focus on the BAO constraints derived from the configuration space measurements shown in the previous section and compare them with the expectations from the mocks \footnote{Additionally, Fourier-space results, detailed in \cref{app:fourier}, further support our finding in configuration space, given the level of agreement between the two conjugate spaces}. For the resulting BAO constraints on the dilatation parameters $\{\alphaiso, \alphaap\}$ and a visualization of the BAO features on the unblinded data, see Table III and Figure 5 of the key paper \citep{DESI.DR2.BAO.cosmo}. To assess the robustness of our BAO measurements, we compare our results against both pre-reconstruction and post-reconstruction fits, validate the uncertainty estimates using mocks, and evaluate statistical goodness-of-fit metrics. This section presents key tests addressing the consistency of our results.

\subsubsection{Pre- vs. Post-Reconstruction BAO Fits}
\label{subsec:pre_post_recon}

\cref{fig:scatteralpha} compares the pre- and post-reconstruction BAO fits for the DESI DR2 data and the 25 \abacussecond DR2 mocks. With the exception of the first panel, which shows the \bgs tracer and employs one-dimensional (1D; $\alphaiso$) fits, all other tracers use two-dimensional (2D; $\alphaiso, \alphaap$) fits. For the 1D BAO fits, we use only the monopole of the correlation function. The data measurements (indicated by stars) fall comfortably within the range spanned by the 25 \abacussecond DR2 mocks. This range serves as our benchmark for defining consistency, given the limited number of mock realizations. 

Reconstruction improves the precision on $\alphaiso$ for all tracers. For DR2, the percentage of improvement compared to the pre-reconstruction fits ranges from 3.7\% for \qso, to 42\% for \lrgelg, varying depending on the signal to noise of the density field and also the severity of degradation to mitigate. $\alphaap$ also benefits greatly from reconstruction, with a gain that ranges from 8.5\% for \qso, to 47\% for \lrgelg. This level of improvement from the data finds good agreement with the distribution of mocks and is also consistent with the DR1 findings, emphasizing the robustness of the reconstruction pipeline. A point to be noted is that in DR1, the \qso saw a small degradation of the constraining power on $\alphaiso$ after reconstruction, which we attributed to a noise fluctuation in the data since this was still consistent with the range of possibilities spanned by the mock catalogs. This is confirmed by the new DR2 results, where the \qso sample shows a small improvement from reconstruction, while now also allowing for a separate constraint on $\alphaap$ due to the promotion of this tracer to a 2D fit. Overall, these results address the unblinding test \#2, showing that the adopted reconstruction settings lead to the expected improvements with respect to the pre-reconstruction constraints. \cref{fig:scatteralpha} also addresses the unblinding test \#6, showing consistency between the best-fit values of the BAO scaling parameters before and after reconstruction. Finally, it also informs unblinding test \#5, demonstrating that the uncertainties on the BAO scaling parameters, both before and after reconstruction, are consistent with the distribution of mocks. This confirms that the relative improvement in individual errors aligns with the average expected trend.

\subsubsection{Robustness of the DESI DR2 Uncertainties}
\label{subsec:uncertainty_validation}

\cref{fig:scatter_abacus} provides a visual check on our uncertainty estimates by comparing the normalized deviations $ (\alpha_{\rm iso} - \langle \alpha_{\rm iso} \rangle)/\sigma_{\alpha_{\rm iso}}$ and $(\alpha_{\rm AP} - \langle \alpha_{\rm AP} \rangle)/\sigma_{\alpha_{\rm AP}}$ from individual realizations (scatter points) to the expected statistical scatter from a multivariate Gaussian. The dashed circles indicate the 1$\sigma$ (68\%) and 2$\sigma$ (95\%) confidence regions for a multivariate Gaussian distribution centered at $(0,0)$. If the uncertainties are well-characterized, we expect the individual realizations to follow the statistical properties of such multivariate Gaussian contours.

Across all tracers and redshift bins, the scatter of values generally aligns with the expected contours. We expect one mock realization to lie outside the 2$\sigma$ contour and 8 outside the 1$\sigma$ contour. The obtained results vary from 1 to 3 and 5 and 10, respectively. The error estimates are thus broadly consistent with statistical expectations. This serves as a complementary consistency check for unblinding test \#5, suggesting that our covariance matrix provides a reasonable description of the statistical scatter within the available set of mock realizations. The limited number of mock realizations (25) and the replication involved in their creation (see \cref{sec:mocks}) prevents this test from being more precise. The most thorough evaluation of the accuracy of uncertainty on BAO measurements recovered when applying RascalC covariance matrices remains the DR1 studies \cite{KP4s7-Rashkovetskyi}, which imply an accuracy of better than 5\%.

\subsubsection{Consistency of LRG and ELG Samples}

\begin{figure}
    \centering
    \includegraphics[width=0.7\linewidth]{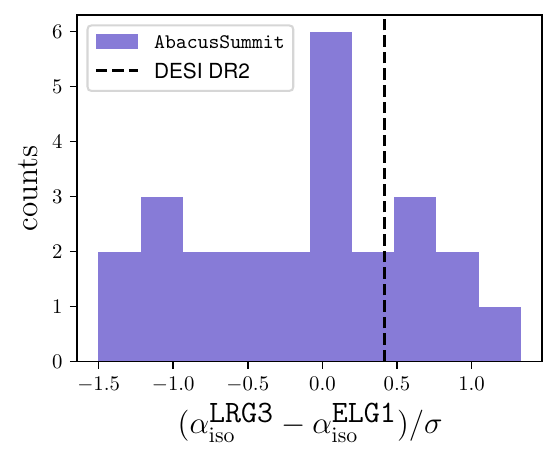}
    \includegraphics[width=0.7\linewidth]{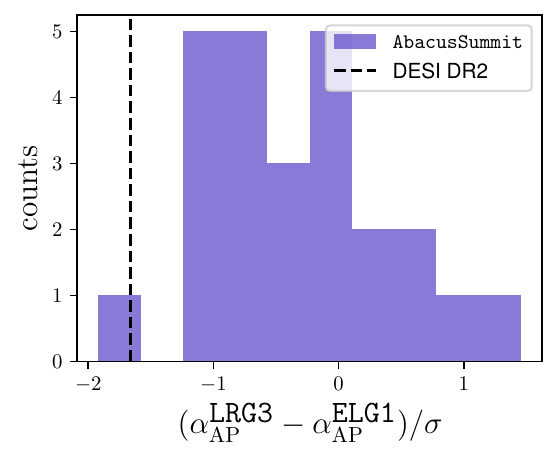}    
    \caption{Offsets in $\alpha_{\rm iso}$ (top) and $\alpha_{\rm AP}$ (bottom) between the \lrgth and \elgo samples for the \desidrtwo data (black) compared to the distribution from 25 \abacussecond DR2 mocks (purple). The observed differences, $\Delta \alphaiso = 0.41\sigma$ and $\Delta \alphaap = -1.66\sigma$, are consistent with noise, validating the combination of \lrgth and \elgo into \lrgelg for cosmological analysis.}
    \label{fig:consistency_lrg_elg}
\end{figure}

All DESI tracers are split by type and redshift when analyzing subsamples in  \cite{DESI.DR2.BAO.cosmo}, with the exception that we use a combination of \lrgelg in the redshift range $0.8<z<1.1$ \citep{Y3BAO-Sanders,KP4s5-Valcin}. In \cref{fig:consistency_lrg_elg}, we have calculated the offset in $\alpha$ constraints between \lrgth and \elgo from DR2 data. We find these to be $\Delta \alphaiso (\lrgth-\elgo) = 0.41\sigma$ and $\Delta \alphaap(\lrgth-\elgo) = -1.66\sigma$. We have explicitly verified that these offsets are compatible with noise fluctuations by comparing them to the same output generated from the 25 mock realizations. This serves as a validation of the consistency of the \lrgth and \elgo samples (unblinding test \#10), enabling their combination into the \lrgelg tracer that is used for the cosmological interpretation.

\subsubsection{Goodness-of-Fit and \texorpdfstring{$\chi^2$}{Chi-squared} Tests}
\label{subsec:chi2_tests}

\begin{table}
    \centering
    \renewcommand{\arraystretch}{1.2}
    \begin{tabular}{lccc}
        \hline\hline
        {Tracer} & {Redshift Range} & {$\chi^2$/dof} & {PTE} \\
        \hline
        \bgs & $0.1 < z < 0.4$ & $26.09 / 16 $ & 0.053 \\
        \lrgo & $0.4 < z < 0.6$ & $40.27 / 33 $ & 0.180 \\
        \lrgt & $0.6 < z < 0.8$ & $23.44 / 33 $ & 0.891 \\
        \lrgth & $0.8 < z < 1.1$ & $53.77 / 33 $ & 0.013 \\
        \lrgelg & $0.8 < z < 1.1$ & $38.44 / 33 $ & 0.237 \\
        \elgo & $0.8 < z < 1.1$ & $26.87 / 33 $ & 0.765 \\
        \elgt & $1.1 < z < 1.6$ & $42.39 / 33 $ & 0.127 \\
        \qso & $0.8 < z < 2.1$ & $22.19 / 33 $ & 0.923 \\
        \hline\hline
    \end{tabular}
    \caption{Summary of $\chi^2$ goodness-of-fit values for the BAO measurements across different tracers and redshift bins using unblinded DESI DR2 data. The probability to exceed (PTE) quantifies the likelihood of obtaining a $\chi^2$ value as large or larger than the observed value, assuming the model provides a good fit to the data. This serves as a diagnostic check to assess whether the BAO fits are statistically reasonable.}    
    \label{tab:chi2_summary}
\end{table}

\paragraph{Validation Using DR2 BAO Measurements:}  
The $\chi^2$ values for the BAO fits, presented in \cref{tab:chi2_summary}, indicate that the models provide statistically reasonable fits to the data. To quantify this, we report the probability to exceed (PTE), which measures the likelihood of obtaining a $\chi^2$ value as large or larger than the observed one, assuming the model is a good fit. As a general guideline, PTE values close to 0 or 1 suggest a poor fit, as they indicate that the observed $\chi^2$ is either much lower or much higher than expected. Conversely, values near 0.5 indicate a fit consistent with statistical expectations.
In our results, the PTE values range from 0.01 (\lrgth) to 0.92 (\qso), with only \lrgth falling below the conventional 0.05 threshold. However, \lrgth is not used in isolation for cosmological analysis; instead, the combined \lrgelg sample is used, which has a PTE of 0.24, well within an acceptable range. The \bgs, \lrgo, \lrgt, \elgo, and \elgt samples all show $\chi^2$ values consistent with statistical expectations, with PTE values between 0.05 and 0.89, indicating no significant deviations. The \qso sample, with a PTE of 0.92, exhibits a slightly lower $\chi^2$/dof ratio, consistent with its lower signal-to-noise ratio and expected statistical fluctuations.
Overall, these results confirm that our BAO measurements are statistically robust and that the reconstructed correlation functions provide a reasonable fit to the data. This directly addresses unblinding test \#1, which is considered satisfactorily passed for all tracers.

\paragraph{Validation Using Mock Catalogs:}
To further assess the reliability of our BAO error estimates, we compare the measured values to expectations derived from mock catalogs. \cref{tab:fit_mocks} presents results from fits to the mean of 25 mock realizations, where the covariance matrix has been rescaled by a factor of 25. This ensures that the errors reflect the combined volume of all realizations, providing a stringent test of the unbiasedness of the mocks.
None of the fits deviate from the expectation of $\alphaiso = \alphaap = 1$ by more than $3\sigma$, reinforcing the reliability of our modeling framework. Pre-reconstruction measurements show a slight tendency for $\alpha$ values larger than one, a well-known effect of non-linear gravitational evolution, which systematically shifts the BAO position to smaller separations (resulting in larger $\alpha$ values). This bias is significantly reduced after reconstruction, confirming that our pipeline effectively corrects for these non-linear shifts.
These findings address unblinding test \#7, demonstrating that our mock-based predictions align well with our observed results.

\subsection{Robustness of the \desidrtwo measurements}
\label{sec:robustness}

\begin{figure*}
    \centering
    \includegraphics[width=\textwidth]
    {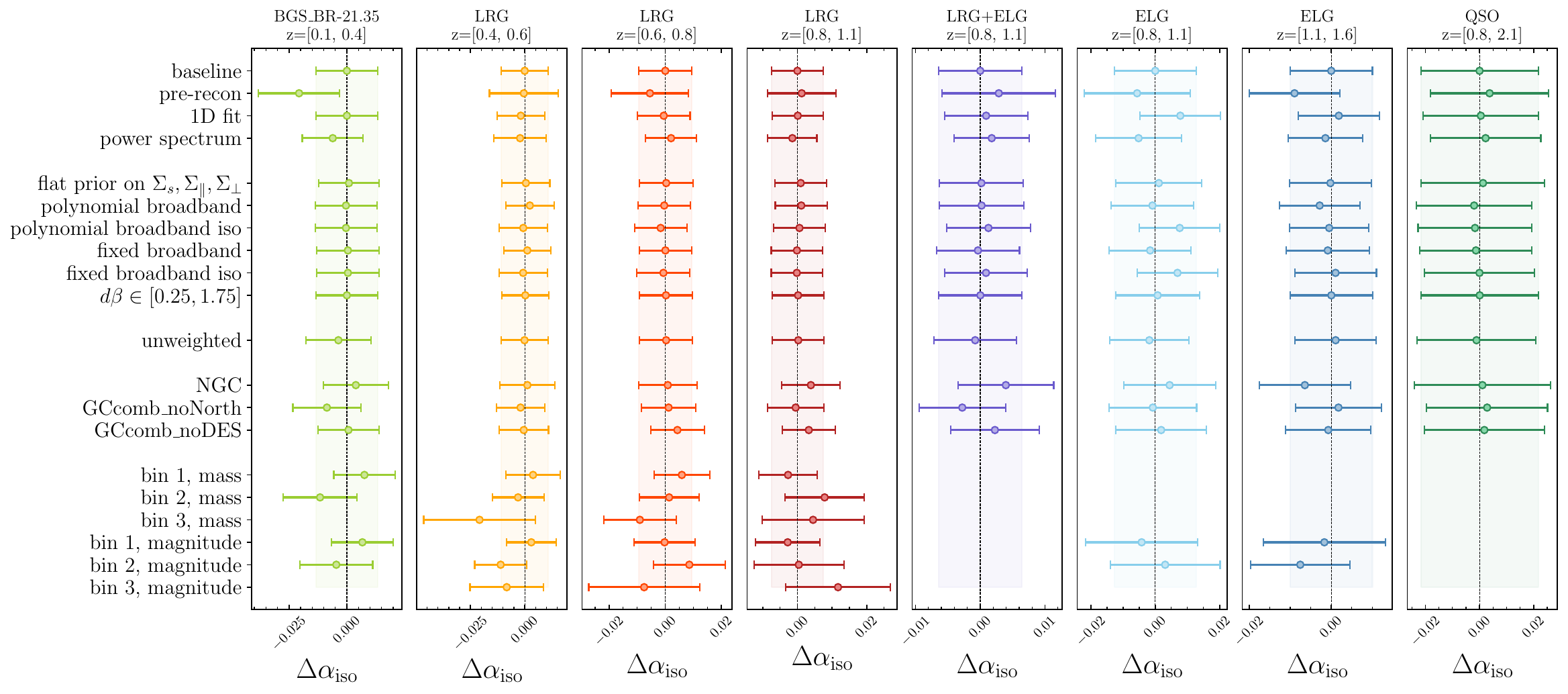}
    \includegraphics[width=\textwidth]{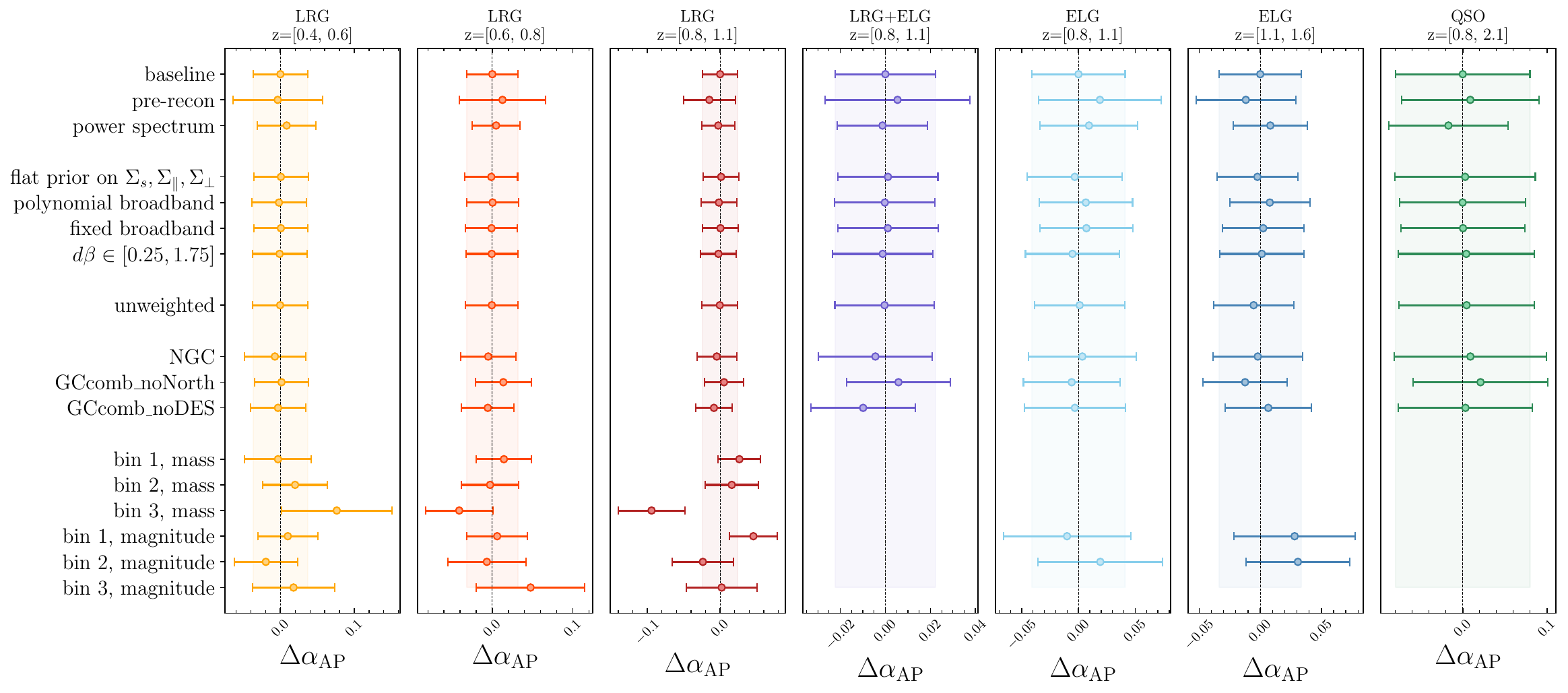}
    \caption{Response of the constraints on the isotropic (top) and anisotropic (bottom) BAO scaling parameters to changes in the data vector or model assumptions, using unblinded \desidrtwo data. The first row represents the difference between the baseline BAO constraints and itself, resulting in values centered at zero, with the shaded regions indicating the statistical uncertainty of the baseline measurements. The remaining rows show the difference $\Delta \alpha$ between the BAO scaling parameters obtained in each test and the baseline value. Specifically, $\Delta \alpha_{\rm iso} = \alpha_{\rm iso}^{\text{case}} - \alpha_{\rm iso}^{\text{baseline}}$ and $\Delta \alpha_{\rm AP} = \alpha_{\rm AP}^{\text{case}} - \alpha_{\rm AP}^{\text{baseline}}$. Note the distinction in the BAO analysis: for the \bgs tracer, one-dimensional (1D) fits are used, whereas two-dimensional (2D) fits are applied for the other tracers.}

    \label{fig:y3unblindedwhisker}
\end{figure*} 

We have extensively examined the impact of several variations around our baseline settings on the BAO constraints, which are shown in \cref{fig:y3unblindedwhisker}. Results are expressed in terms of the absolute difference in $\alphaiso$ and $\alphaap$ relative to the baseline, defined as $\Delta \alpha_{\rm iso} = \alpha_{\rm iso}^{\text{case}} - \alpha_{\rm iso}^{\text{baseline}}$ and $\Delta \alpha_{\rm AP} = \alpha_{\rm AP}^{\text{case}} - \alpha_{\rm AP}^{\text{baseline}}$. Each test corresponds to a different robustness check performed to assess the stability of the BAO constraints under various assumptions.

{\bf Pre-recon:} The first comparison is with the pre-reconstruction correlation function fits. The mean values of the parameters show expected shifts relative to the baseline, with variations that generally do not exceed $1\sigma$, except for the \bgs sample, which deviates by $1.18\sigma$. These shifts are consistent with the nonlinear displacement of the BAO feature, which reconstruction is designed to correct. The level of improvement observed post-reconstruction closely matches expectations from DR2 mocks (see \cref{fig:scatteralpha}).
As expected, parameter precision is reduced in the pre-reconstruction fits compared to the baseline, reconstructed measurements. The gain in precision after reconstruction aligns well with the trends observed in the mocks, reinforcing the robustness of our analysis. This comparison further validates unblinding tests \#2 and \#6.

{\bf 1D BAO fit:} Next, we examine the constraints obtained using a 1D BAO fit, where the only dilation parameter that is varied is $\alphaiso$, with $\alphaap = 1$ fixed. \bgs is the only tracer that uses a 1D fit by default, and therefore, this constraint agrees exactly with the baseline by construction. The $\alphaiso$ constraints from 1D fits show excellent agreement with their 2D counterparts, both in terms of precision and mean values. The deviations from the baseline range from $0.01\sigma$ (\lrgth) to $0.62\sigma$ (\elgo), fully consistent with statistical fluctuations expected from the mocks.

{\bf Power spectrum BAO fit:} 
Constraints from the galaxy power spectrum show good agreement with the configuration-space results when compared to the agreement expected from the mocks. This comparison is explored in more detail in \cref{app:fourier}, where we assess the consistency of power spectrum and correlation function fits under different data vector variations. While these two statistics should, in principle, contain the same amount of information, practical differences arise due to the adopted scale cuts, modeling choices, and particularly the treatment of the covariance matrix. These factors can lead to small variations in the derived constraints. The \bgs sample presents the largest deviation, showing a $0.55\%$ offset, which is slightly larger than the most extreme mock realization but remains within an acceptable range. The overall agreement reinforces the robustness of our methodology and ensures that any differences between the two approaches are well understood. These results directly inform the unblinding test \#8.

{\bf Flat priors on the damping scales:} Our baseline configuration assumes Gaussian priors for the BAO damping and FoG parameters (\cref{eq:damping_fog}), informed from fits to mock catalogs \citep{KP4s2-Chen}. We find that switching these to uniform priors does not significantly affect the constraints on the scaling parameters, with only a very mild degradation in the $\alphaiso$ precision of the \qso. However, fits on multiple realizations of mock catalogs showed more stability of the BAO fits when adopting an appropriately chosen Gaussian prior, ensuring unbiasedness of the parameter constraints without compromising precision \citep{KP4s2-Chen}.

{\bf Polynomial broadband (iso for monopole only):} Our parameterization of the broadband shape of the power spectrum involves a new spline basis \cref{eq:spline_pk} that was introduced in \cite{KP4s2-Chen} for the DR1 BAO analysis. The `polynomial broadband' row in \cref{fig:y3unblindedwhisker} shows constraints obtained using the polynomial-fitting method similar to the approach used in BOSS \citep{BOSS:2016apd}, which are perfectly consistent with the results from the new spline method.

{\bf Fixing broadband (iso for monopole only):} 
We test the impact of broadband modeling on our BAO results by comparing cases where the broadband is fixed for both the monopole and quadrupole of the correlation function (`fixed broadband') versus only for the monopole (`fixed broadband iso'). The results show a very mild impact, with the largest observed shift being 0.55$\sigma$ in $\alphaiso$ for \elgo.

{\bf The effect of the $d\beta$ prior:} The $d\beta$ parameter quantifies the ratio between the true value of the growth rate of structures $f$ and that predicted by the fiducial cosmology. The bounds of the uniform prior that is used as our baseline are [0.7, 1.3], which we deem a reasonably relaxed prior given the constraining power of our data. We tested relaxing those priors even further to [0.25, 1.75], finding a negligible impact on the BAO constraints.

{\bf Unweighted -- imaging systematics:} The correlation between the observed galaxy density and certain imaging properties of the survey in the same sky region produces `imaging systematics' that can bias the clustering measurements. To mitigate them, the LSS catalogs incorporate weights that are added using a regression method that is informed by the maps of different observational properties. Although the clustering itself can be significantly affected by these weights, the `unweighted' row in \cref{fig:y3unblindedwhisker} shows that BAO remains robust even when the weights are completely removed from the catalog. For further validation, \cite{KP3s2-Rosado} specifically demonstrated that DESI BAO measurements for \elgs, the sample that is most susceptible to these effects, remain unaffected by imaging systematics. This directly addresses the unblinding test \#3.

{\bf NGC, noNorth, noDES:} We perform BAO fits under various data splits, described in \cref{subsec:samples_and_splits}. We observe a reduction in precision when restricting the analysis to the NGC region, which is expected given the decreased sky coverage (30\% and 44\% less area for bright-time and dark-time tracers, respectively). However, parameter shifts remain consistent at the 1$\sigma$ level. A smaller reduction in precision is observed when excluding the North (31\% and 20\% of the DR2 sky area for bright and dark time, respectively) or the DES (7\% and 9\%) imaging regions, with posterior mean shifts remaining within 1$\sigma$ of the baseline\footnote{For further details on the impact of using photometric redshifts from the DESI Legacy Imaging Survey on the BAO analysis, see \cite{Saulder:2025dzg}.}. To further quantify the robustness of these survey splits, we performed a Kolmogorov-Smirnov (KS) test on the significance levels ($\sigma$) of the shifts in $(\alpha_{\rm DR2} - \alpha_{\rm sub})/\sigma_{\rm diff}$. The KS test assesses whether a given sample follows an expected probability distribution by comparing its empirical cumulative distribution function (CDF) to a theoretical CDF. In our case, under the null hypothesis, the shifts in $\alpha_{\rm DR2} - \alpha_{\rm sub}$ should be Gaussian-distributed noise, implying that the computed $\sigma$ values should follow a standard normal distribution, $\mathcal{N}(0,1)$. The KS test quantifies the maximum difference between the empirical CDF of the observed $\sigma$ values and the CDF of a unit Gaussian, providing a measure of how well the data matches the expected distribution. To construct the test statistic, we computed the observed shifts as $\alpha_{\rm DR2} - \alpha_{\rm sub}$, which represent the differences in the BAO scale measurements between the full DR2 sample and each subsample. To evaluate the significance of these shifts, we estimated the corresponding uncertainty, $\sigma_{\rm diff}$, directly from the data using:

\[
\sigma^2_{\rm diff} \equiv \sigma^2_{\rm DR2} \left( \frac{N_{\rm DR2}}{N_{\rm sub}} - 1 \right),
\]
where $\sigma^2_{\rm DR2}$ is the variance of $\alpha_{\rm DR2}$, and $N_{\rm DR2}$ and $N_{\rm sub}$ represent the number of galaxies in the full sample and subsample, respectively.\footnote{This formula follows from the standard variance scaling relation for subsample: 
\[
\sigma_{\rm sub}^2 = \sigma_{\rm DR2}^2 \frac{N_{\rm DR2}}{N_{\rm sub}}.
\]
Since the subsample is drawn from the full dataset, the uncertainty difference is set by the sample size ratio, ensuring that $\sigma_{\rm sub}^2 - \sigma_{\rm DR2}^2$ remains positive. This approach properly accounts for the correlation between the full sample and its subsamples, making variance subtraction a valid choice in this specific case.
As a result, the uncertainty associated with the shift in \(\alpha\) can be expressed as:

\[
\sigma=\sqrt{\sigma_{\rm sub}^2-\sigma^2_{\rm DR2}},
\]
which remains well-defined since \(\sigma_{\rm sub}^2\) has been properly rescaled. This formulation assumes \textit{a maximum statistical correlation} between the subsample and the full sample, making it a conservative estimate of the uncertainty.  
}
This definition ensures that the uncertainty accounts for the relative sample sizes and provides a robust estimate of expected fluctuations. By normalizing the observed shifts using $\sigma_{\rm diff}$, we obtained a standardized significance measure that allows direct comparison across different subsamples.

In total, we tested 48 cases, incorporating the \bgs, \lrgo, \lrgt, \lrgth, \lrgelg, \elgo, \elgt, and \qso tracers for both $\alpha_{\rm iso}$ and $\alpha_{\rm AP}$, comparing results for the full sample against the NGC, GCcomb-noDES, and GCcomb-noNorth regions. We obtain a KS statistic of $D = 0.1489$ with a corresponding $p$-value of 0.215, indicating no significant deviation from normality. These results suggest that the observed shifts in $\alpha_{\rm DR2} - \alpha_{\rm sub}$ are consistent with statistical fluctuations and do not indicate systematic biases across the different survey splits.

{\bf Mass and magnitude splits:} The final set of tests assesses the robustness of the BAO measurements against variations in tracer properties by performing mass and magnitude splits, as shown in the last six rows of \cref{fig:y3unblindedwhisker}. Since galaxy bias $b$ depends on stellar mass and luminosity, these properties can influence the amplitude and shape of the two-point correlation function; see, for instance, the differences in clustering amplitudes shown in \cref{tab:amplitude}. To test this, we divide the \bgs, \lrg, and \elg samples into subsamples based on mass and magnitude, following the methodology outlined in \cref{sec:data}. We note that we used the same bias values as in the baseline case, which are depicted in \cref{tab:recon_params}. We merge the mass/magnitude-split subsamples and perform reconstruction, then extract the shifted subsamples by their mass/magnitude bin labeled before the reconstruction.

Although some deviations from the baseline measurements are visually observed in \cref{fig:y3unblindedwhisker}, a quantitative consistency check confirms that the $\alpha_{\rm iso}$ and $\alpha_{\rm AP}$ values remain statistically consistent with our baseline results. Specifically, we performed a KS test, as detailed previously, on all mass and magnitude subsample cases. The results yielded a $p$-value of 0.73, indicating that the observed variations across mass- and magnitude-split subsamples are consistent with statistical fluctuations. To quantify the uncertainty, we estimate the joint uncertainty in the KS test as

\[
\sigma=\sqrt{\sigma_{\rm sub}^2-\sigma^2_{\rm DR2}},
\]
assuming a maximum statistical correlation between the subsamples and the baseline sample, ensuring a conservative uncertainty estimate.
{The maximum difference is in $\alphaap$ of mass bin 3 of \lrgth, ranging from $2.0\sigma$ to $2.4\sigma$, depending on the value of correlation coefficient.}  
These results confirm that BAO measurements are robust to mass and magnitude selection effects. The combination of all data-split tests informs the unblinding test \#4, which is considered satisfactorily passed.
\\

\begin{table}
    \centering
    \renewcommand{\arraystretch}{1.2}
    \setlength{\tabcolsep}{6pt}
    \begin{tabular}{c|cc|cc}
        \hline\hline
        \textbf{Tracer} & \textbf{Mag. bin} & \boldmath{$\xi_0^{s=20}$}  & \textbf{Mass bin} & \boldmath{$\xi_0^{s=20}$} \\
        \hline
        \bgs & $m_{\rm r}$ bin1 & 0.250 & $M_*$ bin1 & 0.276\\
        \bgs & $m_{\rm r}$ bin2 & 0.195 & $M_*$ bin2 & 0.177\\
        \hline
        \lrgo & $m_{\rm W1}$ bin1 & 0.313 & $M_*$ bin1 & 0.339\\
        \lrgo & $m_{\rm W1}$ bin2 & 0.244 & $M_*$ bin2 & 0.259\\
        \lrgo & $m_{\rm W1}$ bin3 & 0.212 & $M_*$ bin3 & 0.180\\
        \hline
        \lrgt & $m_{\rm W1}$ bin1 & 0.294 & $M_*$ bin1 & 0.333\\
        \lrgt & $m_{\rm W1}$ bin2 & 0.241 & $M_*$ bin2 & 0.254\\
        \lrgt & $m_{\rm W1}$ bin3 & 0.223 & $M_*$ bin3 & 0.190\\
        \hline
        \lrgth & $m_{\rm W1}$ bin1 & 0.293 & $M_*$ bin1 & 0.304\\
        \lrgth & $m_{\rm W1}$ bin2 & 0.229 & $M_*$ bin2 & 0.226\\
        \lrgth & $m_{\rm W1}$ bin3 & 0.193 & $M_*$ bin3 & 0.186\\
        \hline
        \elgo & $m_{\rm g}$ bin1 & 0.084 & - & - \\
        \elgo & $m_{\rm g}$ bin2 & 0.079 & - & - \\
        \hline
        \elgt & $m_{\rm g}$ bin1 & 0.091 & - & - \\
        \elgt & $m_{\rm g}$ bin2 & 0.082 & - & - \\
        \hline\hline
    \end{tabular}
    \caption{Summary of clustering amplitudes in mass/magnitude-split subsamples. The clustering amplitudes $\xi_0^{s=20}$ are defined as the two-point correlation function monopole $\xi_0$ at $s=20\mathrm{Mpc}/h$. In each block, the amplitude decreases from bin1 to bin3. The parameters $m_{\rm r}$, $m_{\rm W1}$, and $m_{\rm g}$ correspond to the extinction-corrected apparent magnitudes in the $r$-band, $W1$-band, and $g$-band, respectively, as measured in the Legacy Survey DR9. These magnitudes were used to define the magnitude splits in the \bgs, \lrg, and \elg samples.}
    \label{tab:amplitude}
\end{table}

Overall, our extensive suite of robustness tests demonstrates the stability of the BAO measurements against a range of potential systematics. We performed tests involving variations in modeling assumptions, data-vector choices, survey region splits, mass/magnitude selections, and imaging systematics. The high level of consistency observed across all tests strongly supports the reliability of our measurements. The agreement across these complementary robustness checks suggests that our constraints are not significantly biased by observational systematics, sample selection, or modeling assumptions. This reinforces the robustness of the \desidrtwo BAO analysis and its suitability for cosmological interpretation.

\subsection{Gaussianity of the BAO Posterior}
\begin{figure*}
    \centering
    \begin{tabular}{ccc}
         \hspace{-1cm}\includegraphics[width=0.2\textwidth]{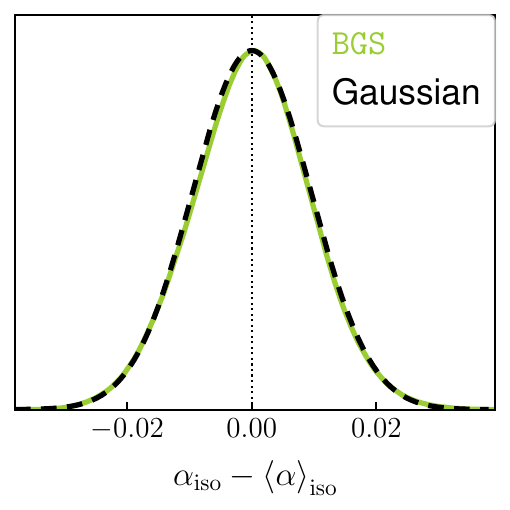}   \\ 
        \includegraphics[width=0.3\textwidth]{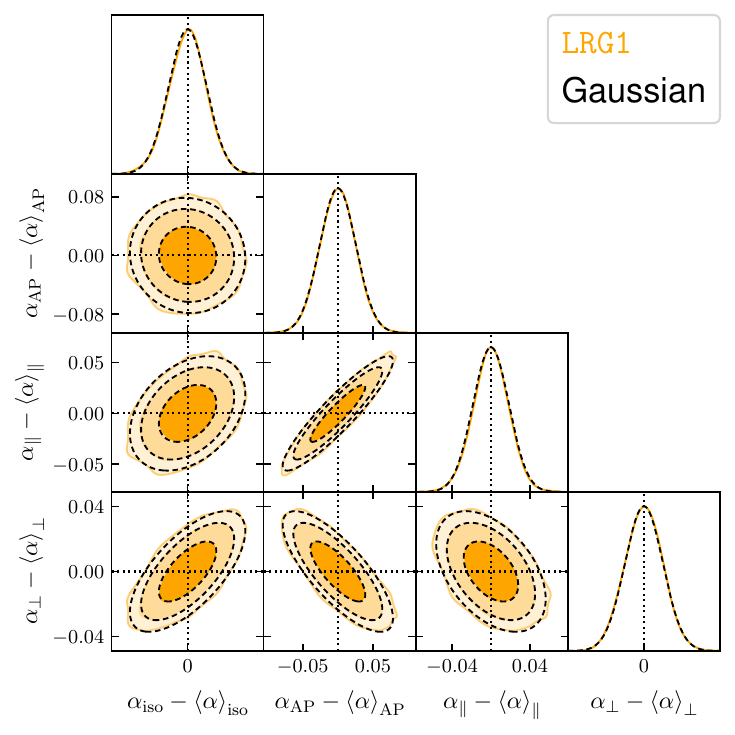} & \includegraphics[width=0.3\textwidth]{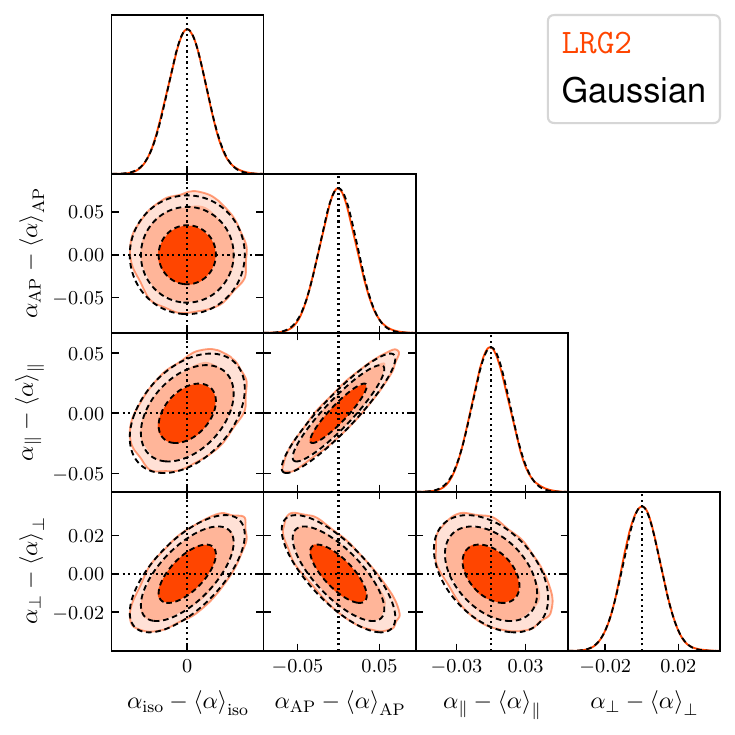} &
        \includegraphics[width=0.3\textwidth]{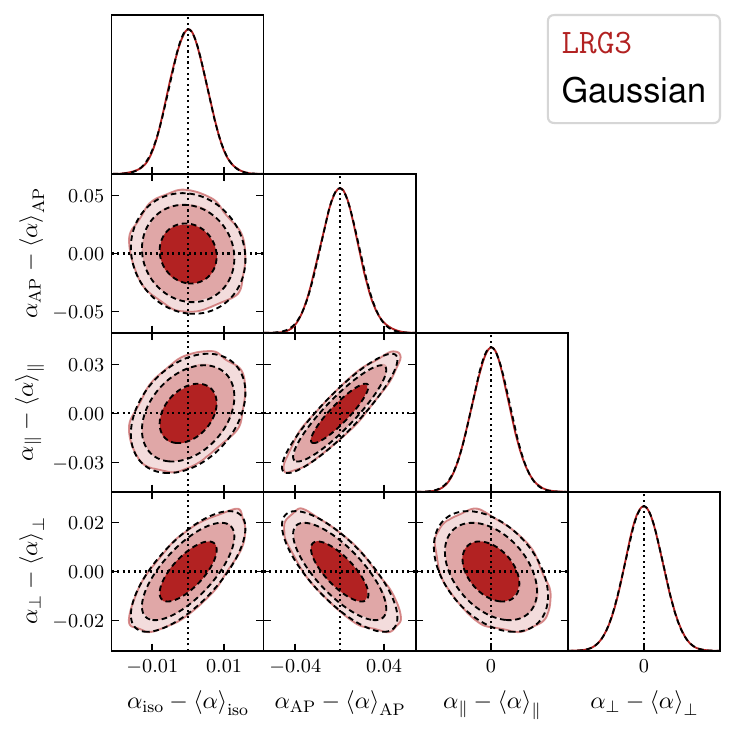}    \\
        \includegraphics[width=0.3\textwidth]{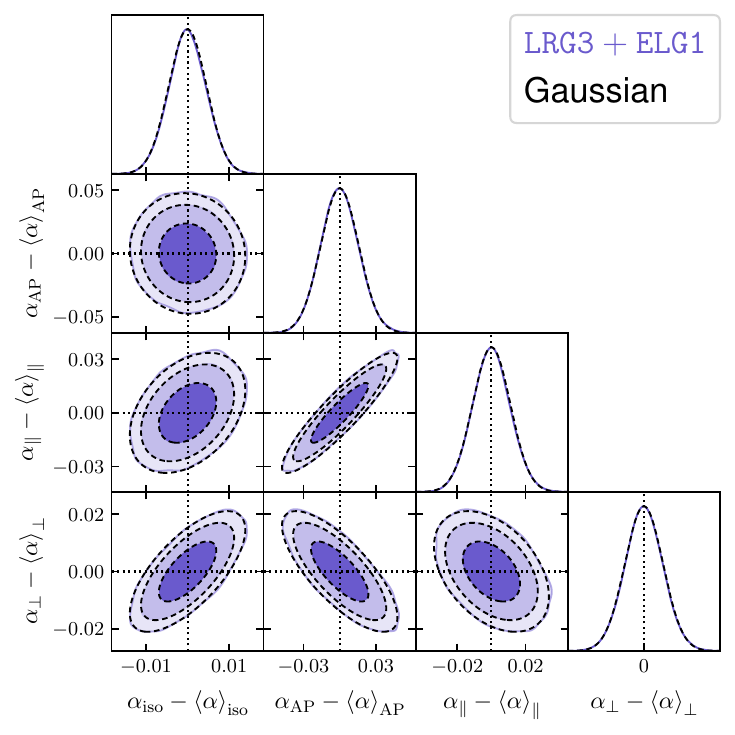} &
        \includegraphics[width=0.3\textwidth]{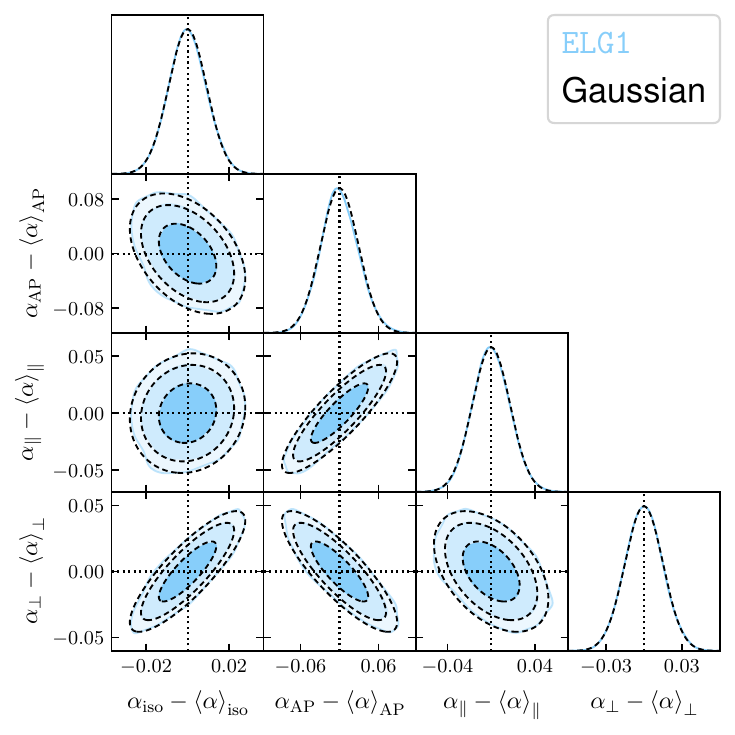} & \includegraphics[width=0.3\textwidth]{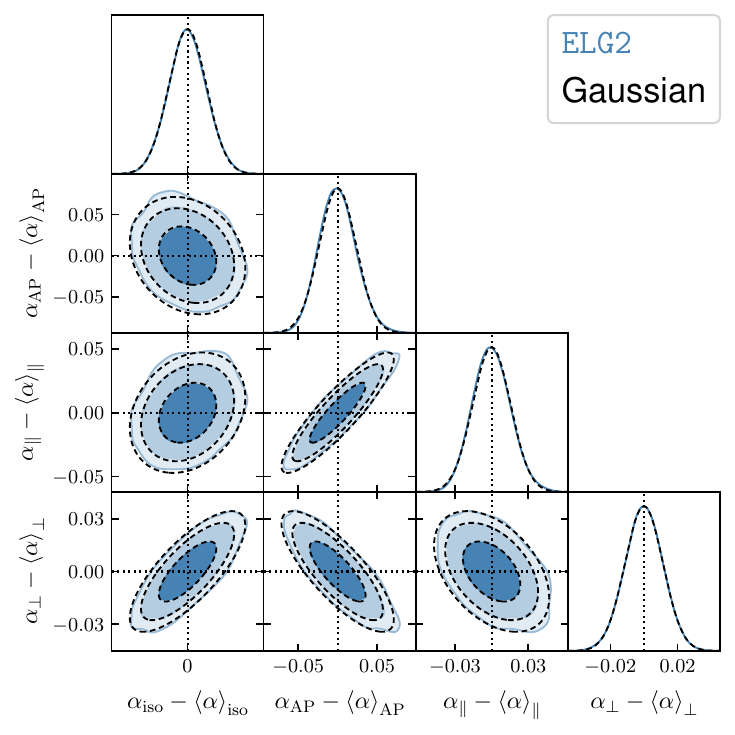} \\
        \includegraphics[width=0.3\textwidth]{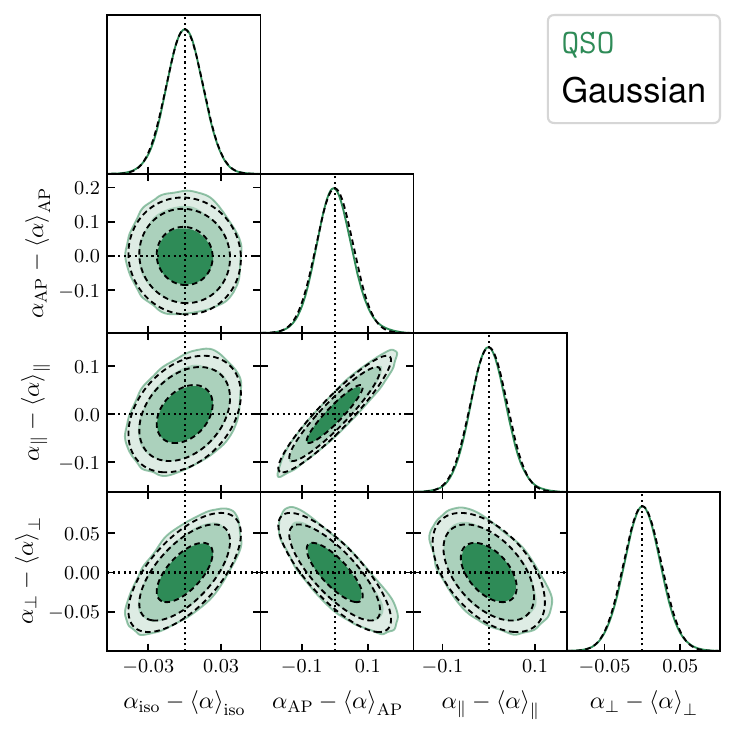}
    \end{tabular}
\caption{Marginalized posterior distributions of BAO scaling parameters in two different bases, compared against a multivariate Gaussian with the same mean and covariance matrix. The inner, middle, and outer contours correspond to the 68\%, 95\%, and 99\% confidence regions, respectively. The results demonstrate that the data posteriors exhibit excellent agreement with Gaussianity for all tracers.}

    \label{fig:gaussianity}
\end{figure*}

Once the posterior distribution of the BAO dilation parameter is determined, we can evaluate our fiducial cosmology in \cref{eq:alphaiso_alphaap} to convert our $\alphaiso$ and $\alphaap$ constraints into distance measurements. When these constraints are passed onto the next stage of the pipeline to begin the inference of cosmological parameters, they are compressed into a Gaussian posterior that is only characterized by its mean and covariance matrix, extracted from the chains. Deviations from Gaussianity could impact the resulting cosmological constraints, so this assumption must be previously examined to ensure consistency.

\cref{fig:gaussianity} shows the posterior distribution of BAO parameters from the \desidrtwo data. The mean of the marginalized posterior of each parameter has been subtracted to ensure the constraints remain blinded. We overlay a multivariate Gaussian distribution with the same mean and covariance as the data. Overall, the data posterior shows excellent agreement with Gaussianity for all tracers. Some small deviations can be spotted at the 3$\sigma$ contours, especially in tracers with lower signal-to-noise ratios such as the \qso. However, we have explicitly verified that most of these small differences can be attributed to sampling noise from the MCMC chains. This figure also highlights the reduced parameter correlation in the $\alphaiso$-$\alphaap$ plane compared to the constraints in the $\alphapar$-$\alphaper$ basis.

\section{Post-Unblinding Tests}
\label{sec:post_unblinding}

While our fiducial choices were determined solely based on the validation process conducted with blinded data, \textit{we also planned a set of post-unblinding tests in advance} to assess the stability of our results. These tests were designed before unblinding to prevent selection biases and ensure they did not influence the baseline analysis choices. These tests offer a complementary assessment of systematic uncertainties that may inform future refinements in the DESI analysis pipeline.  

Below, we summarize the key post-unblinding tests conducted in this study.

\begin{table*}
    \centering
        \begin{tabular}{c|c|c|c|c|c|c}
            \hline
            \hline
            Cosmology & Description &  $r_{\rm d}/r_{\rm d}^{\rm baseline}$  & $D_{\rm V}/ D_{\rm V}^{\rm baseline}$ & $\frac{D_{\rm H}}{D_{\rm M}}/\frac{D_{\rm H}^{\rm baseline}}{D_{\rm M}^{\rm baseline}}$ & $f/f^{\rm baseline}$ & $b_1/b_1^{\rm baseline}$ \\
            \hline
            
            \texttt{c001} & $\Omega_m=0.276$ $\Lambda$CDM & 1.056& 1.002 -- 1.044  & 1.001 -- 1.025 & 0.95 -- 0.99 & 1.05 -- 1.13 \\
            
            \texttt{c002} &   $w_0 = -0.7$, $w_a = -0.5$ (thawing DE) & 0.932& 0.931 -- 0.989  & 0.9713 -- 0.995 & 0.98 -- 1.00 & 0.95 -- 1.07 \\
            
            \texttt{c003} & $N_{\rm eff} = 3.70 $ $\Lambda$CDM & 1.022& 1.001 -- 1.019 & 1.001 -- 1.011 & 0.98 -- 1.00 &0.95 -- 1.07 \\
            
            \texttt{c004} & $\sigma_8 = 0.75$ $\Lambda$CDM & 1.000& 1.000 -- 1.000  & 1.000 -- 1.000 & 1.00 & 1.10 -- 1.20\\
            \hline
            \hline
        \end{tabular}
     \caption{Secondary \texttt{AbacusSummit} cosmologies used to test the impact of the fiducial cosmology. Each column shows the ratio of a key cosmological quantity relative to our baseline fiducial cosmology. The third column quantifies the effect of the fiducial sound horizon ratio on the template, while the fourth and fifth columns display the range of expected geometrical distortions across the redshift interval $0.1<z<2.1$, with values varying as a function of redshift. Specifically, $D_V$ and $D_H/D_M$ are computed using the fiducial cosmology in each case, and their ratios with respect to the baseline cosmology are reported as ranges spanning the redshift interval. The last two columns indicate how the values of the growth rate $f$ and the linear galaxy bias $b_1$—both required for BAO reconstruction—change for each cosmology (again, showing their range across all tracer samples). The largest deviations in fiducial distances occur for the thawing dark energy case, where an isotropic rescaling of up to $\sim7\%$ and an anisotropic distortion of up to $\sim 3\%$ are observed. In contrast, the low $\sigma_8$ cosmology maintains the same fiducial distances as the baseline cosmology, affecting only the galaxy bias assumptions for reconstruction.}     
     
     \label{tab:cosmologies}
\end{table*}

\subsection{Correlation in Theoretical Systematics}
\begin{figure}
    \centering
    \includegraphics[width=1.\linewidth]{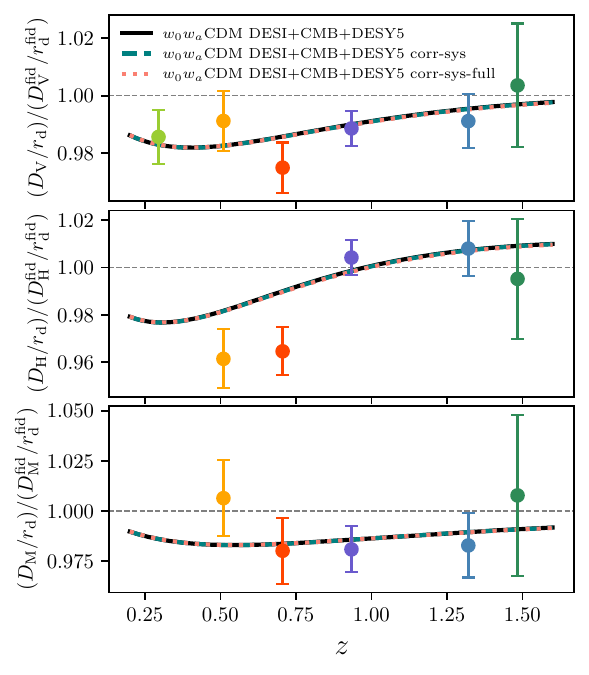}
    \caption{Impact of correlated theoretical systematics on the cosmological inference from BAO distance measurements. Each panel shows the ratio of measured BAO distances relative to the fiducial cosmology, including $D_V/r_d$ (top), $D_H/r_d$ (middle), and $D_M/r_d$ (bottom). The solid black curve represents the best-fit $w_0w_a$CDM model using DESI+CMB+DESY5. The teal dashed curve includes the additional effect of partially correlated systematic uncertainties, while the salmon dotted curve shows the scenario where theoretical systematics are fully correlated across redshift bins. The comparison between these cases quantifies the potential bias introduced by different assumptions about systematic correlations. 
    Error bars include the systematic error budget in each redshift bin.
    }
    \label{fig:hubble_corrsys}
\end{figure}

As in DESI DR1, our {baseline analysis} includes both statistical and systematic uncertainties, with systematic errors treated as uncorrelated across redshift bins and tracers. However, as a \textit{sanity check}, we explore the impact of allowing these systematic uncertainties to be correlated across redshift bins and different tracers, testing whether such correlations significantly affect the inferred BAO constraints.

In both DR1 and DR2 analyses, we adopt a {0.1\% theoretical systematic uncertainty on} $\alpha_{\rm iso}$ {and a 0.2\% uncertainty on} $\alpha_{\rm AP}$, as shown in \cref{tab:systematic_errors}. Since cross-tracer systematic correlations could arise from common theoretical assumptions—such as reconstruction modeling—we assess their impact by introducing correlations \textit{only for theoretical systematics}, while keeping fiducial cosmology and HOD systematics uncorrelated. We evaluate two extreme cases:

\begin{itemize}
    \item \textbf{Fully Correlated:} The theoretical systematic errors are assumed to be completely correlated across redshift bins and tracers, meaning that a systematic shift in one bin applies identically to all others. This represents the strongest correlation scenario we test.

    \item \textbf{Partially Correlated:} We introduce a moderate level of correlation by splitting the theoretical systematic uncertainty into half-correlated and half-uncorrelated components. Specifically, we assign a covariance of $1/2 \times (0.1\%)^2$ between all $\alpha_{\rm iso}$ values across redshift bins and tracers, and similarly, $1/2 \times (0.2\%)^2$ for $\alpha_{\rm AP}$ values. This allows for some degree of coherence in the systematics while maintaining independent components.
\end{itemize}

By comparing these cases to our baseline, where systematic uncertainties remain uncorrelated, we assess whether correlated theoretical systematics could introduce biases in the BAO constraints. This treatment results in a \textit{non-diagonal} systematics covariance matrix across different redshift bins, modifying the overall uncertainty structure when combined with statistical errors.

The effect of allowing correlated theoretical systematics on the inferred cosmology from the BAO distances is shown in \cref{fig:hubble_corrsys}. Note that the location of the data points and the size of the (diagonal) errors do not change in this figure. The solid black curve represents the standard $w_0w_a$CDM fit using DESI+CMB+DESY5, while the red dashed curve includes the impact of correlated systematic uncertainties. We observe a negligible impact on the best fit $w_0w_a$CDM cosmology, which reflects our statistical errors dominating the systematic errors.

\subsection{Galaxy Bias in Reconstruction}
   The blinded analysis limits us from determining an accurate bias for each tracer before unblinding. For DR2, we used the same baseline $b_1$ values as in DR1 for reconstruction, which were validated against Survey Validation (SV) \cite{DESI2023a.KP1.SV,
   rocher2023desi,yuan2024desi,smith2024generating} data and shown to be accurate within a 10\% margin. For \elgs, we perform reconstruction across the entire redshift range (0.8–1.6) with a bias of 1.2. Due to the wide redshift range, the bias of the \elgs evolves slightly, such that the bias for \elgt is close to or above 1.4 \citep{DESI2024.V.KP5}. This represents more than a 10\% offset from the baseline $b_1=1.2$ that we adopted for reconstructing \elgs, i.e., beyond the 10\% tolerance we consider acceptable for optimal reconstruction. 
  
 After unblinding, therefore, we tested \elgt with $b_1=1.4$ in reconstruction; we found only a 0.04\% shift in $\alphaiso$ and a 0.3\% shift in $\alphaap$ with the same precisions as our baseline with $b_1=1.2$. Note that our pre-unblinding test with the \elgt mocks, which had a clustering amplitude similar to the observed \elgt (and therefore a bias similar to that of the data) was also analyzed with $b_1=1.2$ during reconstruction and we demonstrated that the results were unbiased (\cref{tab:fit_mocks}). These tests again prove the robustness of the BAO reconstruction against the assumption of galaxy bias.

\subsection{BAO fits with different fiducial cosmologies}

As described in \cref{sec:modeling}, the fiducial cosmology assumption enters the pipeline in three stages: the redshift-to-distance conversion, leading to potential geometrical distortions; the sound horizon scale, due to fixing the linear power spectrum template; and the linear bias $b_1$ and growth rate $f$ values assumed when running reconstruction. 

In order to test the impact of the assumption of a fiducial cosmology, we repeated our baseline analysis with four additional sets of cosmological parameters. This test is analogous to the one presented in \cite{KP4s9-Perez-Fernandez}, where the fiducial-cosmology systematic error budget was quantified for the DR1 BAO results \cite{DESI2024.III.KP4}.

 \begin{figure*}
    \centering
    \includegraphics[width=0.98\linewidth]{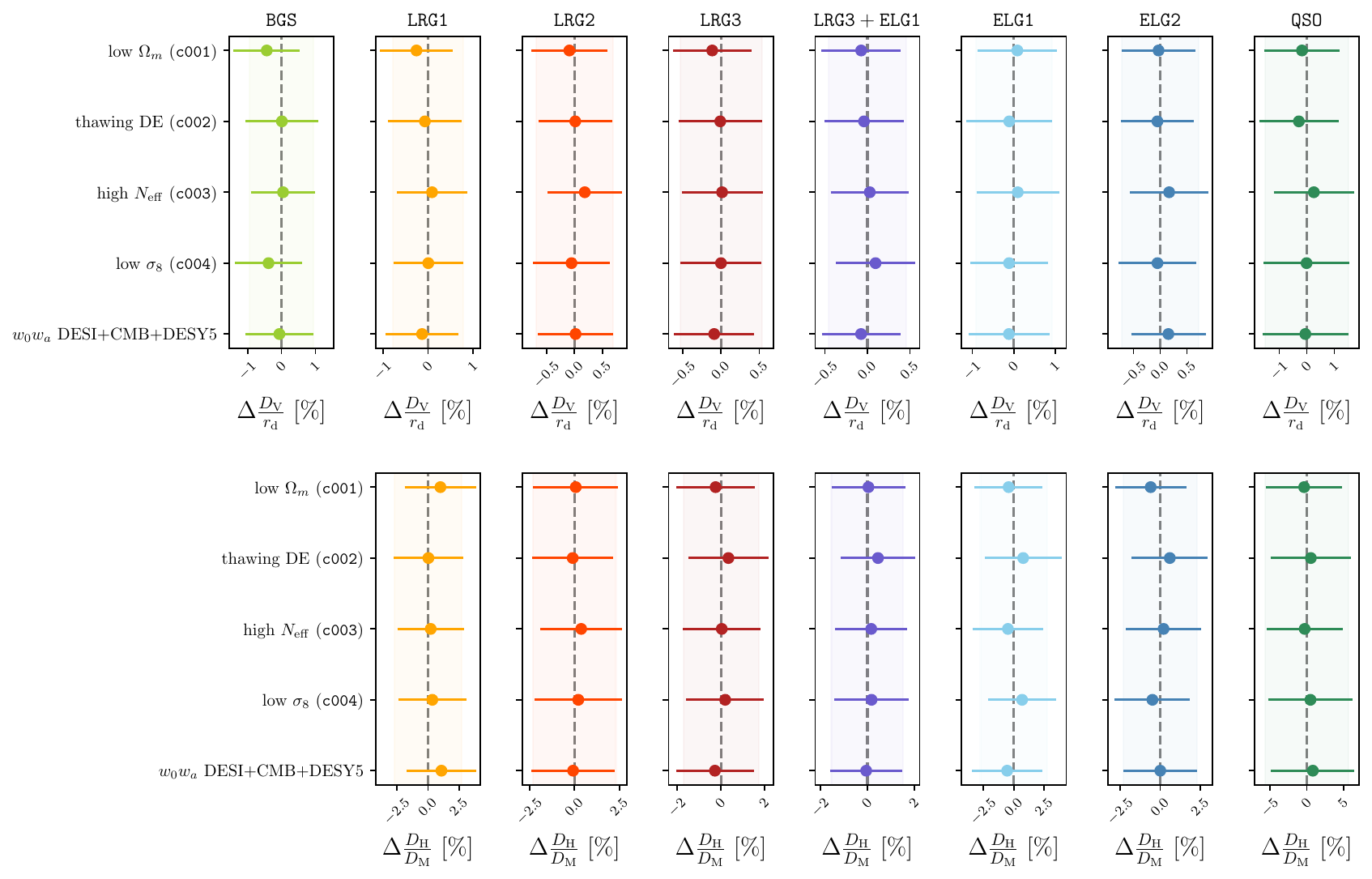}
    \caption{Comparison of the distance measurements obtained by repeating the analysis with different fiducial cosmologies. The differences are taken with respect to the values reported in \cite{DESI.DR2.BAO.cosmo}. {The error bars correspond to the uncertainties in the distances measured for each fiducial cosmology, whereas the shaded regions represent the errors from the baseline analysis. }
    }
    \label{fig:fidcosmo}
\end{figure*}

The tested cosmologies correspond to four alternative models from the \texttt{AbacusSummit} suite \citep{AbacusSummit}, each designed to probe different aspects of the fiducial cosmology assumption. These include a low $\Omega_{\rm m}$ cosmology (\texttt{c001}), a thawing dark energy model with evolving $w(z)$ (\texttt{c002}), a scenario with a higher effective number of relativistic species, $N_{\rm eff}$ (\texttt{c003}), and a variation of our baseline cosmology with a lower matter fluctuation amplitude, $\sigma_8$ (\texttt{c004}). Notably, the BAO analysis has been shown to be robust beyond these models, for instance, in the context of modified gravity \cite{Pan2024JCAP} and early dark energy models \cite{Bernal:2020vbb}.

Each of these cosmologies impacts the BAO analysis in distinct ways, introducing different Alcock-Paczynski distortions across redshift, and modifying the sound horizon scale at the drag epoch due to shifts in the early-universe expansion history. These effects are summarized in \cref{tab:cosmologies}, which provides a quantitative comparison of key cosmological parameters relative to the baseline.

In particular, the thawing-dark-energy cosmology introduces large geometrical distortions, with an expected isotropic contraction of distances of up to $\sim7\%$ and anisotropic distortions of up to $\sim 3 \%$ with respect to the baseline fiducial cosmology.

The high-$N_{\rm eff}$ cosmology has shown to cause shifts of up to $\sim 0.2\%$ in $\alpha_{\rm iso}$ \cite{KP4s9-Perez-Fernandez} due to the effect that changing the number of relativistic species has on the BAO phase shift in the linear power spectrum template. A similar shift had been observed before by \cite{Thepsuriya_2015}, who tested differences of up to $\Delta N_{\rm eff}=2$ between synthetic data and the fitting template. As this effect stems from the assumed rescaling in the AP parameters with the sound horizon ratio (\cref{eq:alphaiso_alphaap}), we regard it as a matter of interpretation.
Nonetheless, if the cosmological analysis had indicated significant deviations from the fiducial $N_{\rm eff}$ value, we would have performed a reanalysis using an updated fiducial cosmology, as planned before unblinding. However, since no such deviations were observed, this was not required.

The low-$\sigma_8$ case, on the other hand, primarily influences the bias used in BAO reconstruction, as this cosmology is otherwise identical to our baseline cosmology except for a low $A_s$ (and consequently a low $\sigma_8$) value. 

For a consistent analysis across different fiducial models, we updated the reconstruction pipeline by adjusting the values of the linear bias ($b_1$) and growth rate ($f$) accordingly. Additionally, we recomputed the \rascalc covariance matrices to reflect the altered redshift-to-distance conversion and reconstruction parameters for each tested cosmology.

\cref{fig:fidcosmo} shows the measurements obtained for $D_{
\rm V}/r_{\rm d}$ and $D_{\rm H}/D_{\rm M}$ in comparison to the values reported in \cite{DESI.DR2.BAO.cosmo}. Across all samples, the deviations from the baseline measurements are relatively small, with the largest differences observed for \bgs. In all cases, the variations are consistent with zero, staying within 0.5$\sigma$. This confirms the stability of the BAO fits against the choice of fiducial cosmology.

\textbf{Pipeline Reanalysis for $w_0w_a$CDM Cosmology:} We repeated the entire analysis pipeline using the best-fit $w_0w_a$CDM cosmology from DESI+CMB+DESY5 data, as reported in Section VII of \cite{DESI.DR2.BAO.cosmo}, to assess the impact of the fiducial cosmology choice on the final BAO measurements. {The results of this reanalysis, shown in \cref{fig:fidcosmo}, demonstrate that the inferred BAO distances remain highly consistent with those obtained using the baseline $\Lambda$CDM fiducial cosmology.} {The largest deviation in $\DH/\DM$ occurs for \lrgo with $0.4\sigma$, while for $\DVrd$, the largest relative difference is $0.2\sigma$, observed in the \elgt bin.} 

\subsection{DESI-SDSS BAO Comparison at $z \sim 0.7$}
We compare the BAO measurements from DESI DR1, DESI DR2, and SDSS at $z \sim 0.7$\footnote{We consider the slight difference on effective redshift ($\Delta z=0.008$) between DESI and SDSS bin by multiplying $D^{\rm fid}(z_{\rm DESI})/D^{\rm fid}(z_{\rm SDSS})$, the ratio of distance in the fiducial cosmology, to the SDSS BAO measurement.}, where the DR1 analysis reported a $2.5\sigma$–$3\sigma$ tension with SDSS. To ensure a consistent comparison, we use the DESI reanalysis of SDSS, which applies the DESI pipeline—including reconstruction, clustering measurements, and BAO fitting to the eBOSS LRG catalog. This reanalysis, presented in Table 17 of \cite{DESI2024.III.KP4}, yields very similar results to the published eBOSS measurements but removes pipeline-induced differences.

\begin{figure}
    \centering
    \includegraphics[width=0.9\linewidth]{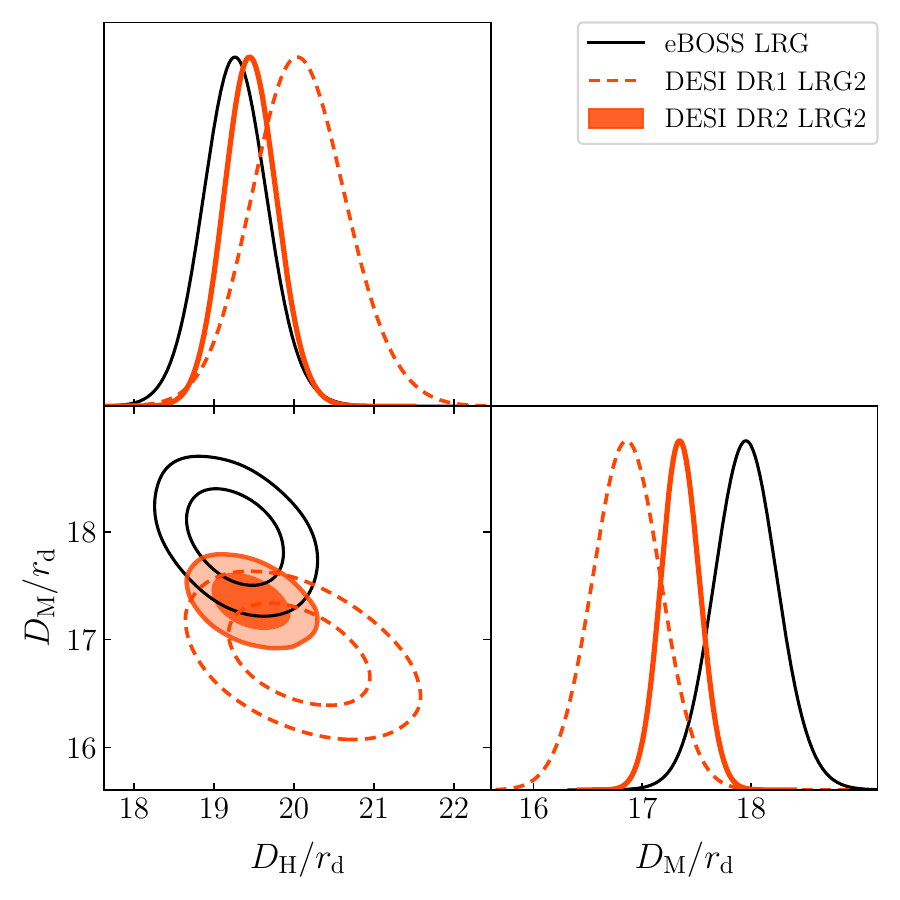}
    \caption{Comparison of BAO constraints from DESI DR1, DESI DR2, and SDSS at $z \sim 0.7$. The plot shows the 68\% and 95\% confidence contours for the LRG2 bin in the $D_H / r_d - D_M / r_d$ plane. The black solid contours represent the eBOSS (SDSS) measurement, the orange dashed contours show the DESI DR1 result, and the filled orange contours correspond to DESI DR2. The DR2 result lies between DR1 and SDSS, supporting the interpretation that the previously observed tension between DR1 and SDSS is likely a statistical fluctuation.}
    \label{fig:sdss_vs_desi}
\end{figure}

As shown in \cref{fig:sdss_vs_desi}, the DR2 measurement lies between DR1 and SDSS, consistent with the interpretation from the DR1 paper that the observed tension is likely due to statistical fluctuations. The improved DR2 precision further supports this conclusion, as it reduces the statistical scatter while remaining broadly consistent with both datasets. A more detailed quantitative comparison can be found in Section III.C.2 of \cite{DESI.DR2.BAO.cosmo}.

\section{Summary and Conclusions}
\label{sec:conclusions}

We have presented the validation of the DESI DR2 measurements of baryon acoustic oscillations from galaxies and quasars. Together with the BAO measurements from the \lya forest \cite{DESI.DR2.BAO.lya}, these constitute the main data set that is used for cosmological inference in \cite{DESI.DR2.BAO.cosmo}. The validation process involved extensive testing with mock galaxy catalogs that mimic the selection and clustering properties of different galaxy tracers, as well as analyses using a blinded version of the \desidrtwo LSS catalogs, followed by complementary post-unblinding checks.

Our analysis pipeline remains largely the same as that used for the DESI DR1 BAO analysis \cite{DESI2024.III.KP4}, with minor modifications to account for the expanded dataset and refinements in analysis choices. These include adjustments to the BGS magnitude cut, an increased minimum fitting scale for BAO measurements, the adoption of 2D BAO fits for select tracers and additional data-splitting tests to assess systematic effects (detailed in \cref{sec:changes_y3}).

We have also outlined a series of rigorous tests to determine whether we were ready to finalize our pipeline settings and unblind the data for the official BAO cosmology analysis. These tests provided key validation steps, ensuring that our results were robust against potential sources of systematic bias (detailed in \cref{sec:unblindin_tests}). This unblinding checklist, detailed in \cref{tab:unblinding_tests}, was successfully completed, leading to the unblinding of the DR2 BAO data during the DESI collaboration meeting in December 2024.

\textbf{Our main findings can be summarized as follows:}

\begin{itemize}

    \item We find strong agreement in clustering amplitude between DR1 and DR2 samples, with DR2 benefiting from a significant reduction in statistical scatter due to its larger volume. The improvement in precision is particularly pronounced for the bright galaxy (BGS) and emission line galaxy (ELG) samples, as shown in \cref{fig:two-function}. This translates to a factor of 2.15 improvement in precision compared to DR1 (\cref{fig:bao_precision}), where the combined BAO precision has improved from $\sim0.52\%$ in DR1 to $\sim0.24\%$ in DR2. Furthermore, DESI DR2 achieves a factor of 2.56 improvement over previous SDSS BAO measurements, which had a precision of $\sim0.62\%$, highlighting the substantial advancement in precision enabled by DESI’s expanded dataset.
    
    \item The \desidrtwo BAO constraints are robust across multiple variations in analysis choices, including different data vectors (correlation function vs. power spectrum, 1D vs. 2D BAO fit), modeling (different broadband parameterizations and parameter priors), and treatment of observational systematics. We tested the response of the constraints across multiple data splits, including redshift bins, sky regions, and magnitude cuts, consistently finding agreement with the baseline (see \cref{fig:y3unblindedwhisker}).

    \item Density-field reconstruction significantly enhances the precision of BAO constraints for all DESI tracers (\cref{fig:scatteralpha}). The uncertainty on the isotropic (anisotropic) BAO scaling parameter $\alphaiso$ ($\alphaap$) is reduced by up to 42\% (47\%) compared to results using unreconstructed catalogs. Compared to DR1, reconstruction efficiency has improved for almost all tracers, likely due to smaller boundary effects and higher survey completeness in DR2, which allow for better displacement field estimation. This level of improvement aligns with expectations from mock catalog analyses and further demonstrates the effectiveness of reconstruction in reducing noise and enhancing the BAO signal.

    \item The posterior distribution of BAO scaling parameters is well approximated by a Gaussian for all tracers (\cref{fig:gaussianity}), enabling the posterior to be fully characterized by a mean and covariance matrix for subsequent cosmological inference.

\item {BAO constraints remain stable under different fiducial cosmologies.} We tested the impact of our fiducial cosmology assumptions by reanalyzing BAO measurements using different cosmologies, including a low-$\Omega_m$ model, a thawing dark energy model, a high-$N_{\rm eff}$ scenario, and a low-$\sigma_8$ cosmology (\cref{fig:fidcosmo}). Across all tracers, deviations in BAO constraints remain within 0.4$\sigma$, confirming that our measurements are not significantly biased by fiducial cosmology choices. The largest variations appear in the \bgs and \qso samples, but even these remain well within statistical uncertainties.

    \item {DESI DR2 provides a significantly improved BAO measurement compared to SDSS.} We examined the reported tension between DESI DR1 and SDSS BAO measurements at $z\sim0.7$, which was previously found to be at the $2.5\sigma$–$3\sigma$ level. Using the DESI pipeline to reanalyze the eBOSS LRG sample, we found that the reanalysis closely matches the published SDSS results, confirming that differences in pipeline methodology were not responsible for the tension. As shown in \cref{fig:sdss_vs_desi}, the DESI DR2 result lies between DR1 and SDSS, supporting the interpretation from the DR1 analysis that the tension was likely due to statistical fluctuations rather than a systematic discrepancy.

    \item {Correlated systematics do not significantly impact the BAO results.} As a post-unblinding test, we investigated the potential impact of correlated systematics across redshift bins and tracers. We considered two extreme scenarios: one where theoretical systematics were fully correlated across bins, and another where they were only partially correlated (\cref{fig:hubble_corrsys}). In both cases, the deviations from the baseline analysis are minimal.
\end{itemize}

The validated DESI DR2 BAO dataset represents a major step forward in precision cosmology. Compared to DR1, it provides more stringent constraints, forming a robust foundation for upcoming studies on dark energy, modified gravity, and the Universe’s expansion history. Future work will extend this analysis by refining systematics control and testing alternative modeling approaches to further enhance the precision and reliability of DESI’s BAO measurements. These results mark a crucial milestone for DESI’s mission, setting the stage for even more precise constraints with future data releases and strengthening the role of BAO as a cornerstone of modern cosmology.

\section{Data Availability}
The data used in this analysis will be made public along the Data Release 2 (details in \url{https://data.desi.lbl.gov/doc/releases/}).

\acknowledgments

UA acknowledges support from the Leinweber Center for Theoretical Physics at the University of Michigan Postdoctoral Research Fellowship and DOE grant DE-FG02-95ER40899. SN acknowledges support from an STFC Ernest Rutherford Fellowship, grant reference ST/T005009/2. H-JS acknowledge support from the U.S. Department of Energy, Office of Science, Office of High Energy Physics under grant No. DE-SC0023241.

This material is based upon work supported by the U.S.\ Department of Energy (DOE), Office of Science, Office of High-Energy Physics, under Contract No.\ DE–AC02–05CH11231, and by the National Energy Research Scientific Computing Center, a DOE Office of Science User Facility under the same contract. Additional support for DESI was provided by the U.S. National Science Foundation (NSF), Division of Astronomical Sciences under Contract No.\ AST-0950945 to the NSF National Optical-Infrared Astronomy Research Laboratory; the Science and Technology Facilities Council of the United Kingdom; the Gordon and Betty Moore Foundation; the Heising-Simons Foundation; the French Alternative Energies and Atomic Energy Commission (CEA); the National Council of Humanities, Science and Technology of Mexico (CONAHCYT); the Ministry of Science and Innovation of Spain (MICINN), and by the DESI Member Institutions: \url{https://www.desi. lbl.gov/collaborating-institutions}.

Any opinions, findings, and conclusions or recommendations expressed in this material are those of the author(s) and do not necessarily reflect the views of the U.S.\ National Science Foundation, the U.S.\ Department of Energy, or any of the listed funding agencies.

The authors are honored to be permitted to conduct scientific research on Iolkam Du’ag (Kitt Peak), a mountain with particular significance to the Tohono O’odham Nation.



\bibliographystyle{mod-apsrev4-2} 
\bibliography{Y1KP7a_references, DESI_supporting_papers, references}


\appendix

\section{BAO measurements in Fourier space}\label{app:fourier}

Parallel to the analysis in CS, we also performed the analysis in FS. In this appendix, we highlight the main results obtained in FS compared to the ones we obtained in the fiducial CS analysis.

\begin{figure*}
    \centering
    \begin{tabular}{cc}
        \includegraphics[width=0.4\textwidth]{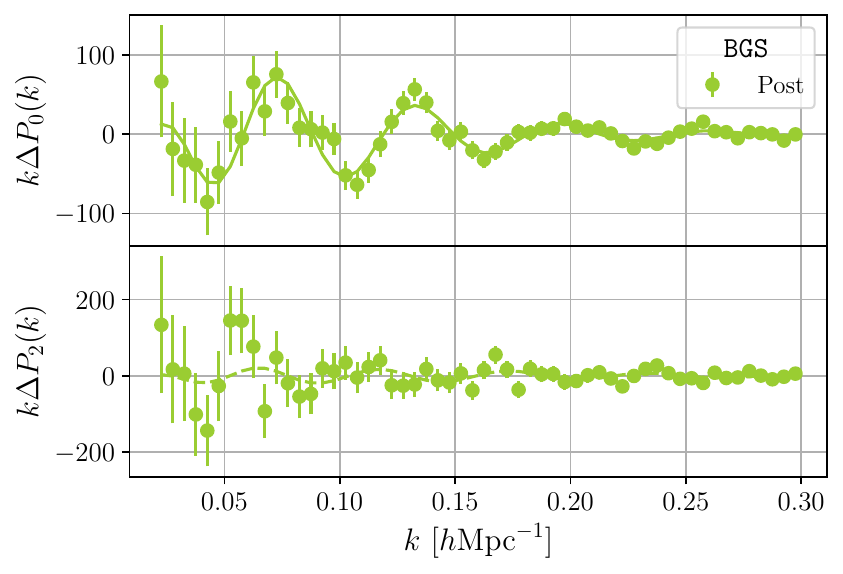} & 
        \includegraphics[width=0.4\textwidth]{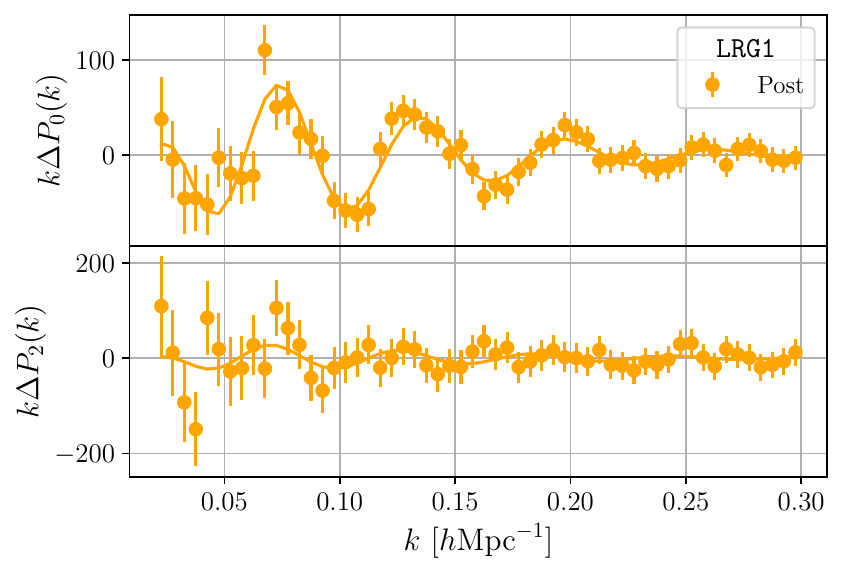} \\ 
        \includegraphics[width=0.4\textwidth]{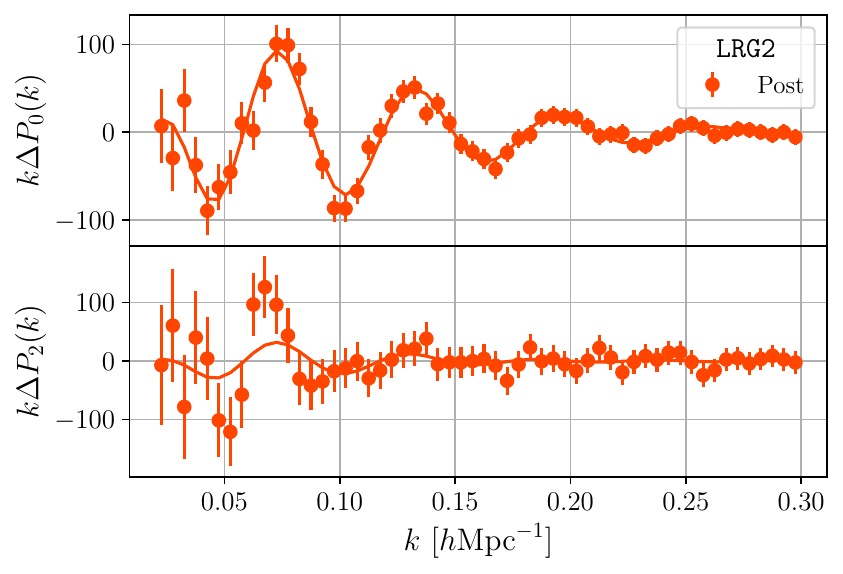} &
        \includegraphics[width=0.4\textwidth]{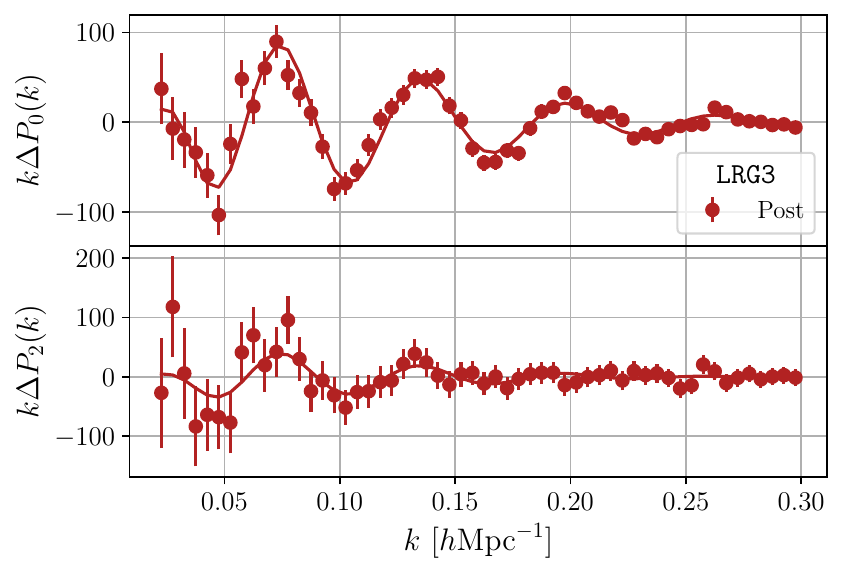} \\
        \includegraphics[width=0.4\textwidth]{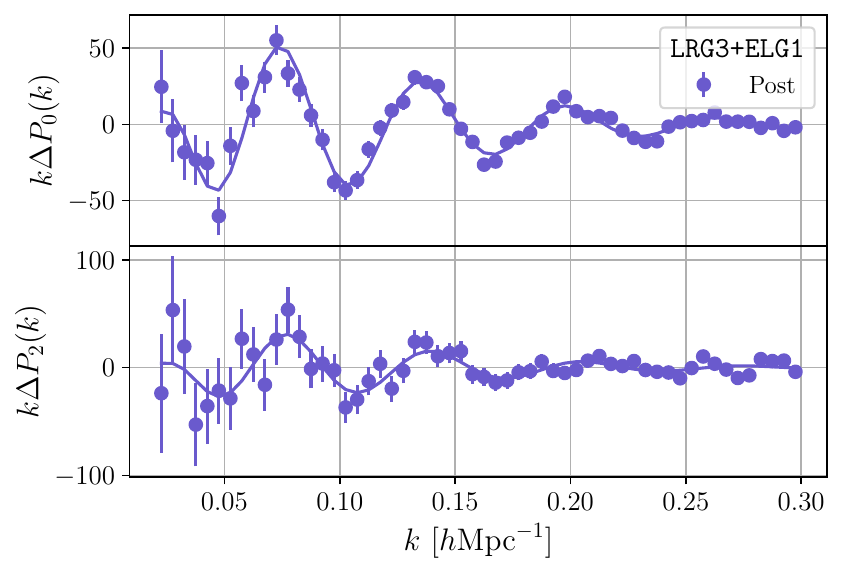} & 
        \includegraphics[width=0.4\textwidth]{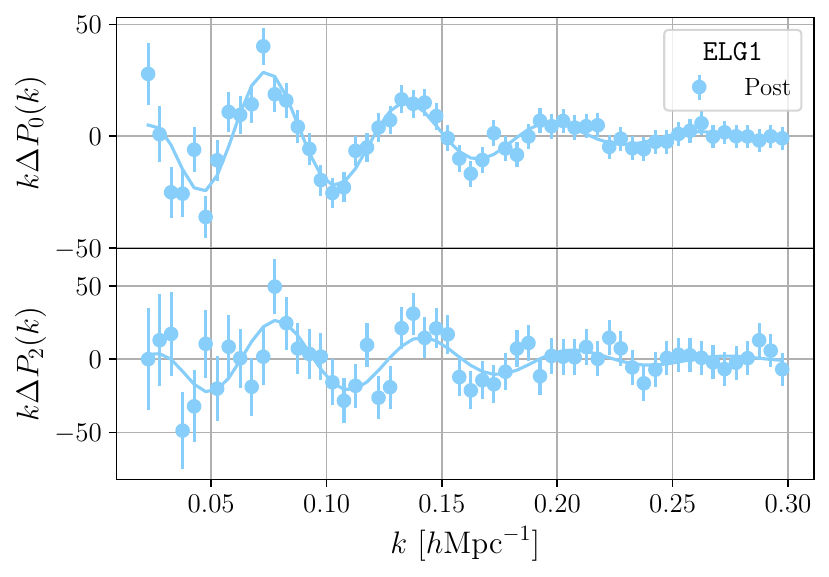} \\
        \includegraphics[width=0.4\textwidth]{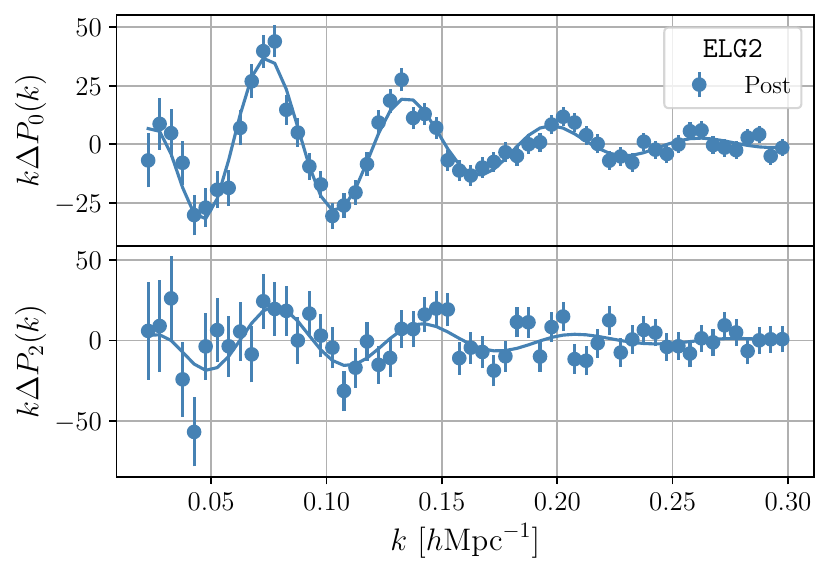} &
        \includegraphics[width=0.4\textwidth]{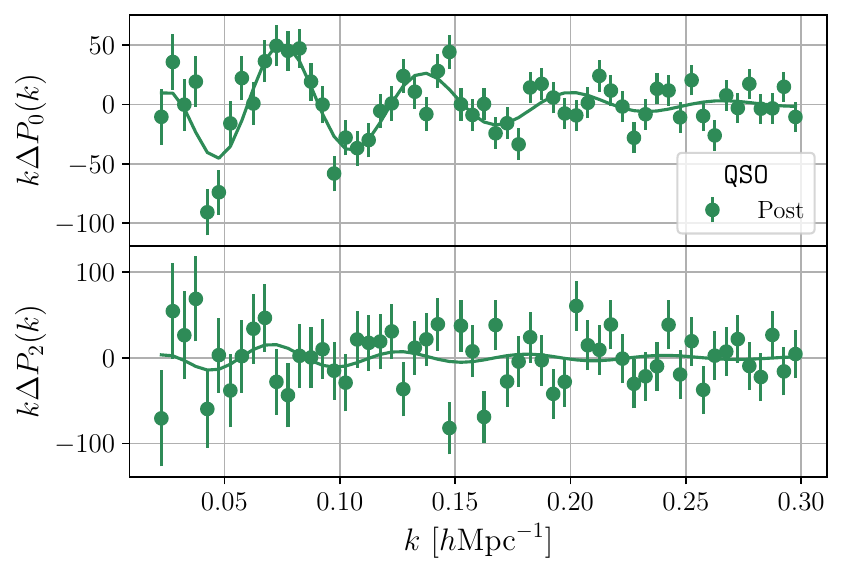} \\ 
    \end{tabular}

\caption{The isolated BAO feature in the monopole and quadrupole moments of the DESI DR2 power spectrum for different tracers. The filled circles represent the post-reconstruction measurements, while the solid lines correspond to the best-fit BAO model. The upper and lower subpanels in each plot display the monopole and quadrupole moments, respectively. For \bgs, the baseline fit is a 1D fit and we show the best-fit model for the quadrupole as a dashed line instead of a solid one. All other tracers use a 2D fit that incorporates both the monopole and quadrupole moments as the baseline settings. Error bars denote 68\% confidence intervals. The y-axis units of $h^{-3}\mathrm{Mpc}^3$ are omitted for clarity.}

    \label{fig:clustering_measurements}
\end{figure*}

In \cref{fig:clustering_measurements}, we show the isolated BAO feature in the multipole moments of the DESI DR2 power spectra of galaxies and quasars. The upper subpanels display the monopole, while the lower subpanels show the quadrupole. The best-fit model is shown as solid lines except for the quadrupole of the \bgs tracer, since for this tracer the baseline fit is a 1D fit and we show it with a dashed line.

After measuring the power spectrum for the different tracers, we proceeded to run the BAO fits. The settings we used are displayed in \cref{tab:prior}. 

\begin{figure*}
    \includegraphics[width=\textwidth]{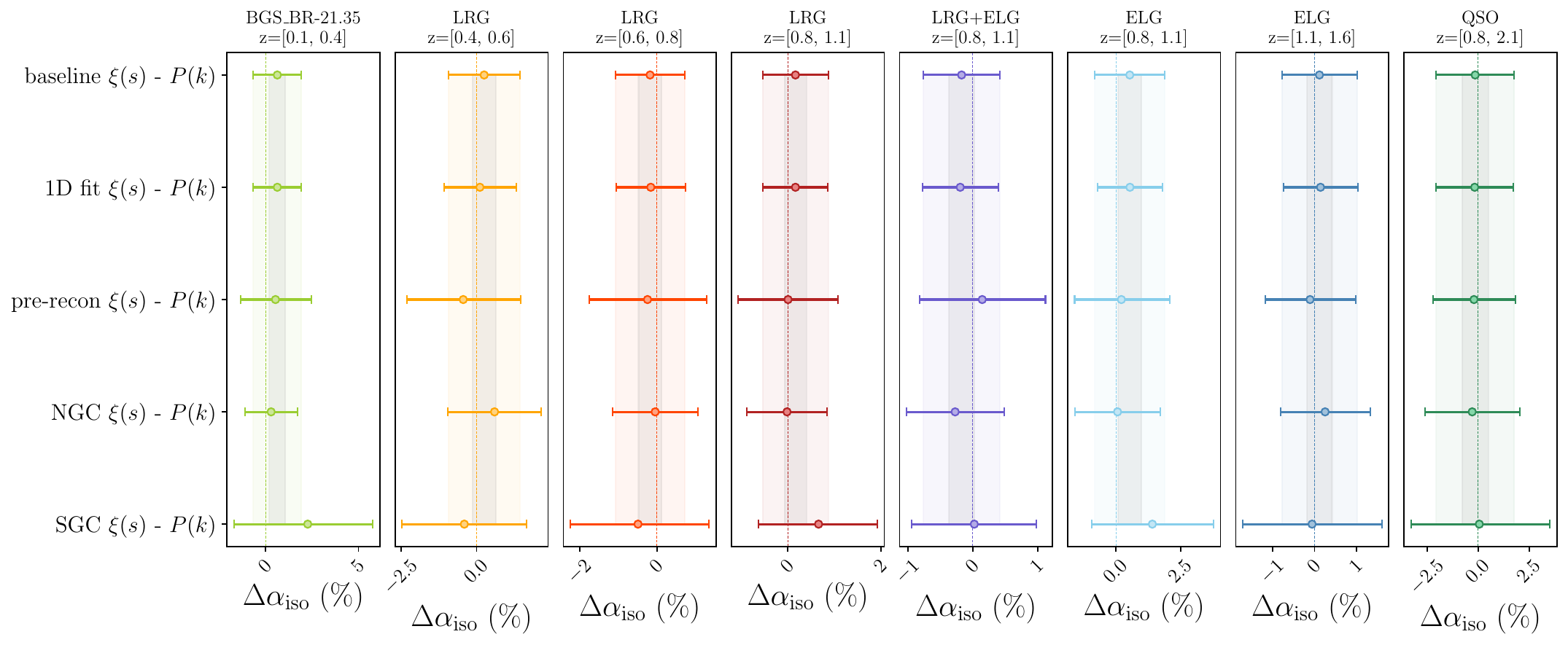}
    \includegraphics[width=\textwidth]{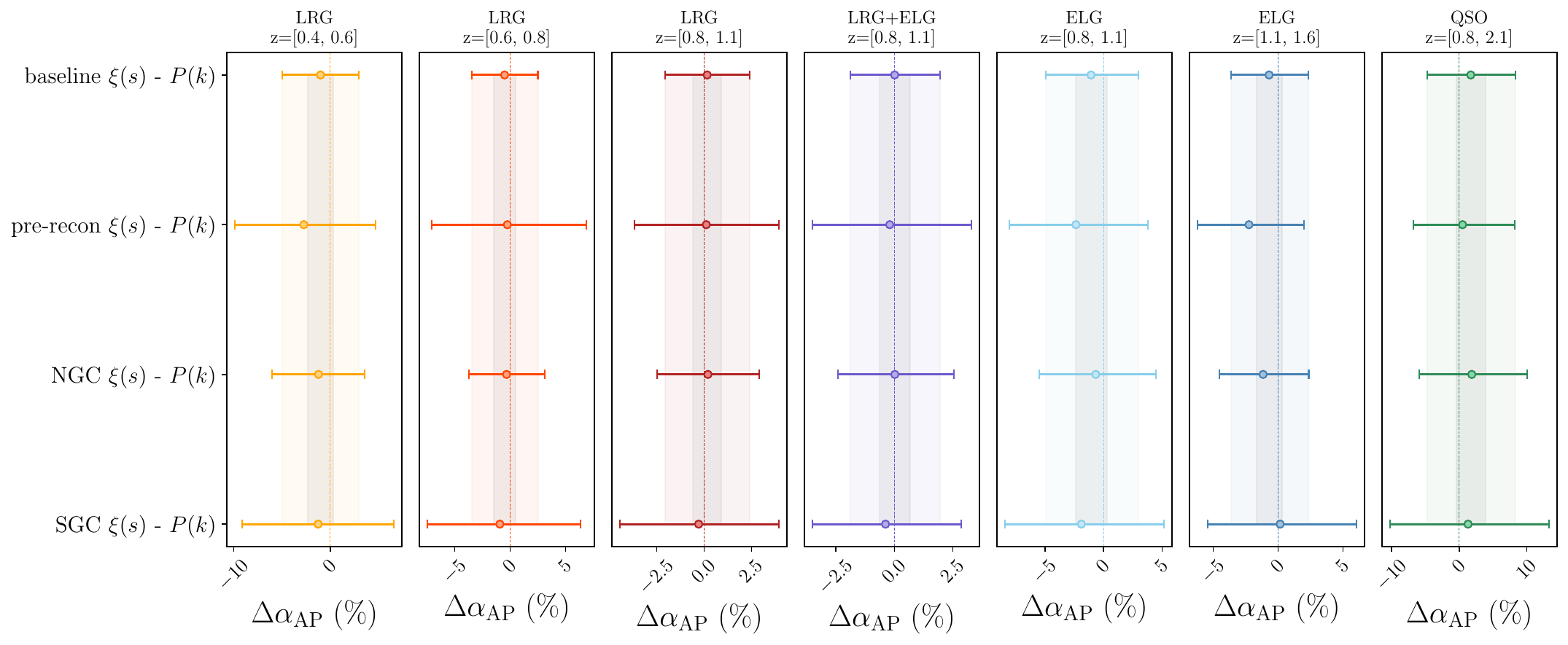}
    \caption{Pairwise comparison of the correlation function and the power spectrum BAO fit results for the unblinded data. In the panel on top, we show differences in $\alphaiso$, and in the one on the bottom, $\alphaap$. We note that the baseline case is a 1D BAO fit for \bgs and a 2D fit for the others (which is the reason why \bgs is not shown in the panel on the bottom). The colored-shaded areas show the $1\sigma$ regions of the baseline case, and the grey-shaded ones show one-third of $1\sigma$.}
    \label{fig:fourier-space}
\end{figure*}

In \cref{fig:fourier-space} we show the pairwise comparison of the CS and the FS BAO-fit results for the unblinded data. In the panel on top, we show differences in $\alphaiso$, and in the one on the bottom, $\alphaap$. We note that the baseline case is a 1D BAO fit for \bgs and a 2D fit for the others (which is the reason why \bgs is not shown in the panel on the bottom). The colored-shaded areas show the $1\sigma$ regions of the baseline case, and the grey-shaded ones show one-third of $1\sigma$. We see that CS and FS results are in very good agreement: the dashed lines at $\Delta\alpha=0$ lie within the grey-shaded areas for most tracers; and the different cases (1D fit, pre-reconstruction, NGC and SGC) are typically consistent with zero (SGC is usually the case that differs the most due to its lower constraining power). The largest differences between CS and FS are for the \bgs tracer, but they are still consistent with zero at the $1\sigma$ level (besides the SGC case, which is slightly outside the $1\sigma$ area of the baseline case).

\begin{figure*}
    \centering
    \begin{tabular}{ccc}
        \includegraphics[width=0.32\textwidth]{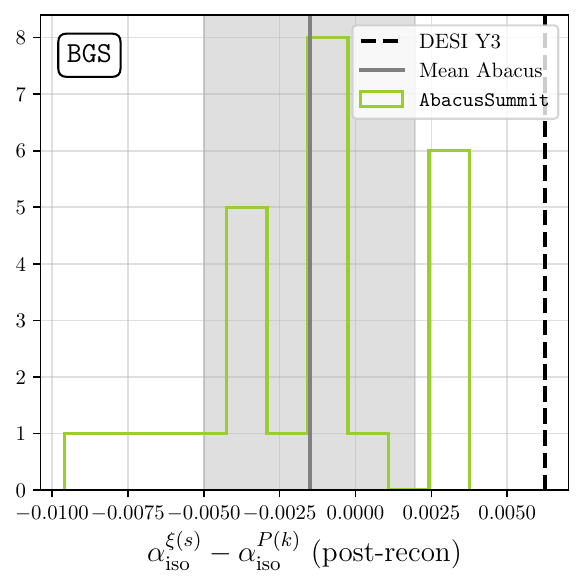} & 
        \includegraphics[width=0.32\textwidth]{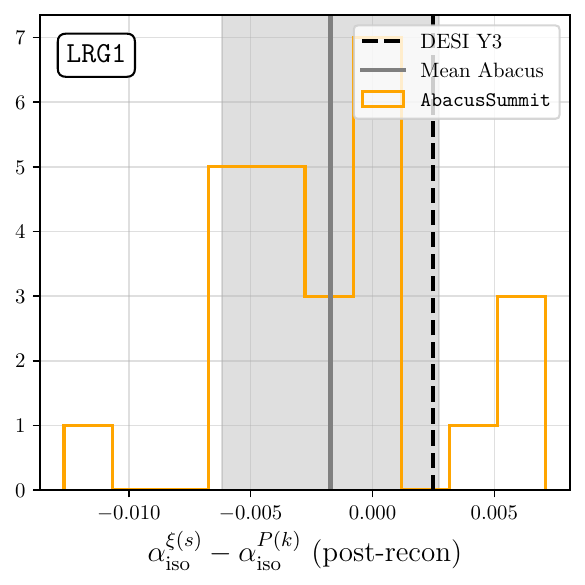} & 
        \includegraphics[width=0.32\textwidth]{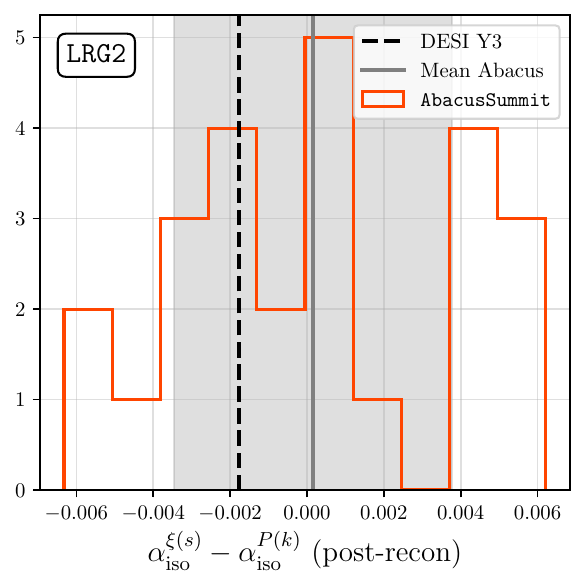} \\
        \includegraphics[width=0.32\textwidth]{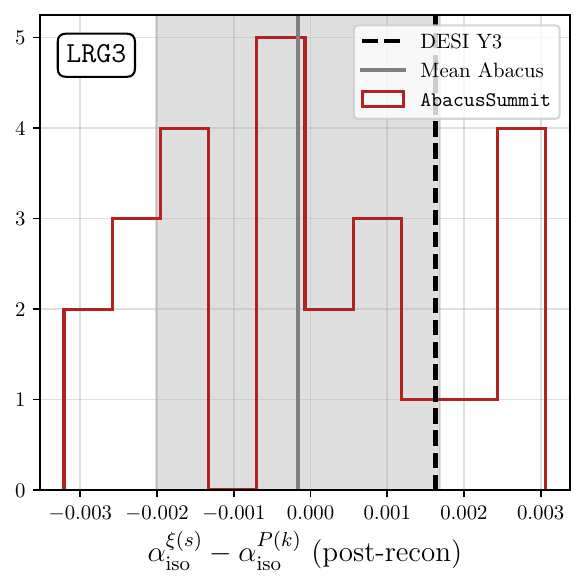} &
        \includegraphics[width=0.32\textwidth]{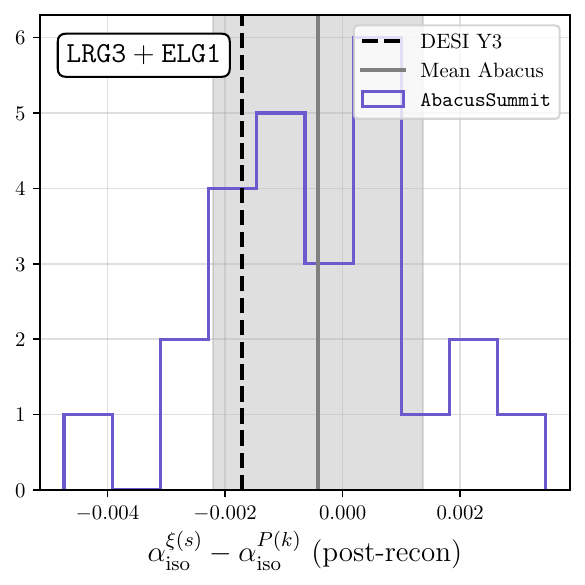} & 
        \includegraphics[width=0.32\textwidth]{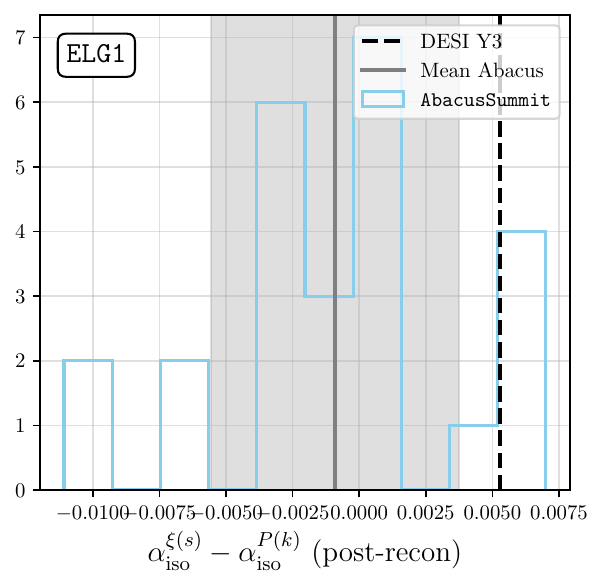} \\
        \includegraphics[width=0.32\textwidth]{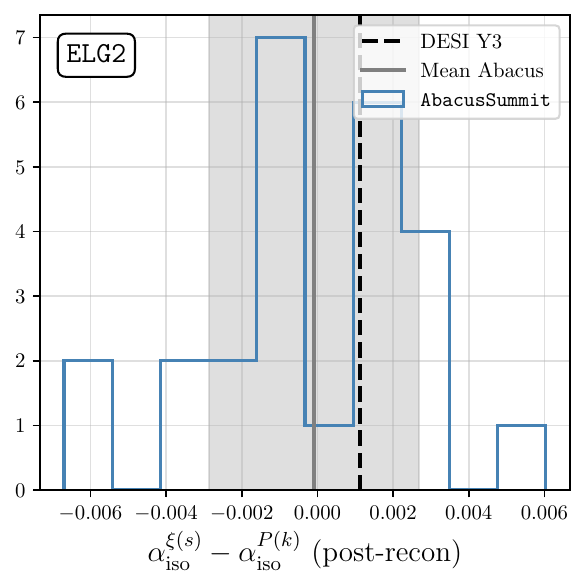} &
        \includegraphics[width=0.32\textwidth]{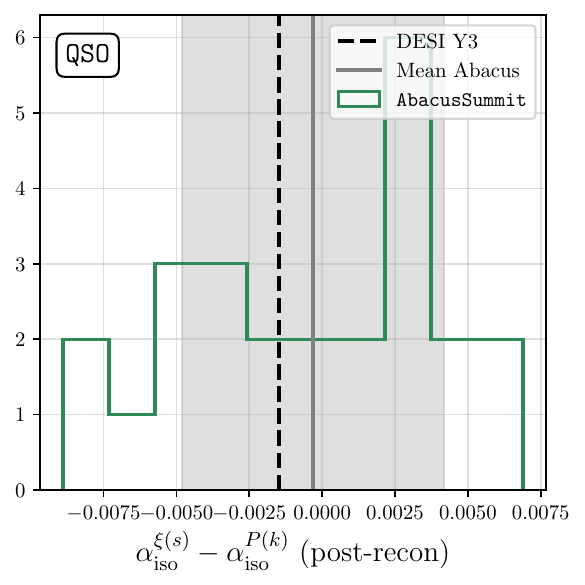} & 
    \end{tabular}
    \caption{The different panels show the distribution of $\Delta\alphaiso$ between CS and FS in the mocks (histograms) and the results obtained on the unblinded DESI DR2 data (dashed black lines). The shaded grey areas show the $1\sigma$ region around the mean of the mocks, and the center is highlighted with a grey line. The first panel shows the results of the 1D BAO fit on the \bgs tracer, whereas we show the results of the 2D fit for all the other tracers.
    }
    \label{fig:scatter_pk_vs_xi}
\end{figure*}

We also ran the BAO fits on the \abacussecond DR2 mocks. This allowed us to check whether the differences between CS and FS found in the unblinded data (which we already showed in \cref{fig:fourier-space}) were consistent with those of our simulations. In \cref{fig:scatter_pk_vs_xi} we show the differences in $\alphaiso$ between CS and FS for the different tracers, for both the mocks and unblinded data. We see that the results in the data (dashed black lines) are usually well within the distribution of the mocks (histograms). {However, this is not the case for \bgs, for which the unblinded data shows a 0.55\% offset, which is a bit larger than the most extreme case from the mocks.} An offset of such magnitude is also observed in other tracers with higher signal-to-noise, such as \elgo. The \bgs mock distribution is somewhat asymmetric, with an extended tail on the left, making such a deviation plausible, especially given the limited number of mocks available for this test. In absolute value, two mocks have larger $\Delta\alphaiso$ than the unblinded data, i.e., two mocks in the left-hand side of the tail of the distribution for \bgs in \cref{fig:scatter_pk_vs_xi} (first panel) have a more extreme value than the one found on the data (the dashed black line). Considering that the offset is within an acceptable range and does not exceed $3\sigma$, we conclude that the \bgs baseline CS BAO fit remains valid.

\section{Assessing 1D vs. 2D BAO Fits for BGS, ELG, and QSO} \label{app:1D_vs_2D}

Baryon Acoustic Oscillation (BAO) measurements can be performed using either one-dimensional (1D) or two-dimensional (2D) fits. A 1D fit constrains only the isotropic BAO scaling parameter, $\alpha_{\rm iso}$, while a 2D fit simultaneously constrains both the isotropic and anisotropic scaling parameters, $\alpha_{\rm iso}$ and $\alpha_{\rm AP}$. While 2D fits provide additional cosmological information by capturing anisotropic distortions, they require sufficiently high signal-to-noise (S/N) and stable posterior distributions to ensure robust results.

In DESI DR1, the BAO fits for \bgs, \elgo, and \qso were performed using 1D fits only, as the statistical precision of the data was not sufficient to justify a 2D fit. With DESI DR2, the increased survey volume and improved statistical precision have enhanced the feasibility of 2D BAO fits for these tracers. Initial mock-based tests suggested that the DR2 dataset has sufficient S/N to perform 2D BAO fits for these samples. However, a detailed examination of the fits was required to ensure their robustness. In DESI DR2, the choice between 1D and 2D fits was evaluated using a combination of mock-based tests and validation metrics on the blinded data.

\subsection{Criteria for Selecting 1D vs. 2D BAO Fits} 

The choice between 1D and 2D fits was guided by the following considerations:
\begin{itemize} 
\item The statistical precision of the sample: A higher S/N allows for stable 2D fits, while a low S/N may introduce bias and unstable parameter estimates.
\item The Gaussianity of the posterior distributions, particularly for the anisotropic parameter $\alpha_{\rm AP}$, avoiding complications of including a weak non-Gaussian constraint on $\alpha_{\rm AP}$. 
\item The correlation between $\alpha_{\rm iso}$ and $\alpha_{\rm AP}$ in the 2D fit should match mock expectations and align with theoretical predictions for optimal BAO isolation (Seo \& Eisenstein).

\item The distribution of $\chi^2$ values when compared to mock realizations. 
\end{itemize}

\subsection{Evaluation of 2D Fits in DR2} 

Tests performed on the \abacussecond DR2 mocks suggested that we have sufficient S/N to apply 2D BAO fits to \bgs, \elgo, and \qso. The improvements in DR2 relative to DR1 suggested that 2D fits could be feasible for these tracers. However, a closer examination of the results led to a more nuanced decision:\\

\paragraph*{BGS:}  

For \bgs, the 2D fit is borderline acceptable—it formally passes validation tests but raises concerns:  

\begin{itemize}

    \item {Non-Gaussianity in $\alpha_{\rm AP}$:} The posterior distribution for $\alpha_{\rm AP}$ exhibits elongation and deviations from Gaussianity, suggesting a more complex likelihood structure that may not be well-approximated by a simple Gaussian assumption.
    
    \item {High $\alpha_{\rm iso}$-$\alpha_{\rm AP}$ Correlation:} The blinded data shows $r = 0.48$, unusually high compared to other tracers. In the $\alpha_\parallel$-$\alpha_\perp$ basis, it lies slightly outside the expected mock distribution ($r = -0.49$ vs. $-0.42$ to $-0.47$). Theoretical predictions \cite{2007ApJ...665...14S} suggest an optimal BAO isolation should yield a correlation value of $r \sim -0.41$, which holds for other tracers but not for \bgs.
\end{itemize}  

At low redshifts ($z < 0.4$), 2D BAO fits are inherently challenging due to:  

\begin{itemize}
    \item {Weak Alcock-Paczynski (AP) distortions:} The AP effect—driven by $D_H/D_M$—is weaker at low $z$, reducing sensitivity to cosmology.
    \item {Lower signal-to-noise (S/N) in anisotropic clustering:} RSD remain strong, but distinguishing AP effects from RSD and statistical noise is harder.
    \item {Limited number density:} The lower \bgs number density further reduces the precision of $\alpha_{\rm AP}$ measurements.
\end{itemize}  

Given these factors, the 1D BAO fit—which measures only $\alpha_{\rm iso}$—is a more stable and robust choice for \bgs. While a higher {S/N} dataset could, in principle, constrain the AP effect, the {limited statistical power and elevated $\alpha_{\rm iso}$-$\alpha_{\rm AP}$ correlation} make the 2D fit less reliable.  

Although the \bgs 2D fit is reasonable and passes most consistency tests, minor deviations and potential {non-Gaussian likelihood effects} led us to conservatively adopt the 1D fit as our default. {Importantly, both the 1D and 2D fits yield consistent results, reinforcing the robustness of the BAO measurement for \bgs.}

\paragraph*{ELG1:} 
The 2D fits for \elgo show hints of similar behavior as \bgs, particularly in the increased correlation between $\alpha_{\rm iso}$ and $\alpha_{\rm AP}$. However, the differences are less pronounced, and the fit remains within the expected distribution of mocks. Since \elgo is not used for cosmological inference (we use the combined \lrgelg sample instead), we do not need to make a strict decision. We provide both 1D and 2D fits for \elgo, but only use the \lrgelg combination for final cosmological constraints.

\paragraph*{QSO:} 
The 2D fits for \qso pass all validation tests, including consistency with mocks, Gaussianity of posteriors, and robustness against data splits. The improvement in S/N at high redshift ensures that $\alpha_{\rm AP}$ is well-constrained, making a 2D fit the preferred choice. Therefore, we use the 2D fit as the default for \qso.

\section{Consistency between \desidrone and DR2 \abacussecond mocks}\label{app:abacus_DR1_vs_DR2}

\begin{figure*}
    \centering
    \begin{tabular}{ccc}
        \includegraphics[width=0.32\textwidth]{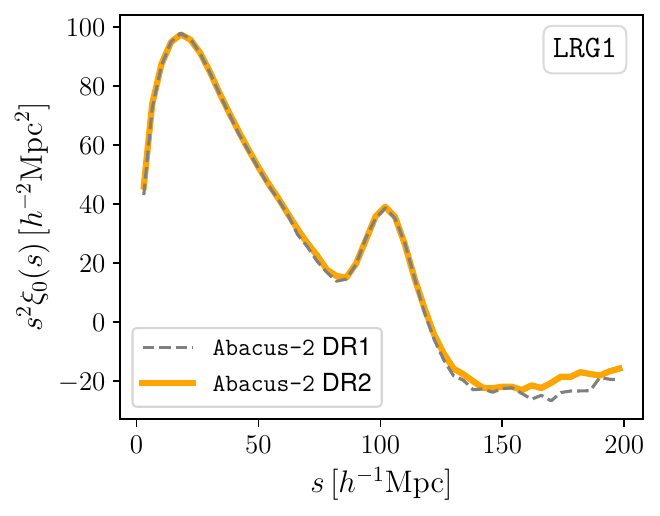} & 
        \includegraphics[width=0.32\textwidth]{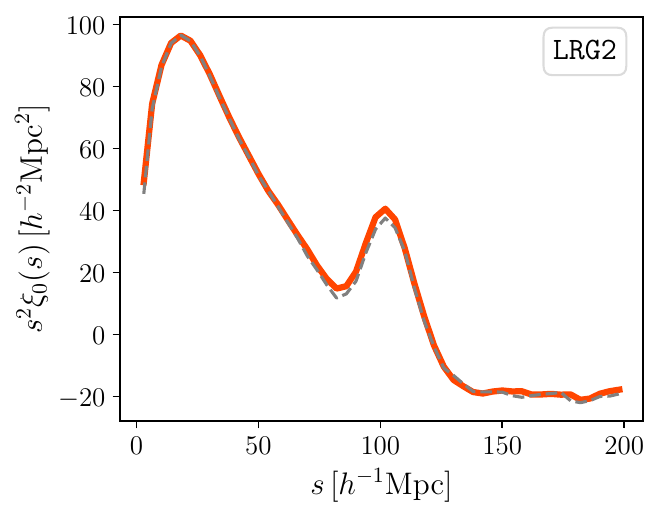} & 
        \includegraphics[width=0.32\textwidth]{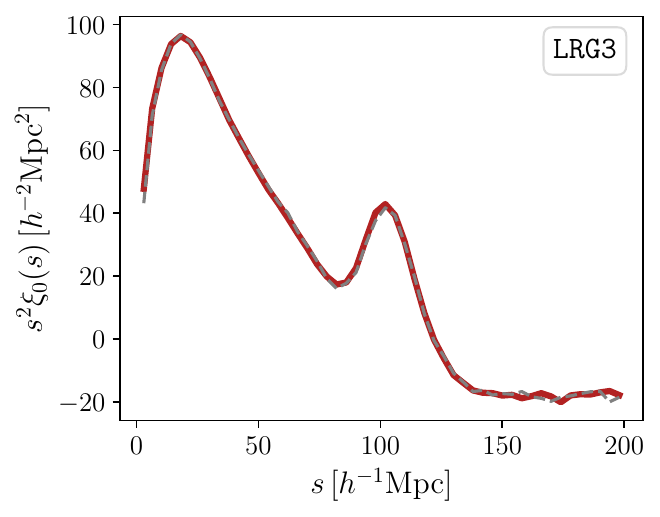} \\
        \includegraphics[width=0.32\textwidth]{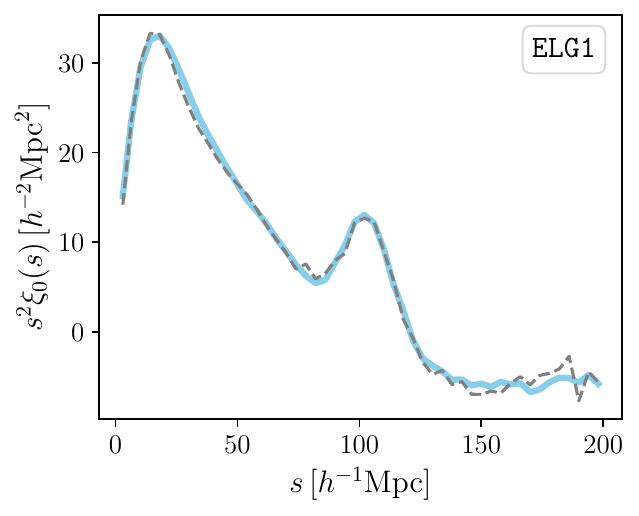} & 
        \includegraphics[width=0.32\textwidth]{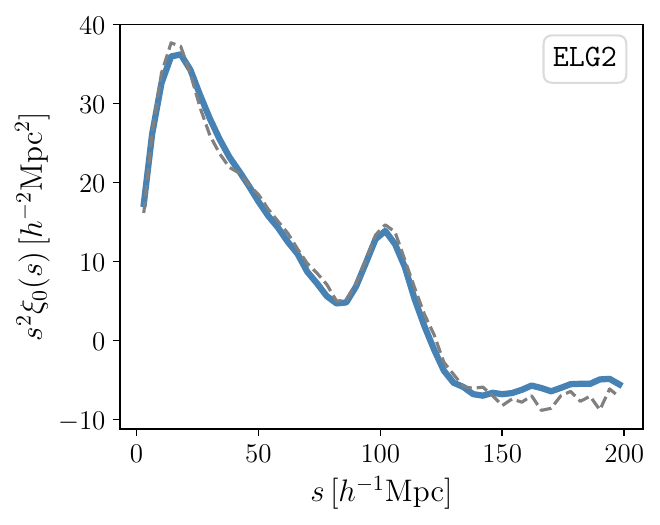} & 
        \includegraphics[width=0.32\textwidth]{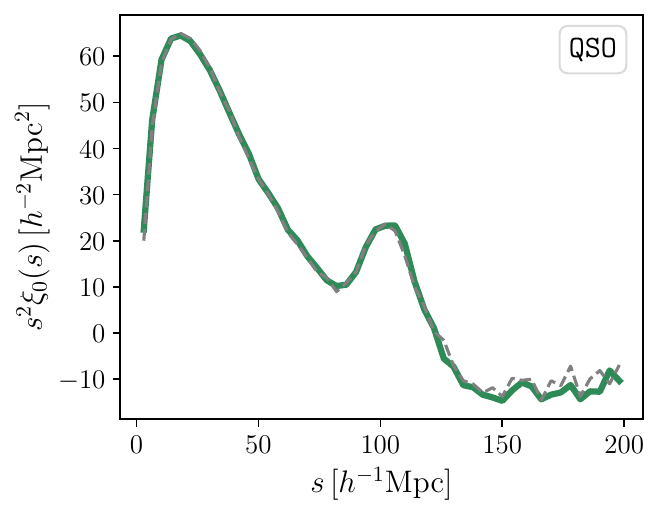} \\
    \end{tabular}
    \caption{Comparison of the monopole of the two-point correlation function, $\xi_0(s)$, between the \abacussecond\ \desidrone mocks (gray) and the \abacussecond\ \desidrtwo mocks (colored) for various tracer samples and redshift ranges (we show the average over the 25 mocks). The panels show results for \lrg ($0.4 < z < 0.6$, $0.6 < z < 0.8$, $0.8 < z < 1.1$), \elg ($0.8 < z < 1.1$, $1.1 < z < 1.6$), and \qso ($0.8 < z < 2.1$).
    }
    \label{fig:abacus_DR1_vs_DR2}
\end{figure*}

In \cref{fig:two-function} we showed the measurements of the two-point correlation functions of the DESI DR2 compared to those of the DESI DR1. In this appendix, we show analogous results but for the \abacussecond mocks. In \cref{fig:abacus_DR1_vs_DR2} we show the monopole of the two-point correlation function measured in the \abacussecond mocks (averaged over the 25 mocks) for the different tracers and redshift bins. Solid-colored lines show the measurements on the \abacussecond DR2 mocks, whereas the shaded-gray ones show DR1. We see that the results are consistent between the two \abacussecond datasets.

\end{document}